\documentclass[]{elsarticle}
\usepackage{lineno,hyperref}
\modulolinenumbers[5]
\makeatletter
\def\ps@pprintTitle{%
 \let\@oddhead\@empty
 \let\@evenhead\@empty
 \def\@oddfoot{}%
 \let\@evenfoot\@oddfoot}
\makeatother
\usepackage{epstopdf}
\usepackage[a4paper, total={6.4in, 9.4in}]{geometry}
\usepackage[a4paper]{geometry}
 
\usepackage{color}
\usepackage[x11names]{xcolor}
\usepackage[utf8]{inputenc}
\usepackage[english]{babel}
\usepackage{amsmath}
\usepackage{pgfplots}
\usepackage{tabularx,ragged2e,booktabs,caption}

\usepackage[labelfont=bf]{caption}
\usepackage[font=footnotesize]{caption}
\usepackage{amssymb,latexsym}
\usepackage{graphicx}
\graphicspath{{./images/}}
\usepackage{booktabs}
\usepackage{amsfonts}
\usepackage{tikz}
\usepackage{array}
\usepackage{mathtools}
 \usepackage{relsize}
\usepackage[T1]{fontenc}
\usepackage{float}
\usepackage{cases}
\usepackage{amsthm}
\usepackage{subcaption}
\pgfplotsset{compat=newest}
\usepackage[toc,page]{appendix}
\newcolumntype{Y}{>{\centering\arraybackslash}X}

\numberwithin{equation}{section}
\usepackage{tabularx,ragged2e,booktabs,caption}

\usepackage[toc]{appendix}
\DeclareMathAlphabet{\mathpzc}{OT1}{pzc}{m}{it}
\newcommand{\lnorm}{\left|\left|}

\newcommand{\rnorm}{\right|\right|}

\newcommand{\Ld}{L^2(\mathcal{D})}
\newcommand{\Lomd}{L^2(\Omega,\mathcal{D})}
\newcommand{\myreferences}{references}
\begin{document}

\begin{frontmatter}
\title{Stochastic turbulence modeling in RANS simulations via Multilevel Monte Carlo}
\address[cwi]{CWI - Centrum Wiskunde \& Informatica, Amsterdam, The Netherlands}
\address[delft]{Faculty of Aerospace Engineering, Delft University of Technology, Delft, The Netherlands.}
\author[cwi,delft]{Prashant~Kumar}
\ead{pkumar@cwi.nl}
\author[delft]{Martin~Schmelzer}
\ead{ m.schmelzer@tudelft.nl}
\author[delft]{Richard~P.~Dwight}
\ead{r.p.dwight@tudelft.nl}

\begin{abstract}
A multilevel Monte Carlo (MLMC) method for quantifying model-form uncertainties associated with the Reynolds-Averaged Navier-Stokes (RANS) simulations is presented. Two, high-dimensional, stochastic extensions of the RANS equations are considered to demonstrate the applicability of the MLMC method. The first approach is based on global perturbation of the baseline eddy viscosity field using a lognormal random field.  A more general second extension is considered based on the work of [\emph{Xiao et al.}(2017)], where the entire Reynolds Stress Tensor (RST) is perturbed while maintaining realizability. For two fundamental flows, we show that the MLMC method based on a hierarchy of meshes is asymptotically faster than plain Monte Carlo. Additionally, we demonstrate that for some flows an optimal multilevel estimator can be obtained for which the cost scales with the same order as a single CFD solve on the finest grid level. 

\end{abstract}
\begin{keyword}
MLMC, RANS, UQ, Random eddy viscosity, Random Reynolds stress tensor
%\texttt{elsarticle.cls}\sep \LaTeX\sep Elsevier \sep template
%\MSC[2010] 00-01\sep  99-00
\end{keyword}

\end{frontmatter}

%\linenumbers
\section{Introduction}
The Reynolds-Averaged Navier-Stokes (RANS) equations combined with turbulence closure models are widely utilized in engineering to predict flows with high Reynolds number. These turbulence closure models are used to obtain an approximate Reynolds stress tensor that is responsible for coupling the mean flow with turbulence. Although many turbulence models exist in the literature, there is no single model that generalizes well to all classes of turbulent flows \cite {pope,ZhaoWei}. Specifically, the performance depends on the modeling assumptions and the type of flow used to calibrate the so-called closure coefficients that are needed as inputs to a turbulence model. 

Since the dominant source of error in the flow prediction comes from the turbulence modeling, a number of approaches have already been developed for the model-form uncertainty quantification (UQ) of RANS simulations, see e.g. \cite{Xiao2018,Duraisamy2018} for recent reviews. The majority of these approaches are based on the perturbation of baseline RANS models. One way to achieve this is by injecting uncertainties in the closure coefficients \cite{Margheri2014,edeling1,edeling2,cheung1} of turbulence models. Another more general physics-based approaches exists, which typically introduces randomness directly into the modeled Reynolds Stress Tensor (RST), either by perturbing its eigenvalues \cite{emory1,emory2,Gorle2013}, tensor invariants \cite{ling, xiao1} or the entire RST field \cite{xiao2}. One can also classify these stochastic models in terms of global and local perturbation (in space). For global approaches, such as in \cite{Margheri2014,edeling1,edeling2,emory2}, the magnitude of perturbations in closure coefficients, eigenvalues of RST, etc. is the same throughout the flow domain. This translates to a low-dimensional UQ problem which can be efficiently handled by deterministic sampling techniques like stochastic collocation or just by simulating flows for limiting states to obtain a prediction interval. Since the error in closure models is not same everywhere, global methods fail to capture the truth in general. On the other hand, local perturbation approaches may be effective due to high-dimensional parameterizations of uncertainties. Some local methods already exist, such as the spatially varying marker functions proposed in \cite{emory1,Gorle2014} or Gaussian random fields \cite{xiao1,xiao2,Wang} as a measure of local variation of the uncertainty. The main bottleneck hampering the development of these local models is the large cost of a forward uncertainty propagation stage. 

The prime objective of this work is to provide a framework for developing a new class of high-dimensional stochastic RANS closures, that were until recently not viable (due to the cost of the propagation), but will be if the work required is within a constant, small factor of the cost of the fine-grid solution procedure. We achieve this using the multilevel Monte Carlo (MLMC) method \cite{MLMC1,ANU:9672986}. In previous works, the potential of the MLMC method has already been demonstrated in context of the inviscid compressible flow in \cite{PISARONI201720} for propagating lower-dimensional geometric and operational uncertainties. In the current work, we use two local stochastic models based on a random eddy viscosity and a random Reynolds stress tensor. The random eddy viscosity is obtained by perturbing the baseline eddy viscosity using Gaussian random fields with some prescribed spatial covariance. This stochastic model is applicable for the quantification of uncertainties arising due to imperfect closure constants. Similarly, the random RST is derived by perturbing the baseline RST. We utilize the algorithm proposed in \cite{xiao2} where the random RST is modeled by means of spatially correlated positive-definite random matrices. This approach is attractive as the random matrix is drawn from a set of positive-definite matrices which automatically guarantees realizable Reynolds stresses. Since, the two stochastic extensions considered are high-dimensional in their random inputs, Monte Carlo (MC) type methods are favorable owing to their dimension-independent convergence. For many UQ problems in fluid dynamics, the computational time and resources required to perform plain MC simulation are prohibitive. Standard MC methods may require thousands of CFD simulations on a fine computational mesh, before the statistical moments of the QoIs converge to some prescribed tolerance. The cost of the forward propagation can be drastically reduced by using the multilevel Monte Carlo method. When estimating the moments by using the MLMC method, samples on a hierarchy of grids are taken in a telescopic decomposition of the expectation. For many problems, the variance in the flow due to random inputs can largely be captured by samples on a very coarse mesh with relatively small computational effort. This coarse estimation can be further refined by adding corrections based on samples from finer meshes. These corrections although computed on finer meshes are small in magnitude, thus only a few simulations are required to gauge the additional details offered by these finer grid levels. While offering large computational speed-up over single level MC, MLMC retains all useful properties of the MC methods like high parallelization potential and integration with the complementary variance reduction techniques.

We propose a standard MLMC method for efficient forward propagation of the uncertainty which is based on a hierarchy of pre-defined grids. For the proposed MLMC estimator, we show that the asymptotic cost does not deteriorate with an increase in the uncertain dimension and is controlled by the mesh convergence properties that further depend on the quality of the mesh and the discretization scheme used. For problems with sufficiently fast decay of the numerical error, we demonstrate a cost scaling of $\mathcal{O}(\varepsilon^{-2})$ to achieve an error tolerance of $\varepsilon$. On the other hand, for problems with a slower error decay rate, we can attain an optimal MLMC estimator, in the sense that the cost grows at the same rate as the deterministic counterpart of the problem.

The other motivation of this work is to show that the considered stochastic models can serve as an accurate Bayesian prior for calibration and data-assimilation involving turbulence models. Using numerical experiments, we show that the two models are sufficiently general and can reliably bound the possible flow behavior. Furthermore, the probability distribution of the random Reynolds stresses also satisfies the maximum entropy principle, a desirable property for a good prior.
 
The paper is organized as follows. In Section \ref{deterministicRANS} we briefly introduce the deterministic RANS equations and two standard deterministic turbulence models. Stochastic RANS models based on the random eddy viscosity and the random Reynolds stress are discussed in Section \ref{stochasticRANS}. A general description of the MLMC method is provided in Section \ref{MCE} along with implementation details that include the construction of appropriate MLMC levels and the quantification of numerical and statistical errors in these estimators. In Section \ref{NE}, numerical experiments are reported based on flow over a periodic hill and fully developed turbulent flow in a square duct.

\section{Deterministic RANS models}\label{deterministicRANS}
Direct numerical simulation of turbulent flow is highly expensive due to a large range of scales. Most engineering applications do not require details of these fine spatio-temporal features but only the effect of turbulence on the mean flow. A system of mean flow equations can be derived by Reynolds averaging, which consists of decomposing the flow into mean components, defined as an average over a large time period $T$, and fluctuations,
\begin{align}\label{velDecom}
\overline{u_i} :=& \underset{T\rightarrow\infty}{\text{lim}} \frac1T\int_0^T u_i \text{d}t,\\
u_i' :=& u_i - \overline{u_i} ,\label{fluct}
\end{align}  
respectively. The quantities $\overline{u_i}$ and $u_i'$ are the mean and the fluctuating components of the instantaneous velocity $u_i$, respectively.  Substituting \eqref{fluct} into the incompressible Navier-Stokes equation and averaging, results in the mean flow equation,
\begin{equation}\label{eq:rans}
\rho (\overline{\mathbf{u}}\cdot \nabla)\overline{u}_i = -\frac{\partial \overline{p}}{\partial x_i} +\frac{\partial}{\partial x_j} \left(\overline{R}_{ij} +R_{ij} \right),\quad i,j=1,2,3.
\end{equation} 
The mean velocity vector is represented by $\overline{\mathbf{u}}=(\overline{u}_1,\overline{u}_2,\overline{u}_3)$, $\overline{p}$ is the time-averaged pressure field and $\rho$ is the (constant) density. Here $\overline{R}_{ij} =\tfrac12\rho\nu \left(\partial \overline{u}_i/\partial x_j + \partial \overline{u}_j/\partial x_i\right)$ denotes the mean stresses (tangential and normal) associated with the molecular viscosity $\nu$. The mean flow is coupled to the turbulence by Reynolds stresses $R_{ij} = \rho\overline{u_i'u_j'}$. The Reynolds stress components $R_{ij}$ appearing in \eqref{eq:rans} are unknown and are modeled using turbulence closure models that can be broadly categorized into Reynolds stress transport models and eddy viscosity models. The former models rely on an approximate set of stress transport equations to compute the Reynolds stress components. Although physically more stringent, stress transport models are not very popular in engineering practice as the discretizations of these coupled set of stress transport equations result in a numerically stiff system that is more expensive to solve. On the other hand, linear eddy viscosity models are popular as they are significantly cheaper and perform reasonably well for a broad range of flows \cite{pope}. However, they are challenged by industrially relevant flows exhibiting separation, impinging, curvature, etc. These models are based on the \emph{Boussinesq approximation} which states that the Reynolds stresses are linearly related to the mean rate-of-strain as
\begin{equation}\label{linearEV}
-\overline{u_i'u_j'}\approx \nu_t \left(\frac{\partial \overline{u}_i }{\partial x_j} + \frac{\partial \overline{u}_j }{\partial x_i} \right) - \frac{2}{3}\delta_{ij} k,
\end{equation}
where $k := \frac12\overline{u_i'u_i'}$ is the turbulent kinetic energy, $\delta_{ij}$ is the Kronecker delta and $\nu_t$ is {the} eddy viscosity. On dimensional grounds the eddy-viscosity is a function of the turbulent velocity and length scale \cite{Leschziner2015}. {These quantities are commonly computed using two-equation turbulence models, such as $k-\epsilon$ or $k-\omega$, that are based on {transport equations} for $k$ and for the turbulence-energy dissipation $\epsilon$ or the specific-dissipation $\omega$. In this article, we use two popular turbulence models, the Launder-Sharma $k-\epsilon$ and a $k-\omega$ model. For both models, a generic $k$ transport equation can be formulated with appropriate terms, listed in Table \ref{tab:terms}, as
\begin{equation}
	\frac{\partial k}{\partial t} + \frac{\partial}{\partial x_j} \left[ k\overline{u}_j - \left(\nu + \frac{\nu_t}{\sigma_k} \right) \frac{\partial k}{\partial x_j} \right] = P - D,\quad \text{and}\quad P = R_{ij} \frac{\partial \overline{u}_i}{\partial x_j}.
\end{equation}

\begin{table}[H]
\begin{center}
\begin{tabularx}{.6\textwidth}{>{\hsize=.2\hsize}X>{\hsize=.4\hsize}X>{\hsize=.1\hsize}X}\toprule[1pt]
Term & Launder-Sharma $k-\epsilon$ & $k-\omega$ \\\midrule\midrule
$D$ & $\epsilon + 2 \nu \left( \dfrac{\partial \sqrt{k}}{\partial x_j} \right)^2$ & $C_\mu \omega k$\\
$\nu_t$ & $C_\mu f_\mu \dfrac{k^2}{\epsilon}$ & $\dfrac{k}{\omega}$\\
$\sigma_k$ & $1$ & $2$\\
$C_\mu$ & 0.09 & 0.09 \\
\bottomrule[1pt]
\end{tabularx}
\end{center}
\caption{Terms and coefficients in the $k$ transport equation for two turbulence models.} \label{tab:terms}
\end{table}

The Launder-Sharma $k-\epsilon$ model is typically employed as a low-Reynolds number model. These kind of models resolve the viscous part of the boundary layer with an appropriately refined mesh instead of utilizing wall functions \cite{Launder1974}. Correct near wall behaviour is achieved by damping functions for the eddy viscosity $f_\mu$ and the dissipation $f_2$ close to a wall. The equation for the dissipation $\epsilon$ reads
\begin{align}
	\frac{\partial \epsilon}{\partial t} + \frac{\partial}{\partial x_j} \left[  \epsilon\overline{u}_j - \left(\nu + \frac{\nu_t}{\sigma_\epsilon} \right) \frac{\partial \epsilon}{\partial x_j}  \right] &= \left(C_{\epsilon_1} P - C_{\epsilon_2} f_2 \epsilon \right) \frac{\epsilon}{k} + 2 \nu \nu_t \left( \frac{\partial^2 \overline{u}_i}{\partial x_j^2} \right)^2, \qquad \text{with}\\
		f_\mu = \exp\left[ \frac{-3.4}{\left(1 + \frac{k^2}{50 \nu \epsilon }\right)^2} \right]&,\;
		f_2 = 1 - 0.3 \exp\left[-\min\left( \left(\frac{k^2}{\nu \epsilon}\right)^2 , 50 \right) \right],
\end{align}

\noindent with $\sigma_\epsilon=1.3$, $C_{\epsilon_1}=1.44$, $C_{\epsilon_2}=1.92$. The other model is the $k-\omega$ model \cite{Wilcox1993}, which uses a specific dissipation $\omega$,
\begin{align}
	\frac{\partial \omega}{\partial t} + \frac{\partial}{\partial x_j} \left[  \omega\overline{u}_j - \left(\nu + \frac{\nu_t}{\sigma_\omega} \right) \frac{\partial \epsilon}{\partial x_j}  \right] &= \gamma \frac{\omega}{k} P - \beta \omega^2,
\end{align}

\noindent with $\sigma_\omega = 2$, $\gamma=0.52$ and $\beta=0.072$. 

These two models are our baseline, to be perturbed in order to obtain stochastic RANS equations. But the method proposed in this article is also applicable to other eddy viscosity models.
%To facilitate understanding, we have used the standard notation from the turbulence modeling literature. In the rest of the article, the symbols $\beta,\gamma$ will be used in the context of the MLMC method.
\section{Stochastic RANS models}\label{stochasticRANS}
We now describe in detail the two stochastic models based on a perturbation of the baseline eddy viscosity field and the baseline Reynolds stress tensor field \cite{xiao2} originating from a deterministic EVM. The former model is mathematically simple and is suitable for quantifying uncertainties that are introduced from a poor choice of RANS closure parameters to compute the eddy viscosity. The latter model is more advanced and is applicable to flows where the assumption of linear stress-strain relation is {insufficient}. When these models are combined with the RANS equations \eqref{eq:rans}, we obtain so-called stochastic partial differential equations (SPDEs) that are solved using the MLMC method.

Before we describe the stochastic models, we clarify our setting. The RANS equations are defined in a bounded domain $\mathcal{D}\subset\mathbb{R}^d$ $(d=1,2,3)$. The complete probability space is denoted by $(\Omega,\mathcal{F},\mathbb{P})$, where $\Omega$ is the sample space with $\sigma$-field $\mathcal{F}$ and probability measure $\mathbb{P}$. Furthermore, the zero-mean Gaussian random field will be denoted by  $Z(\mathbf{x},\omega)$, $\mathbf{x}\in \mathcal{D}$, $\omega\in\Omega$ with a specified positive-definite covariance kernel. Therefore,
\begin{align}\label{eq:GRF}
\mathbb{E}[Z(\mathbf{x},\cdot)]  &=0,\\
\text{Cov}(Z(\mathbf{x_1},\cdot),Z(\mathbf{x_2},\cdot))&= \mathbb{E}[ Z(\mathbf{x_1},\cdot)Z(\mathbf{x_2},\cdot)],\quad \mathbf{x_1},\mathbf{x_2}\in \mathcal{D}.
\end{align}
We will work with a stationary anisotropic squared exponential covariance model, given by
\begin{equation}\label{eq:nonIsotropic}
\text{Cov}(Z(\mathbf{x_1},\cdot),Z(\mathbf{x_2},\cdot)) = C(\mathbf{x_1},\mathbf{x_2}) :=  \sigma_c^2\exp\left(-\frac{(x_1-x_2)^2}{l_x^2}-\frac{(y_1-y_2 )^2}{l_y^2}-\frac{(z_1-z_2)^2}{l_z^2} \right),
\end{equation}
where $C:\mathbb{R}^d\rightarrow\mathbb{R}_+$ with parameters $\sigma_c^2$ the marginal variance; $l_x,l_y$ and $l_z$ correlation lengths along the $x,y$ and $z$ directions, respectively. The realization of $Z$ can be based on the Karhunen-Lo\'eve (KL) decomposition of $Z(\mathbf{x},\omega)$
\begin{equation}\label{eq:KLexp}
Z(\mathbf{x},\omega) = \sum_{j=1}^\infty\sqrt{\lambda_j}\Psi_j(\mathbf{x})\xi_j,\qquad \xi_j\sim\mathcal{N}(0,1).
\end{equation}
Here, $\lambda_j$ and $\Psi_j$ are eigenvalues and eigenfunctions of the covariance kernel ${C}(\mathbf{x_1},\mathbf{x_2})$, obtained from the solution of the Fredholm integral,
\begin{equation}\label{eq:Fred}
\int_\mathcal{D} C(\mathbf{x_1},\mathbf{x_2})\Psi(\mathbf{x_1}) d\mathbf{x_1} =\lambda \Psi(\mathbf{x_2}).
\end{equation}
The sum in \eqref{eq:KLexp} represents an infinite dimensional uncertain field with diminishing contributions of the eigenmodes. The sum is truncated after a finite number of terms, $M_{KL}$, which is usually decided by balancing the KL-truncation error with other sources of error, such as discretization or sampling errors. For Gaussian processes with small correlation lengths and large variances, typically a large number of terms is needed to include all important eigenmodes \cite{RF1}. The evaluation of eigenmodes in the KL expansion is expensive as it requires solving the integral equation \eqref{eq:Fred} for each mode. In case of stationary covariance models, fast sampling of random fields can be achieved via a spectral generator (sometimes referred to as circulant embedding) which employs the discrete FFT (Fast Fourier Transform) \cite{Ravalec2000,RF2,RF4}. A short summary of this technique is provided in Appendix A2. 
\subsection{Random Eddy Viscosity (REV) model}
RANS turbulence models {rely} on transport equations and a set of closure coefficients that are obtained from a calibration against DNS or experimental data. For a given turbulence model, a closure coefficient take different values when calibrated against different types of flow \cite{pope}. Since the model prediction is strongly influenced by the value of the closure coefficients, a common practice is to propagate a joint probability distribution of these closure parameters to obtain uncertainty bounds of the QoIs, see e.g, \cite{Margheri2014,edeling1,cheung1}. These approaches indirectly lead to a globally perturbed eddy viscosity field. Here, one must take into account the fact that the Boussinesq assumption \eqref{linearEV} is in the general case only locally imperfect. Therefore, methods that allow direct local perturbations of the baseline eddy viscosity fields can be effective. A convenient way to achieve this local perturbation is by the means of Gaussian random fields with some prescribed covariance model. Depending on the problem, a covariance model can be designed which induces a high-variability locally in $\nu_t$; around regions where eddy viscosity models are expected to perform poorly. The samples of the random eddy viscosity field $\nu_t(\mathbf{x},\omega)$ can be obtained by perturbing the baseline eddy viscosity field which we now denote by $ \nu_t^{(bl)} (\mathbf{x})$ with the Gaussian random field,
\begin{equation}\label{eq:randEV}
\log \nu_t(\mathbf{x},\omega) = \log \nu_t^{(bl)}(\mathbf{x}) + Z(\mathbf{x}, \omega),
\end{equation}
where $\omega$ denotes the random event in the stochastic domain $\Omega$. The mean field $\nu_t^{(bl)}$ is obtained from a converged deterministic RANS simulation and is based on a baseline turbulence model, or from an average of eddy viscosities obtained from different turbulent models. The above relation is the simplest multiplicative model, $\nu_t(\mathbf{x},\omega) =\nu_t^{(bl)}(\mathbf{x}) e^{Z(\mathbf{x},\omega)}$, that is able to impose positivity of random eddy viscosity samples and values close to zero near the wall region. We point out that Dow and Wang \cite{Wang,Wang2} also explored Gaussian random fields to obtain uncertainty bounds in the mean flow. In their approach the variability of the Gaussian process was based on the discrepancy between eddy viscosities obtained  from the DNS data (known as the "true" eddy viscosity) and those predicted by any turbulence model.

With the random eddy viscosity, we obtain the following SPDE:
\begin{equation}\label{REV_SPDE}
\rho (\overline{\mathbf{u}}\cdot \nabla)\overline{u}_i = -\frac{\partial \overline{p}^*}{\partial x_i} +\frac{\partial}{\partial x_j}\left[ \left(\nu+{\nu_t(\omega)}\right)\left(\frac{\partial \overline{u}_i }{\partial x_j} + \frac{\partial \overline{u}_j }{\partial x_i} \right) \right],
\end{equation}
where $\overline{p}^*:=\overline{p}-\frac23k$. Recall that the above SPDE can be used for quantifying uncertainties due to inconsistencies in the closure parameters of the baseline model and also provide a way to account for the effect of local variations of these parameters in the flow unlike \cite{Margheri2014,edeling1,cheung1}. However, this stochastic model still inherits drawbacks from the Boussinesq hypothesis and is inadequate for quantifying uncertainties associated with turbulence anisotropy. For instance, occurrence of secondary flows as a result of normal stress imbalance (e.g. in a square duct) will remain undetected. Therefore, a more generic stochastic model is also discussed, that involves injection of uncertainties directly into the baseline Reynolds stress tensor.  
 %Although these random eddy viscosity models can be successfully applied for producing prediction intervals for simple shear flows, they , for instance, . 
\subsection{Random Reynolds Stress Tensor (RRST) model}
The RRST model stems from the work by Soize in \cite{soize1,soize2,soize3,soize4} who developed non-parametric probabilistic approaches based on random matrix theory to quantify modeling uncertainties in computational mechanics problems. Soize derived the maximum entropy probability distribution for symmetric positive-definite (SPD) real matrices with a given mean and variance (also known as the \emph{dispersion} parameter, $\delta$) along with a Monte Carlo sampling method. These results with slight modifications can be utilized for the sampling of random Reynolds stress tensors (as physically realizable RSTs are symmetric positive semi-definite matrices). Xiao and coworkers in \cite{xiao2} further extended this approach to incorporate spatial correlation in the Reynolds stress components by the means of Gaussian random fields with a prescribed covariance function. We now briefly outline sampling algorithms for a random SPD matrix that will be utilized later to describe the sampling of the random Reynolds stress tensor fields. We closely follow the description from the original papers \cite{xiao2,soize4,Guilleminot}. 

\subsubsection{Sampling random SPD matrices}
We denote by $\mathbb{M}^{+0}_d(\mathbb{R})$ and $\mathbb{M}^{+}_d(\mathbb{R})$ the set of all $d \times d$ symmetric positive semi-definite and symmetric positive-definite matrices with real entries, respectively. Given a baseline matrix $\mathbf{R}_{(bl)}\in\mathbb{M}^{+}_d(\mathbb{R})$, we wish to sample random matrices $\mathbf{R} \in\mathbb{M}^{+}_d(\mathbb{R})$, such that $\mathbb{E}[\mathbf{R}] = \mathbf{R}_{(bl)}$.  The sampling of $\mathbf{R}$ can be achieved using a \emph{normalized} random SPD matrix $\mathbf{G}\in\mathbb{M}^{+}_d(\mathbb{R})$ with mean $\mathbf{I}_d$ (identity), i.e. $\mathbb{E}[\mathbf{G}] = \mathbf{I}_d$ and the variance parameterized with a dispersion parameter $\delta>0$ defined as
\begin{equation}\label{dispersion}
\delta = \sqrt{\frac1d\mathbb{E}\left[||\mathbf{G}-\mathbf{I}_d||^2_F\right]},
\end{equation}
where $||\cdot||_F$ is the Frobenius norm. A first step is to utilize the Cholesky decomposition $\mathbf{G}=\mathbf{U}^T\mathbf{U}$, where $\mathbf{U}$ is an upper-triangular matrix with positive diagonal entries. Now, the assembly of the random matrix $\mathbf{G}$ boils down to sampling the six entries of $\mathbf{U}$. The non-diagonal entries of $\mathbf{U}$ are sampled by means of
\begin{equation}\label{non_diag}
U_{ij} = \frac{\delta}{\sqrt{d+1}}\xi_{ij},\quad \text{for}\quad i>j,\quad \xi_{ij}\sim\mathcal{N}(0,1). 
\end{equation} 
The diagonal entries are sampled as 
\begin{equation}\label{diag}
U_{ii}  =\frac{\delta}{\sqrt{d+1}}\sqrt{2y_i},\quad \text{for}\quad i=1,2,3,\quad 
\end{equation}
where $y_{i}>0$ is a sample from the gamma distribution $\boldsymbol{\Gamma}(k_i,1)$ with shape parameter $k_i$ and scaling parameter 1, i.e.
\begin{equation}
y_i\sim\boldsymbol{\Gamma}(k_i,1)\quad\text{with} \quad k_i = \frac{d+1}{2\delta^2} + \frac{1-i}{2}.
\end{equation}
The gamma probability density function $f_Y$ is given by:
\begin{equation}\label{gamma}
f_Y(y_i) = \frac{y_i^{k_i-1}\text{exp}(-y_i)}{\Gamma(k_i)}\quad \text{with} \quad k_i = \frac{d+1}{2\delta^2} + \frac{1-i}{2},
\end{equation}
where $\Gamma(\cdot)$ is the standard gamma function. For different diagonal terms, $y_i(\mathbf{x},\cdot)$ will have different marginal PDFs depending on the shape parameter $k_i$. Using $\mathbf{G}$, one can obtain the random matrix $\mathbf{R}$ with mean $\mathbf{R}_{(bl)}$ as:
\begin{equation}\label{rand_mat}
\mathbf{R} = \mathbf{U}_{(bl)}^T\mathbf{G}\mathbf{U}_{(bl)},
\end{equation}
where $\mathbf{U}_{(bl)}$ is an upper-triangular matrix with positive diagonal entries obtained via the Cholesky factorization of the baseline RST $\mathbf{R}_{(bl)} =\mathbf{U}_{(bl)}^T\mathbf{U}_{(bl)}$. Assuming $\mathbf{R}_{(bl)}$ to be positive-definite, the factorization yields a unique matrix $\mathbf{U}_{(bl)}$. Note that in practice $\mathbf{R}_{(bl)} $ is symmetric positive semi-definite, belonging to $\mathbb{M}^{+0}_d(\mathbb{R})$. The RSTs with zero eigenvalues i.e. $\mathbf{R}_{(bl)}\in\mathbb{M}^{+0}_d(\mathbb{R})\backslash \mathbb{M}^{+}_d(\mathbb{R})$ are only encountered when $\text{det}(\mathbf{R}_{(bl)}) =0$, corresponding to  the 2-component turbulence limit \cite{pope}. However, adding an arbitrarily small number to the diagonal will make this tensor symmetric positive-definite. We also point out that, to maintain positive-definiteness of $\mathbf{G}$, the dispersion parameter $\delta$ should be chosen such that $0<\delta<\sqrt{(d+1)(d+5)^{-1}}$, see \cite{soize3} for details. Thus, for $d=3$, we find the constraint $0<\delta<1/\sqrt{2}$.
\subsubsection{Sampling the random tensor field}
The sampling algorithm for SPD matrices can be extended to sample spatially correlated tensor fields. We follow a similar procedure as described in the preceding section but now the entries of the upper-triangular matrix $\mathbf{U}$ are correlated in space. We describe the necessary algorithmic modifications needed to sample these random RST fields.
 
Let the random RST at any point be denoted by $\mathbf{R}(\mathbf{x},\omega) = \mathbf{R}$, the deterministic baseline Reynolds stress tensor field by $\mathbf{R}_{(bl)}(\mathbf{x}) = \mathbf{R}_{(bl)}$ and a spatially varying dispersion field by $\delta(\mathbf{x})$. Furthermore, the entries of the random upper-triangular matrix, $\mathbf{U}(\mathbf{x},\omega) = \mathbf{U}$, are spatially correlated as:
\begin{align}
\text{Cov}\{U_{ij}(\mathbf{x_1},\cdot),U_{ij}(\mathbf{x_2},\cdot) \} = C(\mathbf{x_1},\mathbf{x_2}),\qquad i>j,\\
\text{Cov}\{U_{ii}^2(\mathbf{x_1},\cdot),U_{ii}^2(\mathbf{x_2},\cdot) \} = C(\mathbf{x_1},\mathbf{x_2}),\qquad i=j.\label{diag_2}
\end{align}
As suggested in \cite{xiao2}, we also consider a squared-exponential covariance function for both off-diagonal and for the square of the diagonal terms. Other covariance models, for instance, a periodic or an exponential covariance can also be utilized. For the sake of simplicity, we use $C(\mathbf{x_1},\mathbf{x_2})$ defined in \eqref{eq:nonIsotropic} but with $\sigma_c^2=1$.  
Now, the random tensor field $\mathbf{R}$ is assembled using six independent random fields: $U_{11}(\mathbf{x},\omega)$, $U_{12}(\mathbf{x},\omega)$, $U_{13}(\mathbf{x},\omega)$, $U_{22}(\mathbf{x},\omega)$, $U_{23}(\mathbf{x},\omega)$ and $U_{33}(\mathbf{x},\omega)$. The off-diagonal fields are computed as:
\begin{equation}
U_{ij}(\mathbf{x},\omega) = \frac{\delta(\mathbf{x})}{\sqrt{d+1}}Z_{ij}(\mathbf{x},\omega),\quad \text{for}\quad i>j,\quad Z_{ij}  \sim\mathcal{N}(0,C).
\end{equation}
The Gaussian random fields $Z_{ij}$ are generated in a similar fashion, as described in \eqref{eq:KLexp}. Similar to \eqref{diag}, the diagonal elements are obtained as: 
\begin{equation}
U_{ii}(\mathbf{x},\omega) =\frac{\delta(\mathbf{x})}{\sqrt{d+1}} \sqrt{2y_i(\mathbf{x},\omega)},\quad \text{for}\quad i=1,2,3,
\end{equation}
where $y_i(\mathbf{x},\omega)>0$ denotes a random field with gamma marginal distribution $\boldsymbol{\Gamma}(k_i(\mathbf{x}),1)$ and covariance defined in \eqref{diag_2}. Now, the marginal gamma PDF in \eqref{gamma} is modified to incorporate spatial dependence by $\delta(\mathbf{x})$ as
\begin{equation}\label{diag_element}
f_Y({y_i}(\mathbf{x},\cdot)) = \frac{{y}_i(\mathbf{x},\cdot)^{(k_i(\mathbf{x})-1)} \exp{(-{y}_i(\mathbf{x},\cdot))}}{\Gamma (k_i(\mathbf{x}))},\quad\text{with}\quad k_i(\mathbf{x}) = \frac{(d+1)}{2\delta(\mathbf{x})^2} + \frac{(1-i)}{2}.
\end{equation} 
As the sampling of a non-Gaussian fields using a KL expansion is involved, the authors of \cite{sakamoto} proposed a generalised Polynomial Chaos (gPC) expansion approach which approximates a non-Gaussian field in terms of a weighted combination of Hermite orthogonal polynomials of a standard Gaussian field,
\begin{equation}\label{gPC}
%Y_i(\bfx,\omega)\approx\sum^{N_{PC}}_{n=1} \text{w}_n(\bfx) \mathcal{H}_n(Z(\bfx,\omega)) ,
Y\approx\sum^{N_{PC}}_{n=1} \text{w}_n \mathcal{H}_n(Z) ,
\end{equation}
where $Y$ represents a spatially correlated gamma random field, $N_{PC}$ is the order of the expansion and $\mathcal{H}_n(Z)$ is the Hermite polynomial in $Z$ of order $n$ with weight $\text{w}_n$. Given the orthogonality of Hermite polynomials with respect to the Gaussian measure, we can evaluate the weights as:
\begin{equation}\label{weight}
\text{w}_n  = \frac{\mathbb{E}[Y\mathcal{H}_n(Z)]}{\mathbb{E}[\mathcal{H}_n(Z)^2]}.
\end{equation}
Here the expectation in the denominator has an analytic expression but the expectation in the numerator is not well-defined as the dependence between $Y$ and $Z$ is unknown. Since the distribution of $Y$ is available, one can exploit the fact that $Y = F_Y^{-1}(F_Z(Z))$ and reformulate the numerator in \eqref{weight} as
\begin{equation}\label{weight2}
\mathbb{E}[Y\mathcal{H}_n(Z)]= \int_{-\infty}^\infty F_{Y}^{-1}[F_{Z}(z)]\mathcal{H}_n(z) \text{d}F_Z(z),
\end{equation}
where $F_{Y}(y) = \mathbb{P}\text{rob}(Y\leq y)$ is the cumulative distribution for a gamma random variable $Y$ and $F_Y^{-1}$ represents its inverse. Similarly, $F_{Z}(z) = \mathbb{P}\text{rob}(Z\leq z)$ is the cumulative distribution for a standard Gaussian random variable $Z$. Now, the integral \eqref{weight2} can be numerically computed using any conventional integration technique. With the above weights, the gPC expansion in \eqref{gPC} converges to $Y$ in \emph{weak sense} (convergence in probability distribution) \cite{Dxiu2,Xiu:2010}. Note that $F_Y$ should be appropriately modified according to \eqref{diag_element} to incorporate the spatial dependence in the marginal gamma PDF. It is also pointed out that for a spatially varying dispersion $\delta(\mathbf{x})$ the weights will differ at different spatial locations.

A few remarks are in order. The mean RST field $\mathbf{R}_{(bl)}$ can be directly obtained from the baseline RANS simulation. Also, the value of the dispersion field can be based on expert knowledge and can be set to a large value at locations with high uncertainty. However, to obtain a positive-definite Reynolds stress tensor at each point the dispersion should again be chosen such that $0<\delta(\mathbf{x})<\sqrt{(d+1)(d+5)^{-1}}$. 
 
Using the random Reynolds stress tensor, we can define the stochastic mean flow equation, as follows:
\begin{equation}\label{RRST_SPDE}
\rho (\overline{\mathbf{u}}\cdot \nabla)\overline{u}_i = -\frac{\partial \overline{p}}{\partial x_i} +\frac{\partial}{\partial x_j} \left(\overline{R}_{ij} + R_{ij}(\omega) \right),
\end{equation}
where $\overline{R}_{ij}$ represents mean stress, as defined for the PDE \eqref{eq:rans} and $R_{ij}(\omega)$ represents components of the random tensor field $\mathbf{R}$. In this stochastic model the isotropic eddy viscosity (Boussinesq) assumption is clearly avoided. Furthermore, this model allows us to accommodate different covariance structures for different Reynolds stress components, and thus can represent strongly anisotropic turbulence. We would like to emphasize that the above SPDE is more general than in \eqref{REV_SPDE} as the above formulation allows us to incorporate at most six random fields for each Reynolds stress component and may result in an extremely high-dimensional UQ problem.

\section{The Multilevel Monte Carlo method}\label{MCE}
In this section, we will provide a general description of the single- and multi-level variants of the Monte Carlo method that will be used to solve the SPDEs \eqref{REV_SPDE} and \eqref{RRST_SPDE}.

We assume that the QoIs considered belong to the functional space $\Lomd$, the space of square-integrable measurable functions $u:\Omega\rightarrow L^2(\mathcal{D})$ for the previously defined probability space $(\Omega,\mathcal{F},\mathbb{P})$. These spaces are equipped with the norm
\begin{equation}\label{eq:RMSEnorm}
\lnorm u(\mathbf{x},\omega)\rnorm_{\Lomd} := \mathbb{E}\left[\lnorm u(\mathbf{x},\omega)\rnorm^2_{L^2(\mathcal{D})} \right]^{\tfrac12} =\left(\int_{\Omega}\lnorm u(\mathbf{x},\omega)\rnorm^2_{L^2(\mathcal{D})} \text{d}\mathbb{P} \right)^{\tfrac12}.
\end{equation}
The above $L^2-$ based norm will be used for error analysis of the Monte Carlo estimators in the following.
\subsection{MC estimator}
We will consider the streamwise velocity field $u$ as the QoI for describing the MC estimator. The standard MC estimator for $\mathbb{E}[u_h] $  is obtained by averaging $N$ independent, identically distributed (i.i.d.) samples of the velocity field $\{u_h(\omega_i)\}_{i=1}^N$ on the computational grid $\mathcal{D}_h$ as
\begin{equation}\label{eq:SMC}
\mathbb{E}[u_h] \approx \mathpzc{E}^{MC}_N[u_h] := \frac1N\sum_{i=1}^Nu_h(\omega_i),
\end{equation}
where $\omega_i$ denotes {an} event in the stochastic domain $\Omega$ and $h$ is the largest cell-width in the simulation grid $\mathcal{D}_h$. The above estimator is easy to implement. On a given spatial mesh $\mathcal{D}_h$, we  generate samples of random input and accordingly modify the mean flow equation \eqref{eq:rans}. Then, for each sample, the modified mean flow equation is solved to obtain samples of the QoIs. These samples are averaged to obtain the MC estimate $\mathpzc{E}^{MC}_N[u_h]$. Similarly, the unbiased estimator for the variance is defined 
\begin{equation}\label{eq:MC_var}
\mathcal{V}_N^{MC}[u_h] := \frac{1}{N-1}\sum_{i=1}^N \left(u_h(\omega_i)-\mathpzc{E}_N^{MC}[u_h]\right)^2.
\end{equation} 
\subsubsection{Accuracy of the MC estimator}
Although the standard MC method has been widely used for uncertainty propagation in the context of CFD modeling, a measure of the accuracy for the resulting estimates is rarely reported. Next we derive the error estimates related to the estimator $\mathpzc{E}^{MC}_N[u_h]$. For any deterministic RANS closure model, the errors can be broadly of {three} types: parameter uncertainty, uncertainties due to the form of the model, and discretization error. Obtaining a quantitative measure of the model uncertainties is only possible when a high-fidelity solution (DNS or LES) is available. Discretization error on the other hand, is comparatively easy to quantify for a given computational mesh, as a good reference solution can be obtained by solving the same set of PDEs on a relatively finer mesh. Additionally, for the stochastic RANS models, quantification of the sampling error becomes vital. We will focus on these two errors in our analysis.

Using the triangle inequality, the RMS (root-mean-square) error in  $\mathpzc{E}^{MC}_N[u_h]$ can be bounded by the sum of discretization and sampling errors, as
\begin{equation}\label{eq:MCerror}
\lnorm \mathbb{E}[u] - \mathpzc{E}^{MC}_N[u_h]\rnorm_{\Lomd} \leq \lnorm \mathbb{E}[u] - \mathbb{E}[u_h]\rnorm_{\Ld}+\lnorm \mathbb{E}[u_h] - \mathpzc{E}^{MC}_N[u_h]\rnorm_{\Lomd}.
\end{equation}  
The discretization error can be estimated as:
\begin{equation}\label{numErr}
\lnorm \mathbb{E}[u] - \mathbb{E}[u_h]\rnorm_{\Ld}\leq C_1h^{\alpha},\qquad \alpha>0,
\end{equation}
where $C_1$ is a constant. As the exact solution $\mathbb{E}[u]$ is not available, a relative error measure of the form $\lnorm \mathbb{E}[u_{h}] - \mathbb{E}[u_{2h}]\rnorm_{\Ld}$ can be used to bound the exact discretization error, as
\begin{equation}\label{discErr}
\lnorm \mathbb{E}[u] - \mathbb{E}[u_h]\rnorm_{\Ld} \leq \frac{\lnorm \mathbb{E}[u_{h} - u_{2h}]\rnorm_{\Ld}}{2^\alpha -1}.
\end{equation}
The above relation can be easily derived using the reverse triangle inequality and \eqref{numErr}. The rate $\alpha$ depends on the regularity of the QoI in the stochastic and physical space and the order of the discretization scheme used to solve the PDE. It is possible to approximate the right-hand side term in \eqref{discErr}, numerically using the MC method, which serves as an indicator of numerical error.

From the central limit theorem, the sampling error due to $N$ samples is given as
\begin{equation}\label{eq:sampErr}
\lnorm\mathbb{E}[u_h] - \mathpzc{E}^{MC}_N[u_h]\rnorm_{\Lomd} = \sqrt{\frac{\lnorm\mathcal{V}[u_h]\rnorm_{\Ld} }{N}},
\end{equation}
where $\lnorm\mathcal{V}[u_h]\rnorm_{\Ld}$ is the $L^2-$ based variance approximated as  
\begin{align}\label{eq:fieldVar}
\lnorm\mathcal{V}[u_h]\rnorm_{\Ld} :=& \int_\mathcal{D}\int_\Omega \left(\mathbb{E}[u_h(\mathbf{x},\cdot)] - u_h(\mathbf{x},\omega)\right)^2 \text{d}\mathbb{P}\text{d}\mathbf{x},\nonumber\\
\approx &  \frac{1}{N-1}\sum^{N}_{j=1} \int_{\mathcal{D}}\left(\left(\frac{1}{N}\sum^{N}_{i=1} {u}_h(\mathbf{x},\omega_i)\right) - {u}_h(\mathbf{x},\omega_j) \right)^2\text{d}\mathbf{x}.
\end{align}
To obtain an optimized MC estimator for a given mesh $\mathcal{D}_h$, the sampling error \eqref{eq:sampErr} should be equilibrated with the discretization error \eqref{discErr} yielding the optimal value of $N$,
\begin{equation}\label{MCsamp}
N=\mathcal{O}(h^{-2\alpha}).
\end{equation}
Note that with the above criteria, the RMS error in the estimator $\mathpzc{E}^{MC}_N[u_h]$ reduces to $\mathcal{O}(h^\alpha)$ which is the best possible accuracy which can be achieved on this grid. Further, if the computational cost of obtaining one sample of the QoI (including costs for sampling the random field, CFD simulation and post-processing) is expressed as $\mathcal{O}(h^{-\gamma})$ where $\gamma\geq d$ is the rate at which the cost of one sample grows with grid refinement and $d$ is the spatial dimension. The asymptotic cost of the standard MC estimator can then be expressed as
\begin{equation}\label{MC_comp}
\mathcal{W}^{MC}_{h,N} = \mathcal{O}(Nh^{-\gamma}) = \mathcal{O}(h^{-2\alpha-\gamma}).
\end{equation}
Finally, one can express "accuracy-versus-work", as:
\begin{equation}\label{eq:workVsErr}
\lnorm \mathbb{E}[u] - \mathpzc{E}^{MC}_N[u_h]\rnorm_{\Lomd} \lesssim\left(\mathcal{W}^{MC}_{h,N}\right)^{\tfrac{-\alpha}{2\alpha+\gamma}} 
\end{equation}
The rates $\alpha$ and $\gamma$ can be empirically determined if they are not known a-priori. It is pointed out that the cost of the estimator can be reduced by using a higher-order discretization scheme (by increasing $\alpha$) or by an optimal CFD solver for which $\gamma \approx d$. Obtaining such solvers is difficult in fluid dynamics, and in general the solver performance deteriorates with increase in the Reynolds number.
\subsection{MLMC estimator}
A multilevel Monte Carlo (MLMC) estimator is derived by generalising the standard MC method to a hierarchy of grids. Consider a hierarchy of grid levels $\{\mathcal{D}_{\ell}\}^L_{\ell=0}$ for the spatial domain $\mathcal{D}$ with the largest cell-width for level $\ell$ defined as 
\begin{equation}\label{eq:hmax}
h_\ell = \mathcal{O}(s^{-\ell}h_0),
\end{equation}
where $M_\ell$ is the total number of cells in the mesh $\mathcal{D}_{\ell}$, $h_0$ is largest cell-width on the coarsest mesh $\mathcal{D}_0$ and $s>0$ represents a grid refinement factor. Now, using the linearity of the expectation operator, one can define the expected value of a QoI on the finest level $L$ by the following telescopic sum:
\begin{equation}\label{eq:telescope}
\mathbb{E}[u_L] = \mathbb{E}[u_0] + \sum_{\ell =1}^L \mathbb{E}[u_\ell - u_{\ell-1}].
\end{equation}
In terms of the computational cost, it is cheap to approximate $\mathbb{E}[u_0]$ as the samples are computed on the coarsest mesh. Furthermore, the correction term, $\mathbb{E}[u_\ell - u_{\ell-1}]$, can be accurately determined using only a few samples as the level-dependent variance, $\mathbb{V}[u_\ell - u_{\ell-1}]$, is small compared to the sample variance, $\mathbb{V}[u_\ell]$. To approximate $\mathbb{E}[u_L]$, a multilevel estimator $\mathpzc{E}^{ML}_{L}$ can be constructed using a sum of standard MC estimators:
\begin{align}\label{MLMCestimator}
\mathbb{E}[u_L] \approx \mathpzc{E}^{ML}_L[u_{L}] : =&\sum^L_{\ell=0} \mathpzc{E}^{MC}_{N_\ell}[{u}_{\ell}-{u}_{{\ell-1}}] ,\\
=&\sum^L_{\ell=0}\frac{1}{N_\ell}\sum^{N_\ell}_{i=1} ({u}_{\ell}(\omega_i)-{u}_{{\ell-1}}(\omega_i)),
\end{align}
where $u_{{-1}} =0$ is used for notational convenience. The number of MLMC samples $N_\ell\in\mathbb{N}$ forms a decreasing sequence for increasing $\ell$. In order to keep the variance of the correction terms small, the MC samples $ u_\ell(\omega_i) - u_{\ell-1}(\omega_i)$, should be based on the same random input $\omega_i$ for simulation on two consecutive levels $\ell$ and $\ell-1$. We will discuss this in detail in Section \ref{sameRF}. 

As each of the expectations in the above estimator is computed independently, the variance of the multilevel estimator is the sum of the variances of individual estimators, i.e. 
\begin{equation}\label{eq:MLMCVar}
\mathbb{V}\left[\mathpzc{E}^{ML}_L[u_{L}] \right]=\sum^L_{\ell=0}\frac{\mathcal{V}_\ell}{N_\ell},
\end{equation}
with the level-dependent variance $\mathcal{V}_\ell$ defined as
%\begin{equation}
%:= \lnorm\mathcal{V}[{u}_{\ell}-{u}_{{\ell-1}}]\rnorm_{\Ld}$ is computed as in 
\begin{equation}
\mathcal{V}_\ell: = \lnorm \mathcal{V}[{u}_{\ell}-{u}_{{\ell-1}}]\rnorm_{\Ld}= \lnorm \mathbb{E}[u_{\ell}(\mathbf{x},\cdot)-u_{\ell-1}(\mathbf{x},\cdot)] - (u_{\ell}(\mathbf{x},\omega) -u_{\ell-1}(\mathbf{x},\omega))\rnorm_{\Lomd}^2,
\end{equation}
which can be approximated as in \eqref{eq:fieldVar}. Further, we assume that the level-dependent variance also decays with grid refinement with a positive rate $\beta$, thus $\mathcal{V}_\ell = \mathcal{O}(h_\ell^\beta)$. Similar to $\alpha$, the rate $\beta$ also depends on the regularity of $u(\mathbf{x}, \omega)$ w.r.t. the spatial and stochastic space. For sufficiently smooth solutions, typically $\beta = 2\alpha$. 

The multilevel estimator for the variance can be defined as
\begin{equation}\label{ML_var}
\mathcal{V}_L^{ML}[u_L] := \sum_{\ell=0}^L \mathcal{V}^{MC}_{N_\ell}[u_\ell]-\mathcal{V}^{MC}_{N_{\ell}}[u_{\ell-1}],
\end{equation}
where at level $\ell$, both variances $\mathcal{V}^{MC}_{N_\ell}[u_\ell]$ and $\mathcal{V}^{MC}_{N_{\ell}}[u_{\ell-1}]$ are computed as in \eqref{eq:MC_var} using samples computed from the same random inputs $\{\omega_i\}_{i=1}^{N_\ell}$. In the following section, we discuss the error associated with the MLMC estimator $\mathpzc{E}^{ML}_L[u_L]$. A detailed analysis of the multilevel variance estimator can be found in \cite{bierig}.
\subsubsection{Accuracy of the MLMC estimator}

The MLMC estimator $\mathpzc{E}^{ML}_L[u_L]$ is obtained by two approximations,
\begin{equation}
\mathbb{E}[u]\approx \mathbb{E}[u_L]\approx\mathpzc{E}^{ML}_L[u_L].
\end{equation}
Therefore, the MSE (mean-squared-error) in $\mathpzc{E}^{ML}_L[u_L]$ can be quantified as
\begin{align}\label{eq:MSEMLMC}
\lnorm\mathbb{E}[u]-   \mathpzc{E}^{ML}_L[u_L]\rnorm_{\Lomd}^2
\leq & \lnorm\mathbb{E}[u]  -   \mathbb{E}[u_L]\rnorm_{\Ld}^2 +\lnorm\mathbb{E}[u_L]-  \mathpzc{E}^{ML}_L[u_L] \rnorm_{\Lomd}^2,\\
=& (C_1h_L^{\alpha})^2 + \sum^L_{\ell=0}\frac{\mathcal{V}_\ell}{N_\ell},
\end{align} 
where $C_1$ is a constant. The first term at the right-hand side corresponds to the discretization bias whereas the second term is the sum of sampling errors due to $L+1$ MC estimators used in the MLMC approximation. Similar to a single-level MC method, the sampling error is balanced with the discretization error. For this, the number of level-dependent samples $N_\ell$ can be chosen such that each term $\frac{\mathcal{V}_\ell}{N_\ell}$ is reduced to the order $\mathcal{O}(h_L^{2\alpha})$. Assuming a uniform grid refinement, $h_{\ell-1} = 2h_{\ell}$, we can define a sample sequence as
\begin{equation}\label{eq:Nl}
N_\ell = \lceil N_L2^{\beta(L - \ell)}\rceil,
\end{equation}
where $N_L$ is fixed and is used as  a tuning parameter \cite{mishra2012sparse}. Ideally, the value of $N_L$ should be chosen such that a balance $\mathcal{V}_L/N_L = \mathcal{O}(h_L^{2\alpha})$ is achieved. In practice, the value $N_L$ is often very small $\sim \mathcal{O}(1)$ and can be chosen heuristically. It is also pointed out that the sampling error on the coarsest level $\frac{\mathcal{V}_0}{N_0}$ does not depend on $\beta$ and may require a larger number of samples than given by the formula \eqref{eq:Nl}.

For a given tolerance $\varepsilon$, one can also solve an optimization problem that minimizes the total cost of the MLMC estimator \cite{MLMC1}. In this approach, the optimal choice of the level-dependent sample $N_\ell$ requires a-priori values of the MLMC rates $\alpha, \beta$ and $\gamma$. In most cases, these rates are not available and have to be computed using a few "warmup samples and levels". The implementation of this approach is slightly involved and non-trivial to parallelize. On the other hand, with the sampling approach $\eqref{eq:Nl}$, the number of samples on all levels is fixed in advance and can be parallelized easily. Also, the rates $\alpha,\beta $ can be determined from the baseline RANS simulations. We will numerically demonstrate the advantage of this approach.

The total cost of the MLMC estimator is
\begin{equation}\label{eq:MLMC_work}
\mathcal{W}^{ML}_{L} =\sum^L_{\ell=0}N_\ell \mathcal{W}_{\ell},
\end{equation}
where $\mathcal{W}_{\ell} = \mathcal{O}\left(h_{\ell}^{-\gamma}\right)$ corresponds to the cost of one sample on level $\ell$. We can conveniently express $\mathcal{W}^{ML}_L = \mathcal{O}\bigg(\sum_{\ell=0}^L2^{(\gamma-\beta)\ell}\bigg)$ leading to three cases. When the level-dependent variance $\mathcal{V}_\ell$ decays at a faster rate than the cost $\mathcal{W}_\ell$ with levels (so, when $\beta>\gamma$), the dominant cost of the estimator comes from the coarsest level. For $\beta=\gamma$, all levels contribute equally in terms of the cost. Finally, if $\beta<\gamma$, the dominant cost comes from the finest level. The authors in \cite{MLMC1,MLMC2,mishra2012sparse} have estimated the asymptotic work versus error for the MLMC estimator.  We directly state the accuracy versus work estimate without going into the detailed derivations:
\begin{equation}\label{MLMC_comp}
\lnorm \mathbb{E}[u] - \mathpzc{E}^{ML}_L[u_{L}]\rnorm_{\Lomd}  \lesssim
\begin{cases}
\left({\mathcal{W}^{ML}_{L}}\right)^{-\tfrac{1}{2}}\qquad\qquad\qquad\quad \text{if}\quad \beta>\gamma,\\
\left({\mathcal{W}^{ML}_{L}}\right)^{-\tfrac{1}{2}} \log\left({\mathcal{W}^{ML}_{L}}\right)^{\tfrac12}\quad \text{ if}\quad \beta=\gamma,\\
\left({\mathcal{W}^{ML}_{L}}\right)^{\tfrac{-\alpha}{2\alpha+\gamma-\beta}}\quad\qquad\qquad \text{if}\quad \beta<\gamma.
\end{cases}
\end{equation}
 Notice that for all these cases, the MLMC estimator has a better asymptotic cost than the standard Monte Carlo method $({\mathcal{W}^{MC}_{h,N}})^{-\alpha/(2\alpha+\gamma)}$ derived earlier. Moreover, a high-order discretization scheme may increase $\alpha$ and $\beta$ leading to a reduced number of levels and a faster decay of the number of samples with level, respectively. Lastly, if we have $\beta=2\alpha$, the third case in \eqref{MLMC_comp} reduces to $\big(\mathcal{W}^{ML}_L\big)^{-\alpha/\gamma}$ which is the same as the accuracy versus work estimate for a deterministic version of the problem. Thus, the multilevel estimator obtained in this way is sometimes regarded to be \emph{optimal}, as the asymptotic cost is same as one deterministic solve on the finest level in the hierarchy.

\subsubsection{Computation of $ \mathpzc{E}^{MC}_{N_\ell}[u_\ell - u_{\ell-1}]$}\label{sameRF}
While computing samples at different levels for the MLMC estimator, it is important to ensure that the telescopic identity \eqref{eq:telescope} is not violated. Essentially, one needs to confirm that the  random samples $u_\ell$ while estimating $\mathbb{E}[u_{\ell+1} - u_{\ell}]$ and $\mathbb{E}[u_{\ell} - u_{\ell-1}]$ have the same expected value, i.e. 

\begin{equation}\label{eq:teleViolate}
\mathbb{E}[u_\ell]^{(coarse)} = \mathbb{E}[u_\ell]^{(fine)}\qquad \text{for}\qquad \ell\in\{0,1,2,..., L-1 \}. 
\end{equation}
Therefore, a correct treatment of the random input on each two levels is required. More precisely, when computing the sample $u_{\ell}(\omega_i) - u_{\ell-1}(\omega_i)$, the same realization of the eddy viscosity field $\nu_t(\omega_i)$ or the random Reynolds stress tensor $\mathbf{R}(\omega_i)$ should be used for the simulation on the meshes $\mathcal{D}_\ell$ and $\mathcal{D}_{\ell-1}$. A common practice is to first generate the random field on $\mathcal{D}_\ell$ and then use a locally averaged random field for the coarser grid $\mathcal{D}_{\ell-1}$. However, caution must be taken while performing this local averaging step as the upscaled versions of these random fields may not exhibit the same covariance structure as the finer level sample, violating \eqref{eq:teleViolate}. There are several ways to upscale the random inputs without changing their statistical properties. One way is to use the same random vector $\{\xi_j\}^{N_{KL}}_{j=1}$ in the truncated KL expansions at both levels:
\begin{eqnarray}
\log \nu_t^\ell (\mathbf{x}_\ell,\omega_i) =& \log \nu_t^{(bl)}(\mathbf{x}_\ell) + \sum^{N_{KL}}_{j=1} \sqrt{\lambda_j}\Psi_j(\mathbf{x}_\ell)\xi_j,\\
\log \nu_t^{\ell-1} (\mathbf{x}_{\ell-1},\omega_i) = &\log \nu_t^{(bl)}(\mathbf{x}_{\ell-1}) + \sum^{N_{KL}}_{j=1} \sqrt{\lambda_j}\Psi_j(\mathbf{x}_{\ell-1})\xi_j.
\end{eqnarray}
This approach can be computationally expensive if the truncation dimension $N_{KL}$ is large. If the sampling meshes of the random field for the fine $\ell$ and coarse $\ell-1$ level are nested (which is true for vertex-centred grids), this problem can be trivially circumvented by injecting the random field from a fine to a coarse grid without performing any type of averaging. For cell-centred grids, where the sampling nodes are non-nested,  sampling on a vertex-centred grid twice as fine as finest level $\ell$ can be use to produce same random field on levels $\ell$ and $\ell-1$ \cite{Kumar_2017}. For instance, a sample of the discrete random field which is generated on a vertex-centred $129\times129$ grid can give valid random fields on $64\times64$ and $32\times32$ grids, which corresponding to levels $\ell$ and $\ell-1$, respectively.  These injection based workarounds are very convenient to implement but can be computationally expensive for 3D flow problems, as the cost of sampling may become comparable to CFD simulations. A third possibility is the \textit{covariance upscaling} method as proposed in \cite{mishra2016multi}, which is also utilized in this paper (see Appendix A2). This method is efficient for large-scale problems where the cost of sampling these random fields becomes significant or comparable to the cost of a CFD simulation. 

\subsection{MLMC-RANS implementation}\label{MLMCRANS}
The MLMC-RANS framework is developed in MATLAB and interacts with the OpenFOAM (Open source Field Operation And Manipulation) CFD package \cite{openfoam}. It is available from the authors upon request. MATLAB based programs are responsible for the generation of random inputs (eddy viscosity fields and Reynolds stress tensors), invoking OpenFOAM with random inputs, the collection of samples of the QoI and post-processing. Within OpenFOAM, schemes for computation of the gradients and divergence are based on second-order finite volume (FV) approximations.  The baseline solution of the turbulence models is obtained using the simpleFoam solver available in OpenFOAM, and to propagate the random eddy viscosity and random Reynolds stresses different solvers were implemented for the stochastic momentum equations \eqref{REV_SPDE} and \eqref{RRST_SPDE}, respectively.

While the propagation of random eddy viscosity is straightforward and doesn't require modification of the solver in general, the propagation of random Reynolds stresses is numerically more challenging. To achieve numerically stable performance of the solver, we adopt a blending of the random Reynolds stress, which we wish to propagate, and a contribution based on the Boussinesq assumption \cite{Basara2003}. While the latter alters the propagated effective Reynolds stress, it promotes numerical convergence of the solver. The momentum equation \eqref{RRST_SPDE} is modified accordingly,

\begin{equation}\label{RRST_SPDE_mod}
\rho (\overline{\mathbf{u}}\cdot \nabla)\overline{u}_i = -\frac{\partial \overline{p}}{\partial x_i} +\frac{\partial}{\partial x_j} \left(\overline{R}_{ij} + (1-\xi) R^{(bl)}_{ij}  + \xi R_{ij}(\omega) \right),
\end{equation}

\noindent in which the linear eddy viscosity contribution $R^{(bl)}_{ij}$ is given in \eqref{linearEV}. The production of turbulent kinetic energy is modified accordingly. The blending parameter $\xi \in[0,1]$ quantifies the amount of $R^{(bl)}_{ij}$ to increase numerical stability. For $\xi =1$, we achieve the full propagation of the random tensor field. This is possible in case of simpler flows, for e.g.,  flow in a square duct. Also, the value of $\xi$ is linearly increased with the number of iterations (ramping) to a constant value. Note that a value of $\xi<1$ indirectly corresponds to a lower variance, than specified for a given dispersion $\delta$.

To facilitate the analysis, our implementation of the MLMC method is based on a pre-defined geometric hierarchy of meshes such that the largest cell width follows $h_{\ell-1} \approx 2h_{\ell}$.  In general, an MLMC estimator can be constructed with any hierarchy for which the accuracy and cost increase with the levels. The quality of the mesh at any given MLMC level $\ell$ is assessed using the dimensionless wall distance, defined as $y^{+1}_\ell=h^{cc}_\ell u^{*}_\ell/\nu$ where $h^{cc}_\ell$ denotes the distance of the cell-centers adjacent to the wall, $u^*_{\ell}$ is the friction velocity defined as $u^*_{\ell} = \sqrt{\tau^w_\ell/\rho}$ with $\tau^w_\ell = \mu (\partial u/\partial y)_{y=0}$. Standard notation $\nu$ and $\mu$ is used for kinematic and dynamic viscosities, respectively. For resolving the viscous sublayer, {the} {$y^{+1}_\ell$} value should be less than one, however, this criterion can be relaxed for coarser levels in the MLMC hierarchy provided that the RANS solution results in a meaningful flow field. Furthermore, we check that the level-dependent variance should be strictly less than the pure sample variance of the quantity of interest, i.e., $\lnorm \mathcal{V}[{u}_{\ell}-{u}_{{\ell-1}}]\rnorm_{\Ld} < \lnorm \mathcal{V}[{u}_{\ell}]\rnorm_{\Ld}$. Violation of this condition may result in an MLMC estimator which is more expensive than a standard MC estimator. 

As this work involves stationary covariance models, we use a spectral generator for the fast sampling of the Gaussian random fields. It is pointed out that with this algorithm the computational cost of sampling a random field is of  the order $\mathcal{O}(M_{\ell}\log M_{\ell})$, where $M_{\ell}$ is the number of mesh points on any level $\ell$ and is negligible compared to the cost of one CFD solve at that level. Additionally, the random fields generated using  spectral methods  are exact on the sampling mesh. In case of the KL expansion based sampling, one needs to quantify the error incurred due to the truncation of the eigenmodes.
\section{Numerical experiments}\label{NE}
We use two test problems, a fully developed turbulent flow in a square duct and a flow over a periodic hill, to study the performance of the MLMC method. A bulk Reynolds number $Re=1100$ is considered for the square duct flow with benchmark data available from Huser et al. (1993) \cite{huser_1993}. This problem has become a standard test case to demonstrate the inability of linear eddy viscosity models to predict the secondary flows that arise from the normal stress imbalance. Linear eddy viscosity models assume equal normal stresses and completely fail to predict secondary flow features, resulting in parallel flow.  We only employ the random Reynolds stress model for this test case as the random eddy viscosity model suffers from the same drawback as the deterministic linear eddy viscosity model and fails to produce any secondary flows. For the periodic hill problem, we use $Re=2800$ with the DNS data from Breuer et al. (2009) \cite{breuer}. This is a complex benchmarking test problem, offering a number of flow features such as anisotropy, strong streamline curvature, a recirculating zone and free shear layer, that are challenging for RANS turbulence models. Both stochastic models are analyzed for the periodic hill flow.

\subsection{Flow in a square duct}
A schematic representation of the square duct flow is presented in Fig. \ref{SD_schema} (left) showing the eight-vortex pattern with counter-rotating vortices in each quadrant. Due to symmetry, we choose to simulate the flow only for the top-right quadrant on a domain of size $[0,H]\times[0,H]$, where $H=1$ is the half-height of the square duct. 
\begin{figure}[H]
 \begin{subfigure}[b]{0.49\textwidth}
\hbox{\hspace{-1cm}{\includegraphics[clip, trim=0cm 0cm 0cm 0cm,scale=0.35]{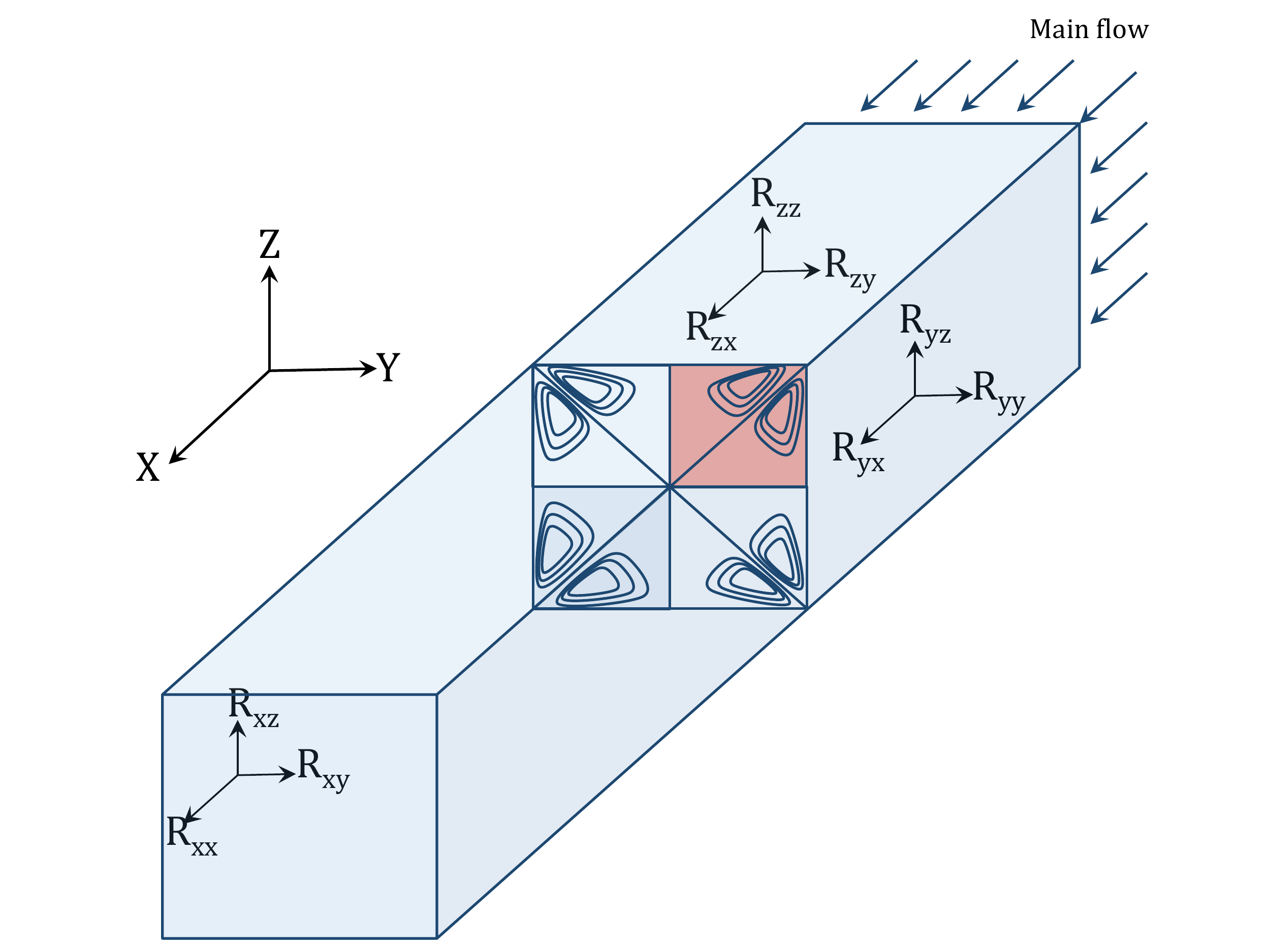}}}
\end{subfigure}
 \begin{subfigure}[b]{0.49\textwidth}
\hbox{\hspace{1cm}{\includegraphics[clip, trim=0cm 0cm 0cm 0cm,scale=0.24]{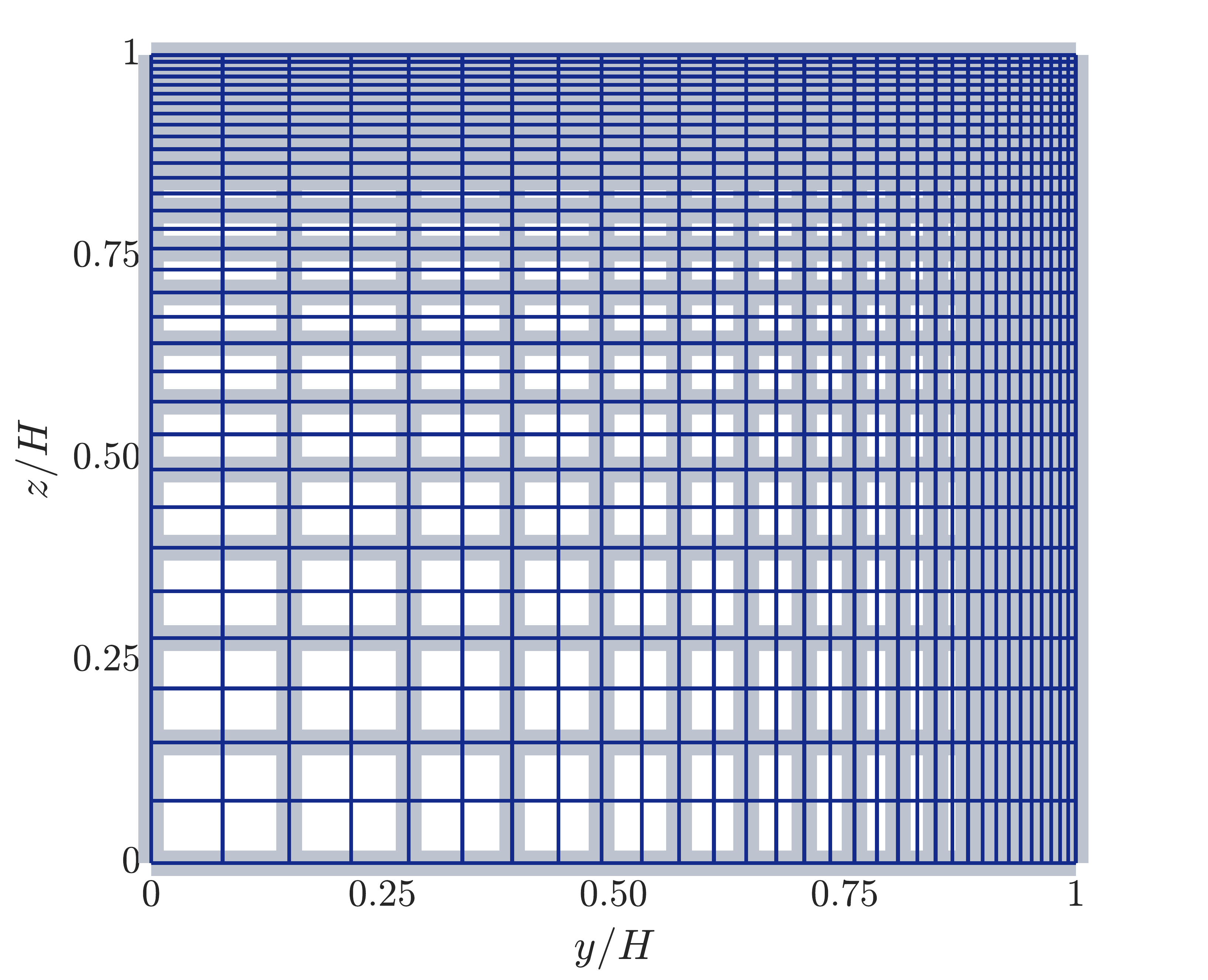}}}
\end{subfigure}
\caption{Schematic representation of time-averaged flow in a square duct (left) showing the 8-vortex pattern with each quadrant exhibiting vortices of alternating sign. (Right) Nested meshes $\ell=0$ (light blue)  and $\ell=1$ (dark blue) used to simulate the flow in the top-right quadrant, grading near the walls.}\label{SD_schema}
\end{figure}
We use a separate grid hierarchy for the OpenFOAM simulations and for sampling the random fields, denoted by OF and RF meshes, respectively, with specifications listed in Table \ref{sqDuct}. For the OF meshes, each grid level is graded with finer cells along the top and right walls to resolve boundary layers, see Fig. \ref{SD_schema} (right). In the case of RF meshes, the random fields are first sampled on a uniform Cartesian mesh in the domain $[0,1]^2$ and are then interpolated to the cell-centers of the RANS simulation mesh. The CPU times on a serial machine required to obtain one sample on each level is also provided in Table \ref{sqDuct}. For the considered combination of numerical schemes, the CPU times scale roughly as $\mathcal{O}(h_\ell^{-3})$ (in other words, $\gamma\approx 3$). This is due to the fact that the convergence rate of the solver deteriorates with grid refinement, therefore, the number of iterations required to reach a fixed residual tolerance also grows with levels. Additionally, the residual tolerance also needs to be reduced with grid refinement in order to obtain a converged solution upto the discretization accuracy, and on the finest levels one sample takes about eight CPU hours to obtain a residual reduction of $\mathcal{O}(10^{-8})$.
\begin{table}[H]
\begin{center}
\begin{tabularx}{\textwidth}{@{}l|YYYY@{}}\toprule[1pt]
Level $(\ell)$ & OF mesh &$h_{\ell}$  & RF mesh  & CPU time (sec)\\\midrule\midrule
$0$ &$16\times16$ &0.16 &$8\times8$ &$0.24\times10^2$\\
$1$ &$32\times32$ &0.08 &$16\times16$&$0.68\times10^2$ \\
$2$ &$64\times64$ &0.04 &$32\times32$ &$4.20\times10^2$\\
$3$ &$128\times128$ &0.02 &$64\times64$&$2.86\times10^3$ \\
$4$ &$256\times256$ &0.01 &$128\times128$&$2.93\times10^4$\\
\bottomrule[1pt]
\end{tabularx}
\end{center}
\caption{Specifications of the MLMC grid hierarchy for the square duct test case. ``OF mes'' denotes the simulation mesh in OpenFOAM and ``RF mesh'' the grid used for the generation of the random Reynolds stress tensor. CPU time is the total time for one sample.}\label{sqDuct}
\end{table}
\subsubsection{MLMC with RRST model}
We begin by analyzing the statistics of the random Reynolds stress tensors for two sets of parameters (Case 1 and Case 2) as specified in Table \ref{SD_params}. Here, we can regard Case 1 as an ``easy'' parameter set, with a low dispersion and large correlation lengths and Case 2 as ``more complex'' with a large dispersion and small correlation lengths. For both cases a 5th order gPC expansion is used such that errors in approximating the random field are negligible compared to the discretization and sampling errors. In this work, we will only consider cases with a constant dispersion, but a more general approach can be based on a spatially varying dispersion based on available data and expert knowledge as in \cite{xiao1,xiao2}. For both cases, a full propagation of the random Reynolds stress (i.e. $\xi=1$) is considered. 

In Fig. \ref{Ex_RRST}, we present examples of the first three Reynolds stress components, $R_{11},R_{12},R_{13}$, generated using the two parameter sets along with the baseline Reynolds stress tensors $\mathbf{R}_{(bl)}$ (derived from the $k-\omega$ model). Firstly, we verify the constraint $\mathbb{E}[\mathbf{R}] = \mathbf{R}_{(bl)}$ by computing the empirical probability distribution using around $1.6\times10^4$ samples on the coarsest $16\times16$ grid level. The empirical PDFs for the first three components of the Reynolds stress at a location inside one of the vortices $(y/H ,z/H)=(0.52,0.21)$ are presented in Fig. \ref{PDF} for the two cases. The PDFs of other components of the Reynolds stress tensor exhibit similar behaviour, and are omitted. We observe that the sample mean is very close to the baseline value and for Case 2, due to a larger $\delta$, a slight deviation ($\sim 5\times10^{-4}$) from the baseline is observed, consistent with the sampling error. The state of the anisotropy resulting from the samples of the random Reynolds stresses is visualized using the barycentric triangle \cite{Banerjee} in Fig. \ref{bary_PDF}. Again the probability density contours are based on $1.6\times10^4$ samples at location $(y/H ,z/H)=(0.52,0.21)$ for each case. The procedure to construct these contours is explained in Appendix {A1}. We observe that the distance between the state of anisotropy obtained from the baseline simulation and the sample mean is sensitive to the dispersion parameter. For a larger dispersion, many samples fall away from the baseline state but due to the positive-definite constraint they are restricted until the edges of the barycentric triangle. Thus, the sample mean is located far from the baseline anisotropy state, see \cite{xiao2} for details. The effect of this constraint is mild for a smaller dispersion and the mean anisotropy state is very close to the baseline.
\begin{table}[H]
\begin{center}
\begin{tabularx}{.8\textwidth}{>{\hsize=.5\hsize}X>{\hsize=2\hsize}X>{\hsize=.4\hsize}X>{\hsize=.4\hsize}X}\toprule[1pt]
Parameter & Description & Case 1 & Case 2  \\\midrule\midrule
$l_y/H,l_z/H$ & Correlation length along $y/ z$-direction & 2&1\\
$\sigma^2_c$ & Variance of log-normal random field&1&1\\
$\delta(\mathbf{x})$ & Dispersion parameter &0.1 & 0.4\\
$N_{PC}$ & Order of polynomial chaos expansion & 5&5\\
$\xi$& Blending factor&1&1\\
\bottomrule[1pt]
\end{tabularx}
\end{center}
\caption{Parameter sets to generate random Reynolds stress tensor for the square duct flow.}\label{SD_params}
\end{table}

\begin{figure}[H]
\begin{subfigure}[b]{0.03\textwidth}
\rotatebox{90}{Baseline}
\vspace{2cm}
\end{subfigure}
\begin{subfigure}[b]{0.32\textwidth}
\caption*{$R_{11}$}
{\includegraphics[clip, trim=0cm 0cm 0cm 0cm,scale =0.18]{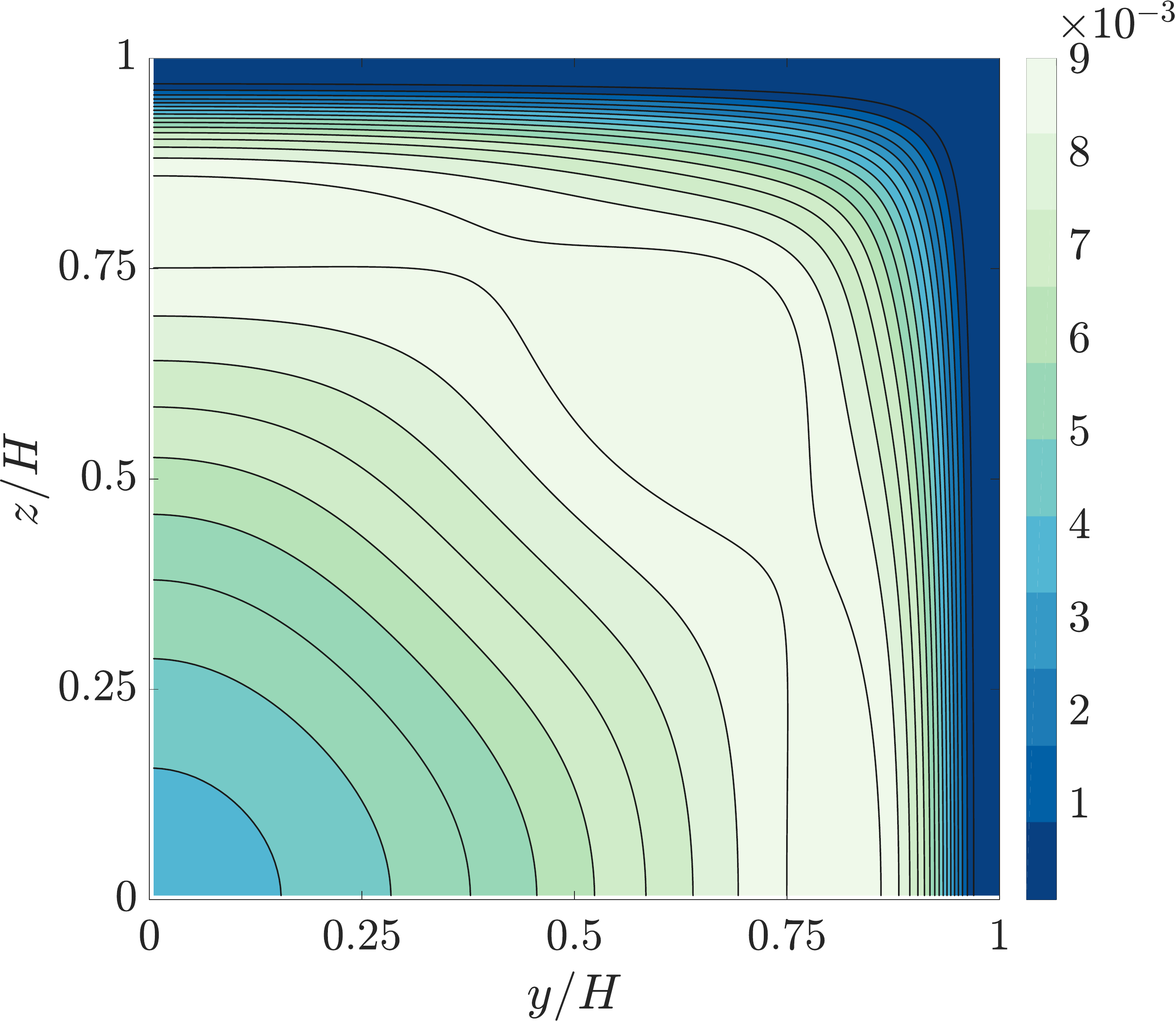}}
\end{subfigure}
\begin{subfigure}[b]{0.32\textwidth}
\caption*{$R_{12}$}
{\includegraphics[clip,  trim=0cm 0cm 0cm 0cm,scale=0.18]{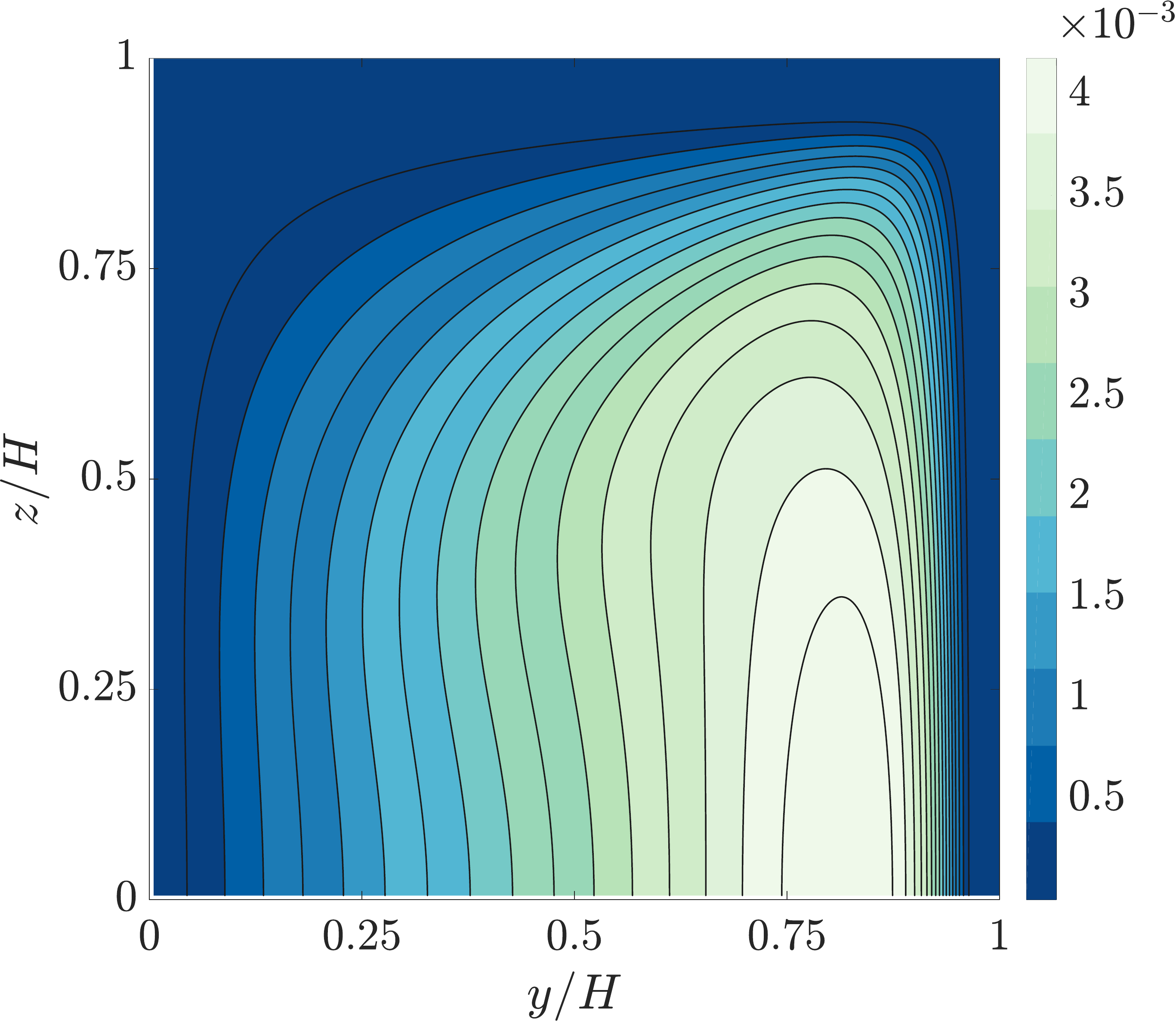}}
\end{subfigure}
\begin{subfigure}[b]{0.31\textwidth}
\caption*{$R_{13}$}
{\includegraphics[clip,  trim=0cm 0cm 0cm 0cm,scale=0.18]{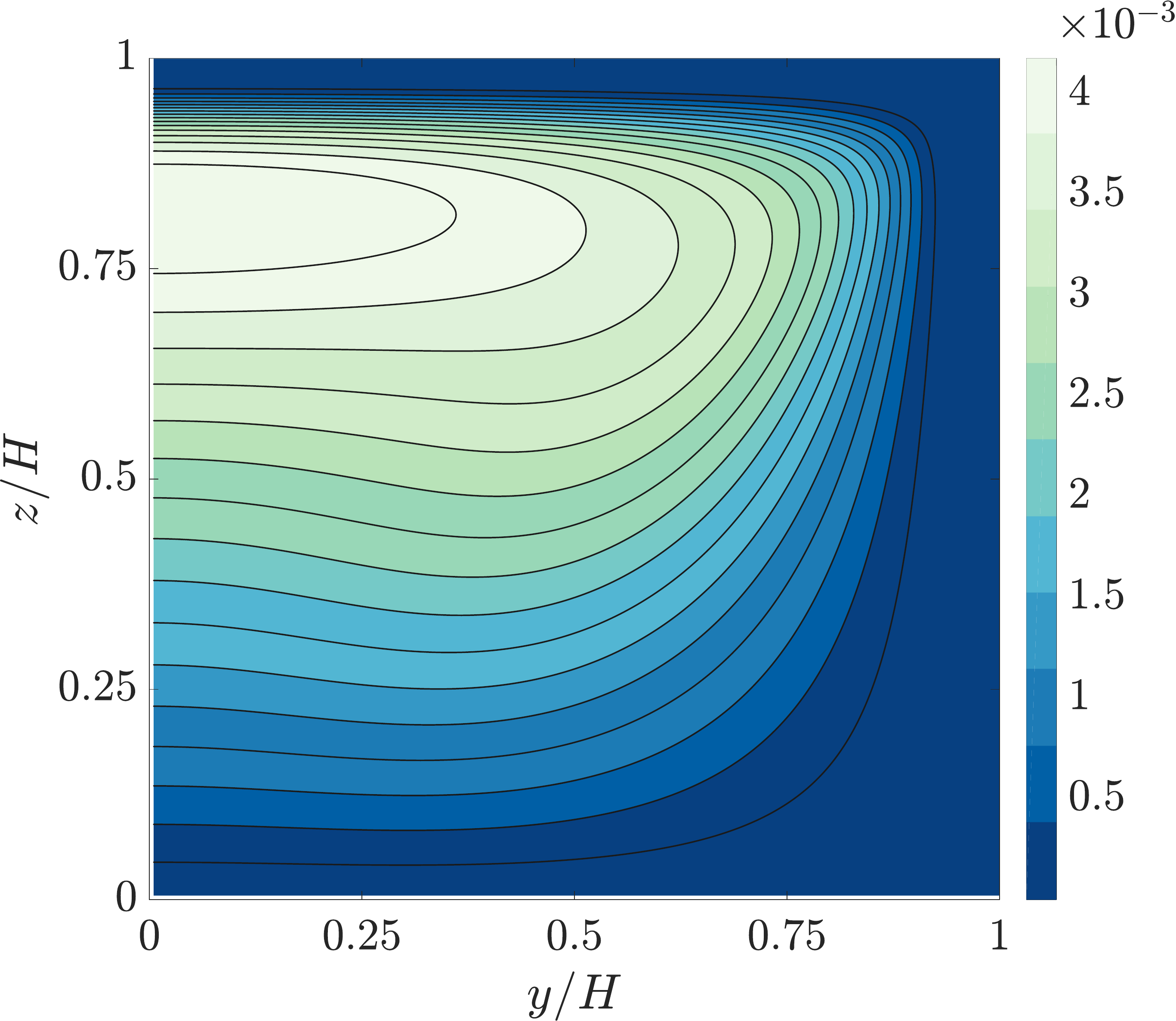}}
\end{subfigure}

\begin{subfigure}[b]{0.03\textwidth}
\rotatebox{90}{Case 1}
\vspace{2cm}
\end{subfigure}
\begin{subfigure}[b]{0.32\textwidth}
{\includegraphics[clip, trim=0cm 0cm 0cm 0cm,scale =0.18]{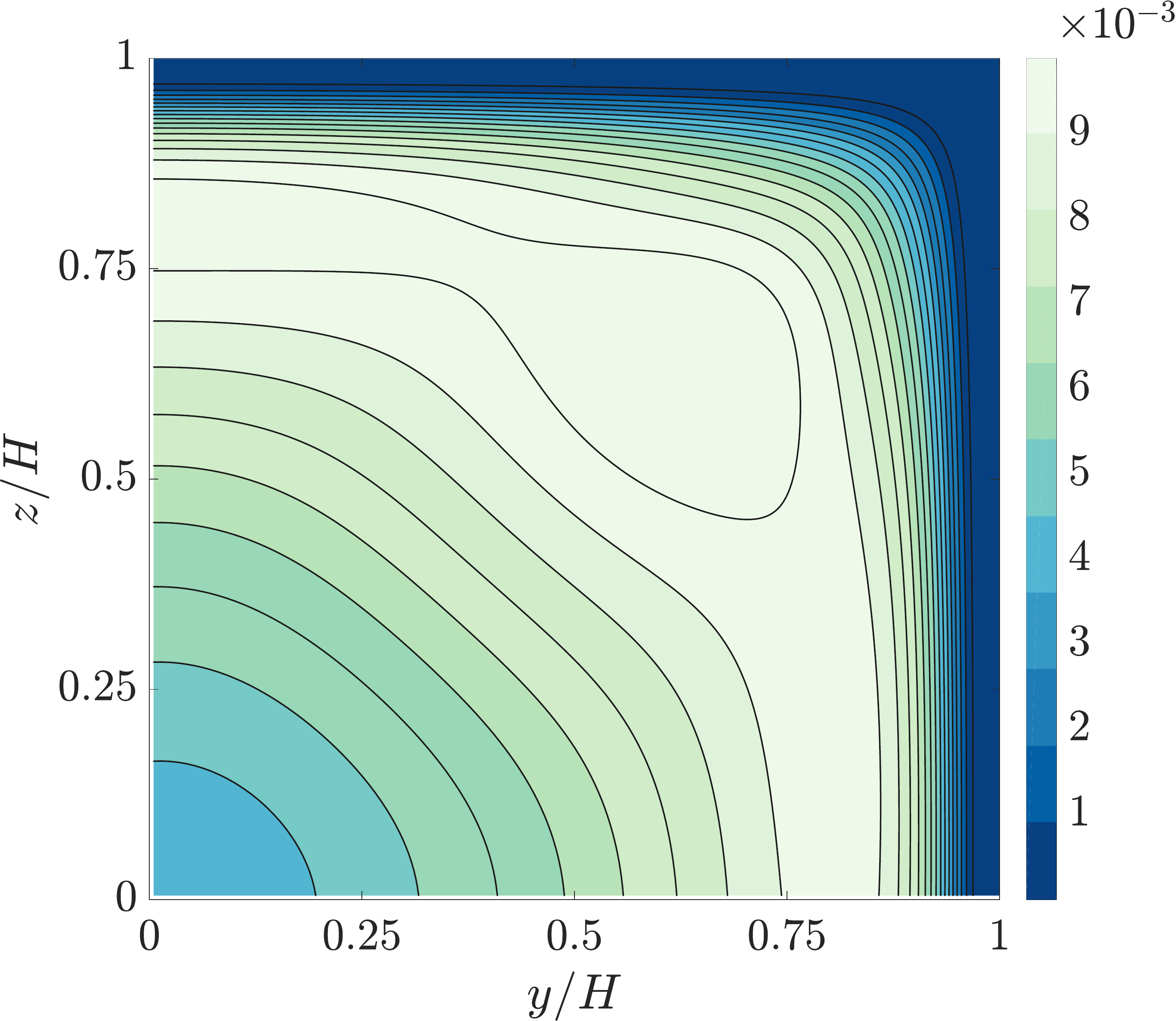}}
%\caption{Case 1}
\end{subfigure}
\begin{subfigure}[b]{0.32\textwidth}
{\includegraphics[clip,  trim=0cm 0cm 0cm 0cm,scale=0.18]{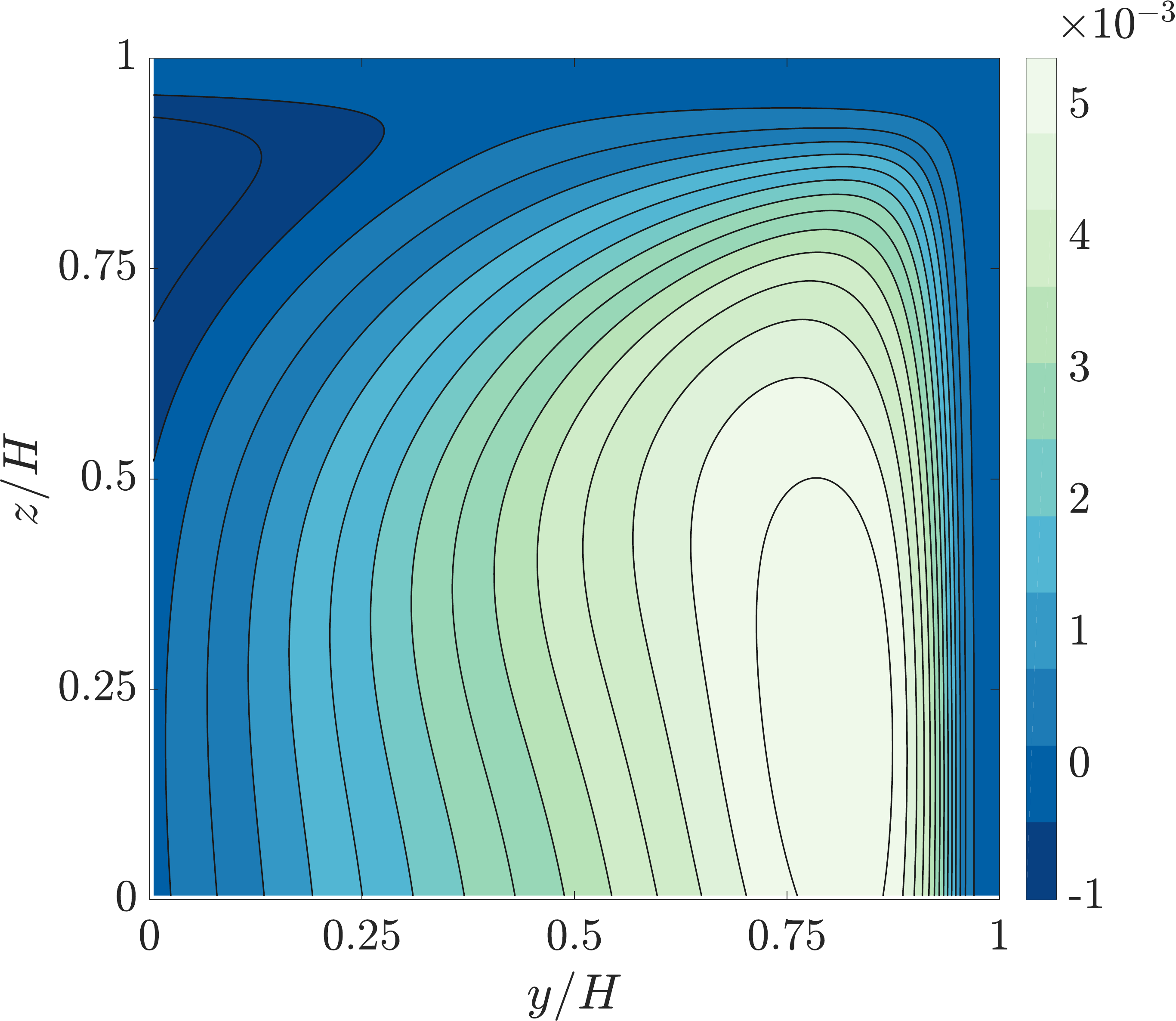}}
%\caption{Case 2}
\end{subfigure}
\begin{subfigure}[b]{0.31\textwidth}
{\includegraphics[clip,  trim=0cm 0cm 0cm 0cm,scale=0.18]{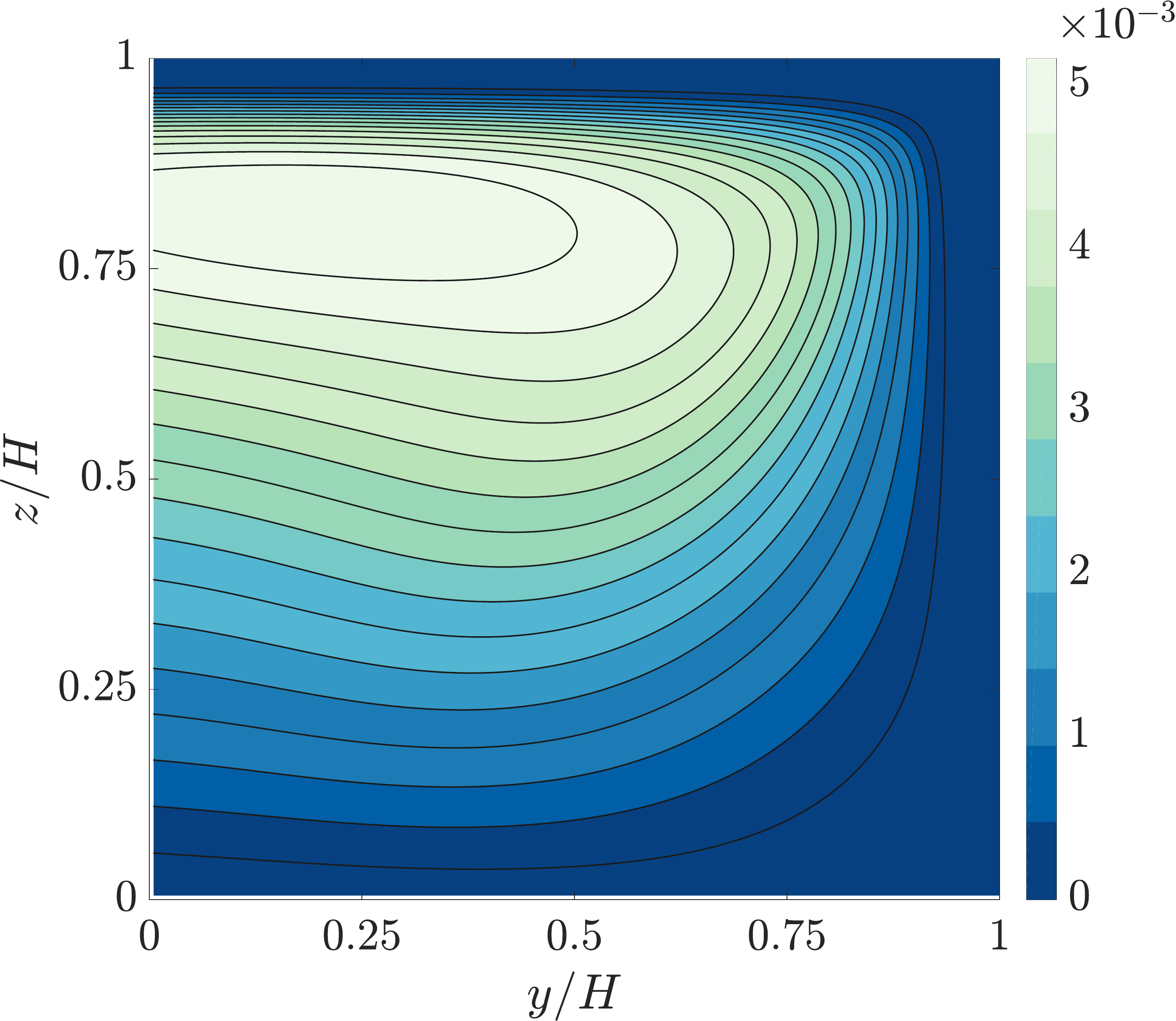}}
%\caption{DNS}
\end{subfigure}

\begin{subfigure}[b]{0.03\textwidth}
\rotatebox{90}{Case 2}
\vspace{2cm}
\end{subfigure}
\begin{subfigure}[b]{0.32\textwidth}
{\includegraphics[clip, trim=0cm 0cm 0cm 0cm,scale =0.18]{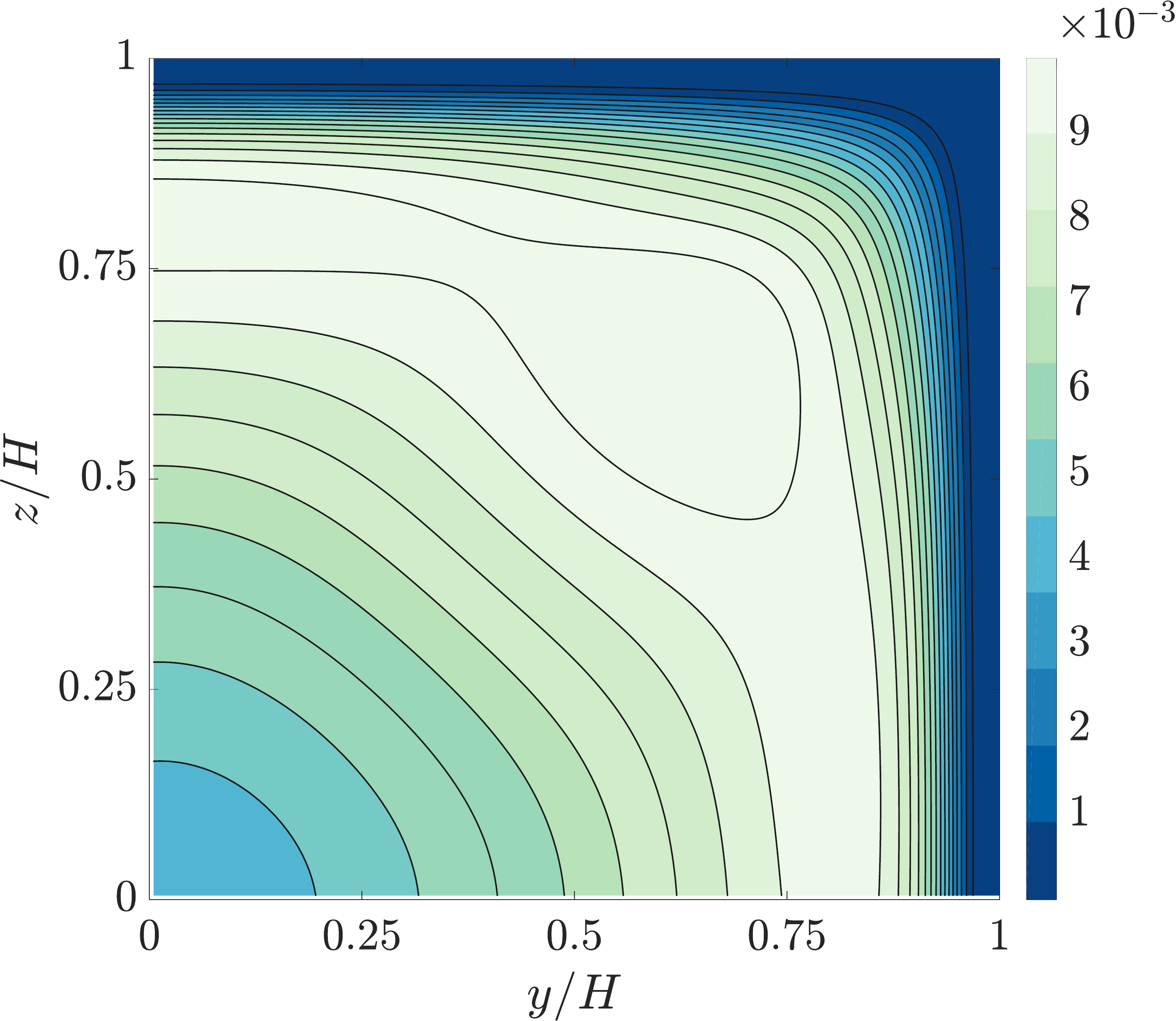}}
%\caption{Case 1}
\end{subfigure}
\begin{subfigure}[b]{0.32\textwidth}
{\includegraphics[clip,  trim=0cm 0cm 0cm 0cm,scale=0.18]{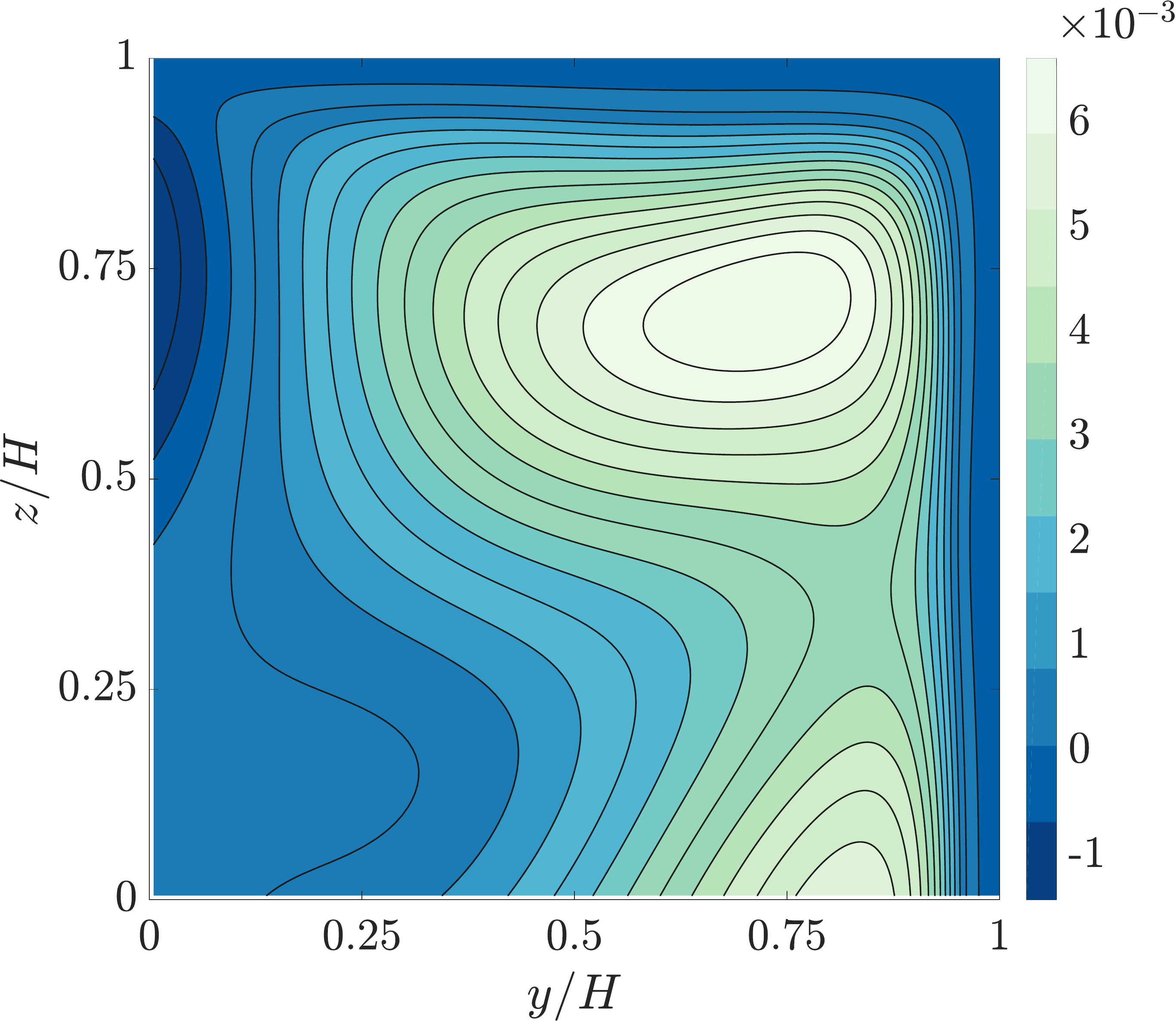}}
%\caption{Case 2}
\end{subfigure}
\begin{subfigure}[b]{0.31\textwidth}
{\includegraphics[clip,  trim=0cm 0cm 0cm 0cm,scale=0.18]{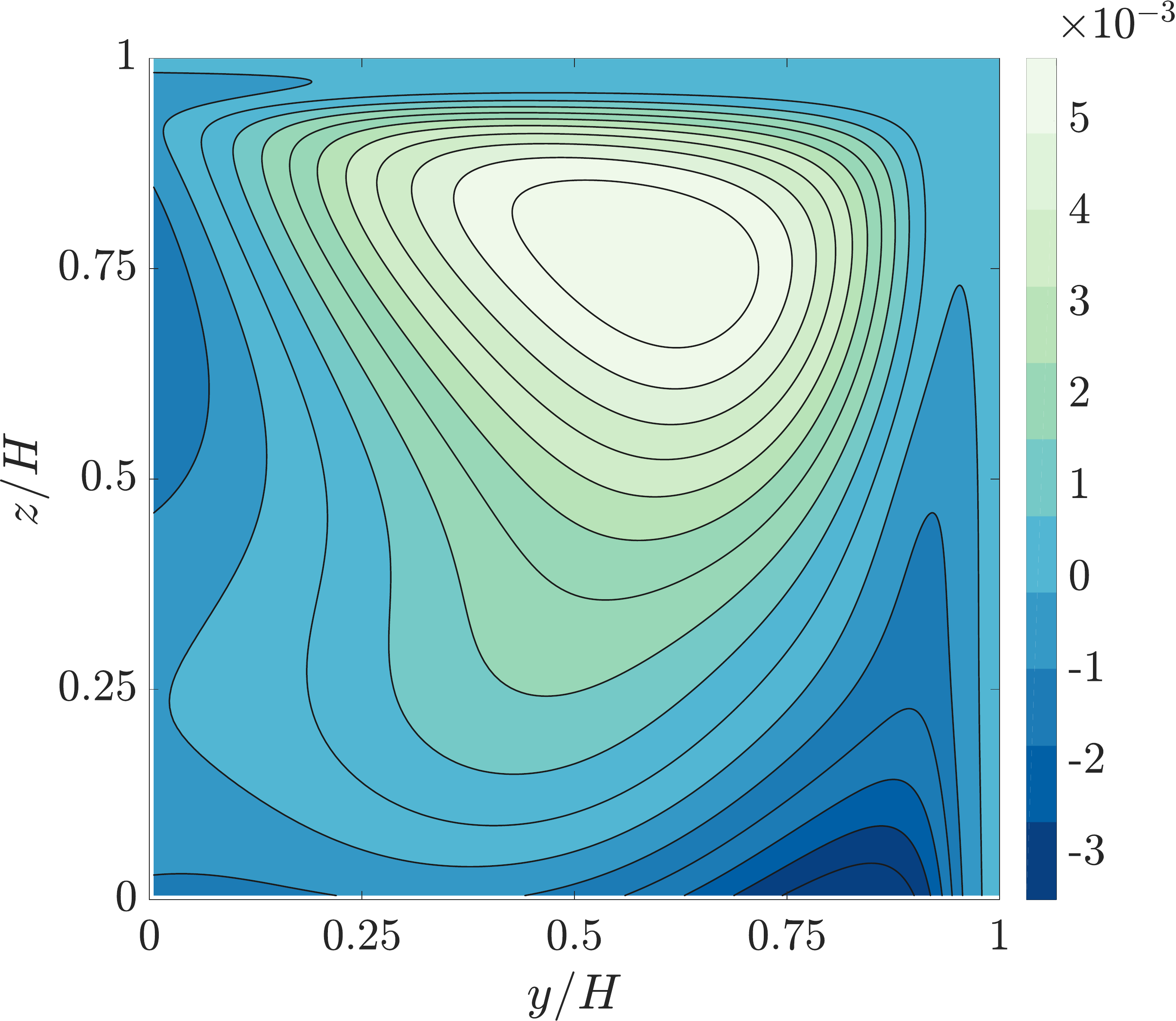}}
%\caption{DNS}
\end{subfigure}
\caption{Reynolds stress components, $R_{11},R_{12},R_{13}$, obtained from the baseline $k-\omega$ model (top row) and an example of perturbed random Reynolds stresses generated from Case 1 (middle row) and Case 2 (bottom row).}\label{Ex_RRST}
\end{figure}

\begin{figure}[H]
\begin{subfigure}[b]{0.32\textwidth}
{\includegraphics[clip, trim=0cm 0cm 0cm 0cm,scale =0.18]{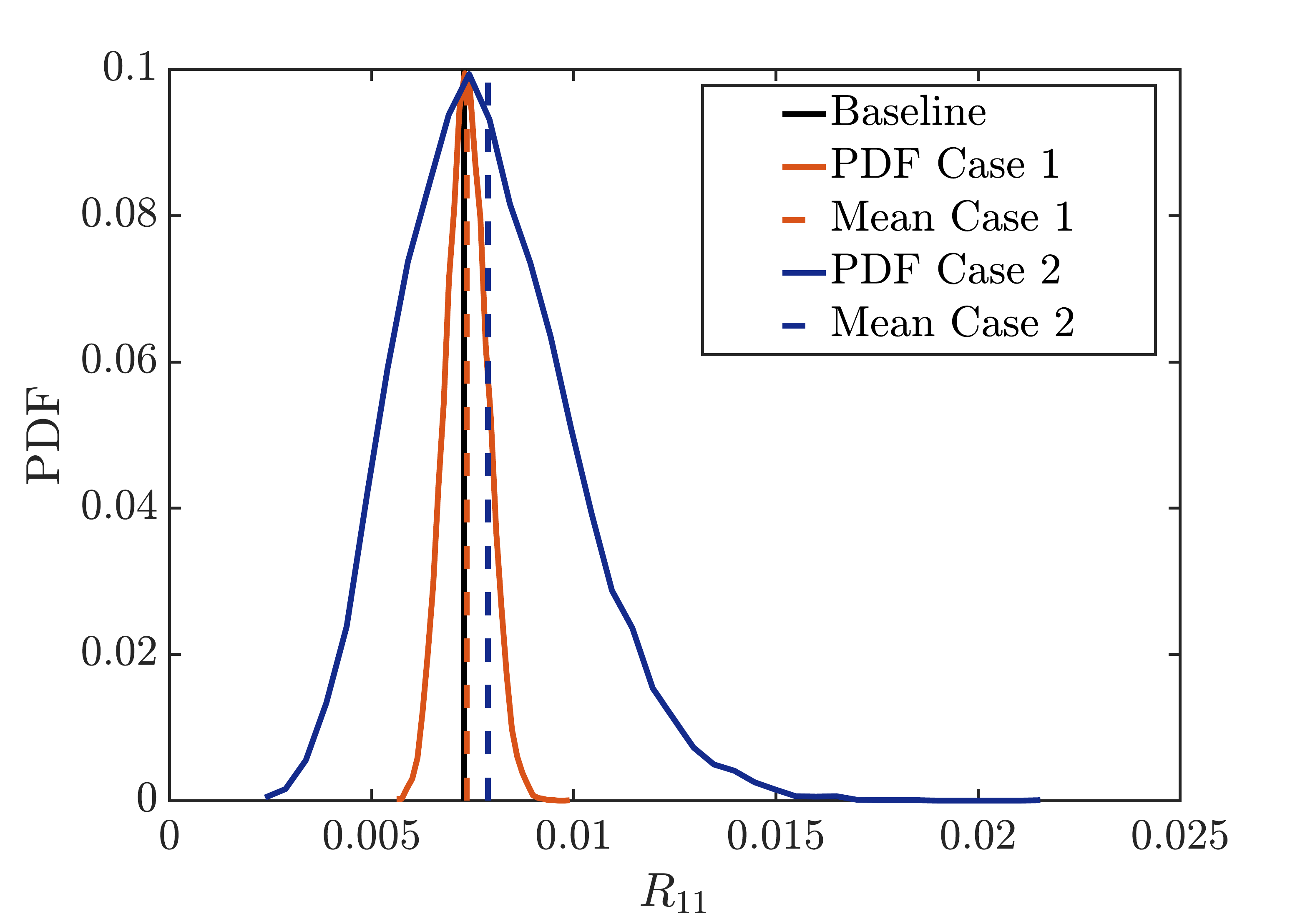}}
\caption{$R_{11}$}
\end{subfigure}
\begin{subfigure}[b]{0.32\textwidth}
{\includegraphics[clip,  trim=0cm 0cm 0cm 0cm,scale=0.18]{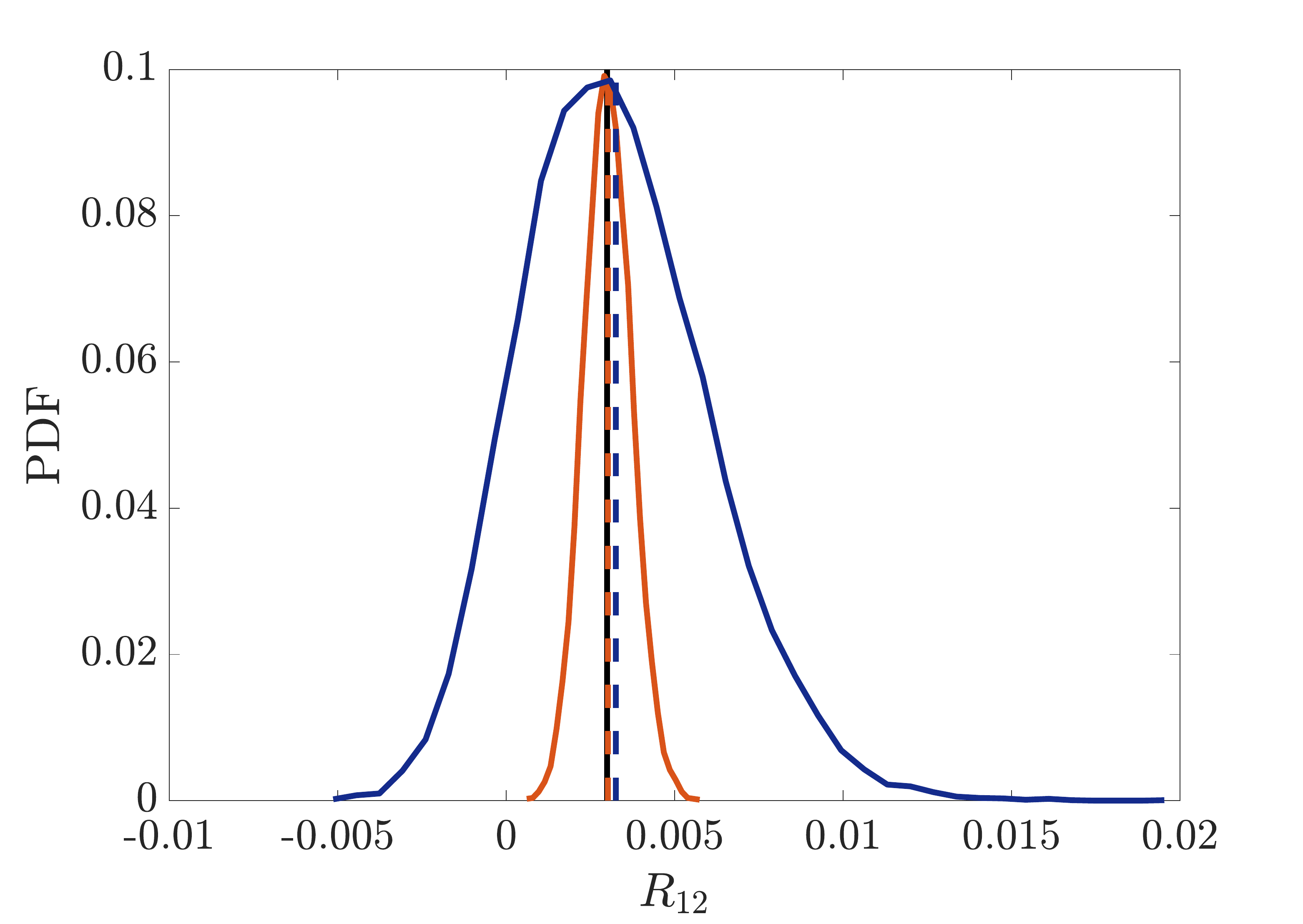}}
\caption{$R_{12}$}
\end{subfigure}
\begin{subfigure}[b]{0.32\textwidth}
{\includegraphics[clip,  trim=0cm 0cm 0cm 0cm,scale=0.18]{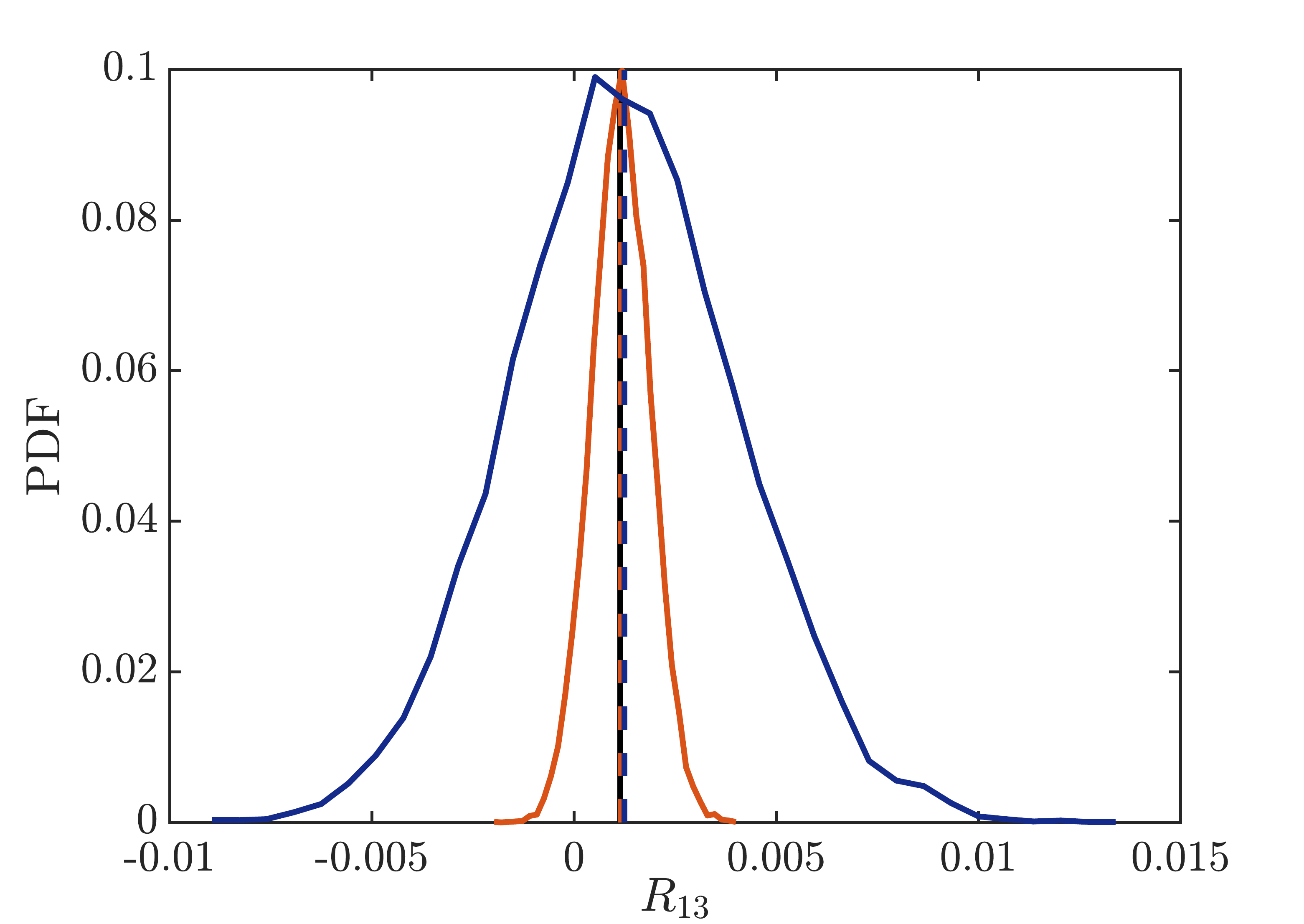}}
\caption{$R_{13}$}
\end{subfigure}
\caption{Empirical PDF of the Reynolds stress components at location $(y/H,z.H) = (0.52,0.21)$ for Case 1 ($\delta =0.1$) and Case 2 ($\delta =0.4$). For the diagonal component $R_{11}$, a gamma marginal distribution is obtained and for the off-diagonal components $R_{12},R_{13}$, Gaussian distributions are observed.}\label{PDF}
\end{figure}
\begin{figure}[H]
\begin{center}
\begin{subfigure}[b]{0.49\textwidth}
\hbox{\hspace{0cm}{\includegraphics[clip, trim=0cm 0cm 0cm 0cm,scale=0.27]{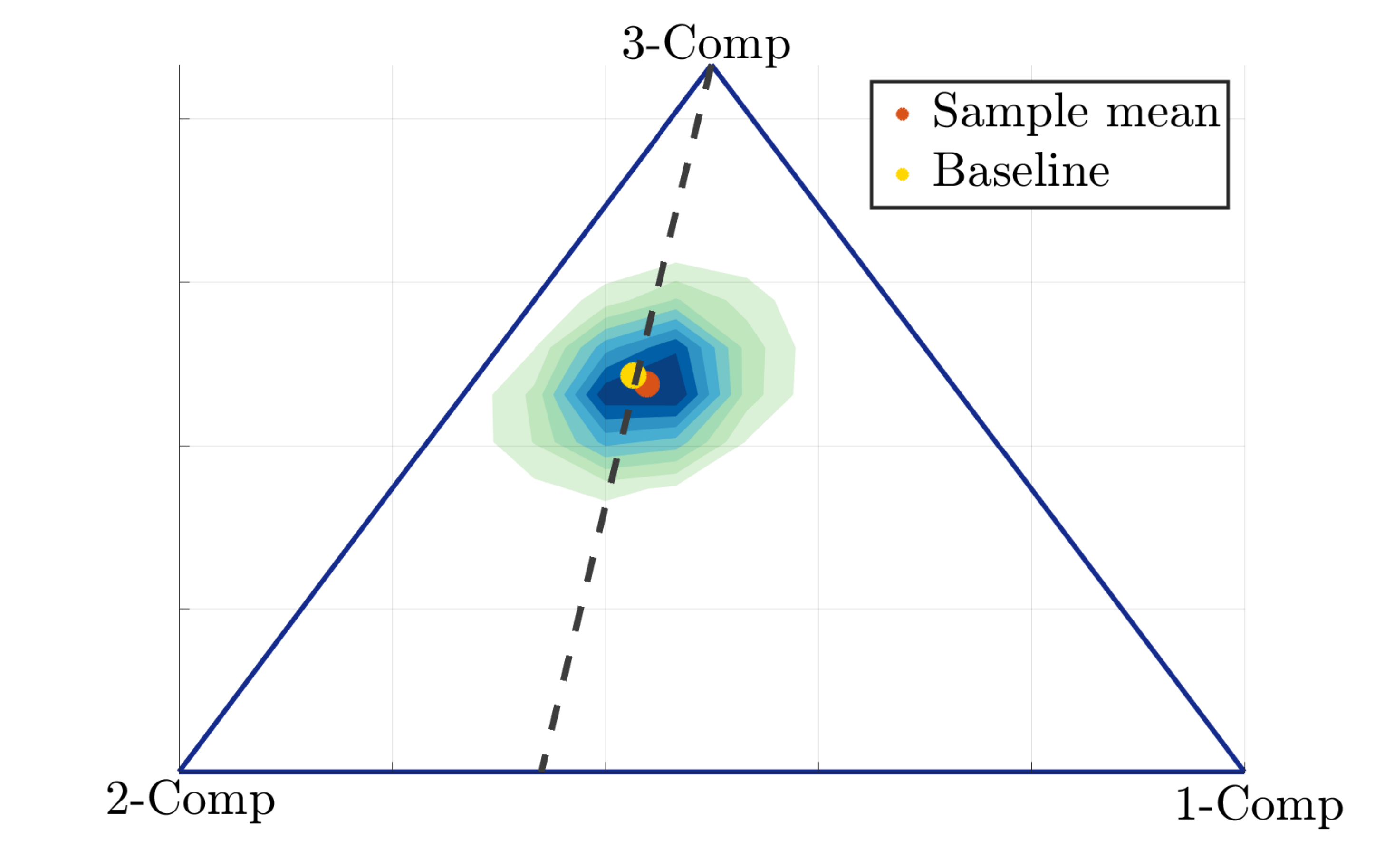}}}
\caption{Case 1, $\delta =0.1$}
\end{subfigure}
 \begin{subfigure}[b]{0.49\textwidth}
\hbox{\hspace{0cm}{\includegraphics[clip, trim=0cm 0cm 0cm 0cm,scale=0.27]{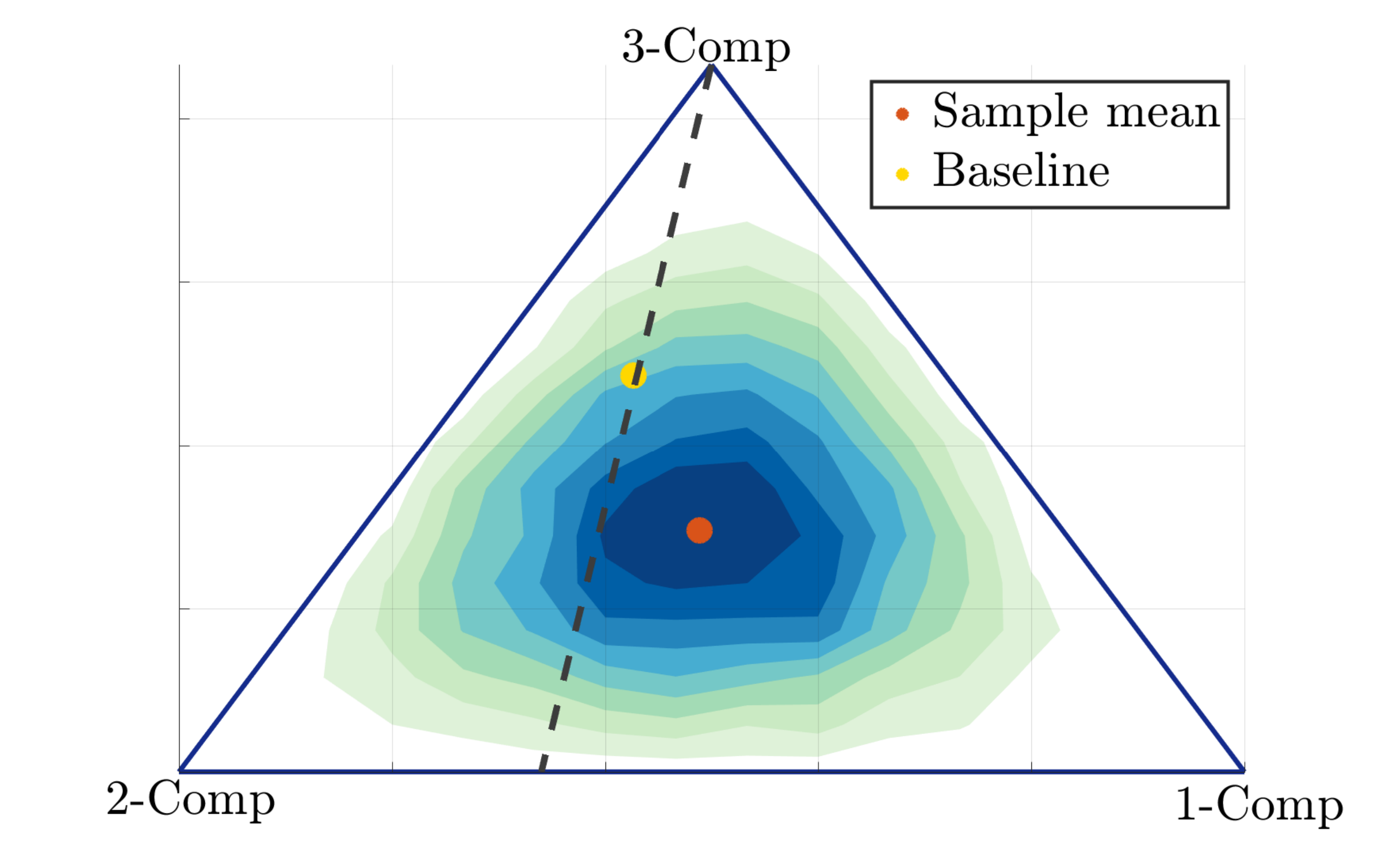}}}
\caption{Case 2, $\delta =0.4$}
\end{subfigure}
\end{center}
\caption{Probability density contours for random Reynolds stresses at location $(y/H, z/H)  = (0.52,0.21)$ projected to the barycentric triangle based on $1.6\times10^4$  samples on $\ell=0$.}\label{bary_PDF}
\end{figure}

%\subsection{Numerical convergence analysis}
We begin by studying the FV error convergence for Case 1 and Case 2. We will only consider the $u$ and $v$ components of the velocity as $w$ has similar characteristics as $v$. In Fig. \ref{bias_SD}, we show the relative error $||u_\ell - u_{\ell -1}||_{\Lomd}$  along with the FV errors from the deterministic RANS simulations (based on the $k-\omega$ model) plotted against the maximum cell width $h_\ell$. The relative error for $v$ is also presented in Fig. \ref{bias_SD} (right). As the deterministic RANS simulation predicts $v=0$, we again use the deterministic error in $u$ for comparison of the FV convergence rates. These relative errors are computed with a sufficient number of samples such that sampling errors on each level are less than the FV bias. We observe a convergence of $\mathcal{O}(h_\ell^{1.5})$ (rounded to one decimal place) for the deterministic simulations and further note that the stochastic version of the FV error also decays at a similar rate. Here, we remark that although we use second-order accurate schemes, a slightly slower error convergence is obtained, most likely due to the non-uniformity of the meshes used. Also, deterministic simulations on the finest $256\times256$ grid, OpenFOAM has convergence issues. Interestingly, this is not observed for the stochastic simulations. Further, due to a higher value of the dispersion parameter $\delta$ for Case 2, compared to Case 1, we see a larger absolute numerical error, but it decays at a similar rate. These plots are important in order to determine the number of levels that should be included in the MLMC hierarchy to reduce the RMSE to a given tolerance $\varepsilon$. For the standard Monte Carlo simulation, the error associated with a particular mesh is utilized to determine the number of samples needed on that mesh, to equilibrate the sampling error with the discretization error, as in \eqref{MCsamp}.  

The convergence of the level-dependent variance $\lnorm \mathcal{V}[\cdot]\rnorm_{\Ld}$ is shown in Fig. \ref{var_SD}. For reference, an $\mathcal{O}(h_\ell^3)$ convergence line is plotted to emphasize $\beta\approx 2\alpha$. The significance of these plots is that they can be used to assess the sampling variance at different levels and extract the rate $\beta$ used to determine the MLMC sample sequence in the formula \eqref{eq:Nl}. We observe a higher variance for larger dispersion from Case 2 compared to Case 1, as expected.
\begin{figure}[H]
\begin{subfigure}[b]{0.49\textwidth}
{\includegraphics[clip, trim=1cm 0cm 0cm 0cm, scale =0.27]{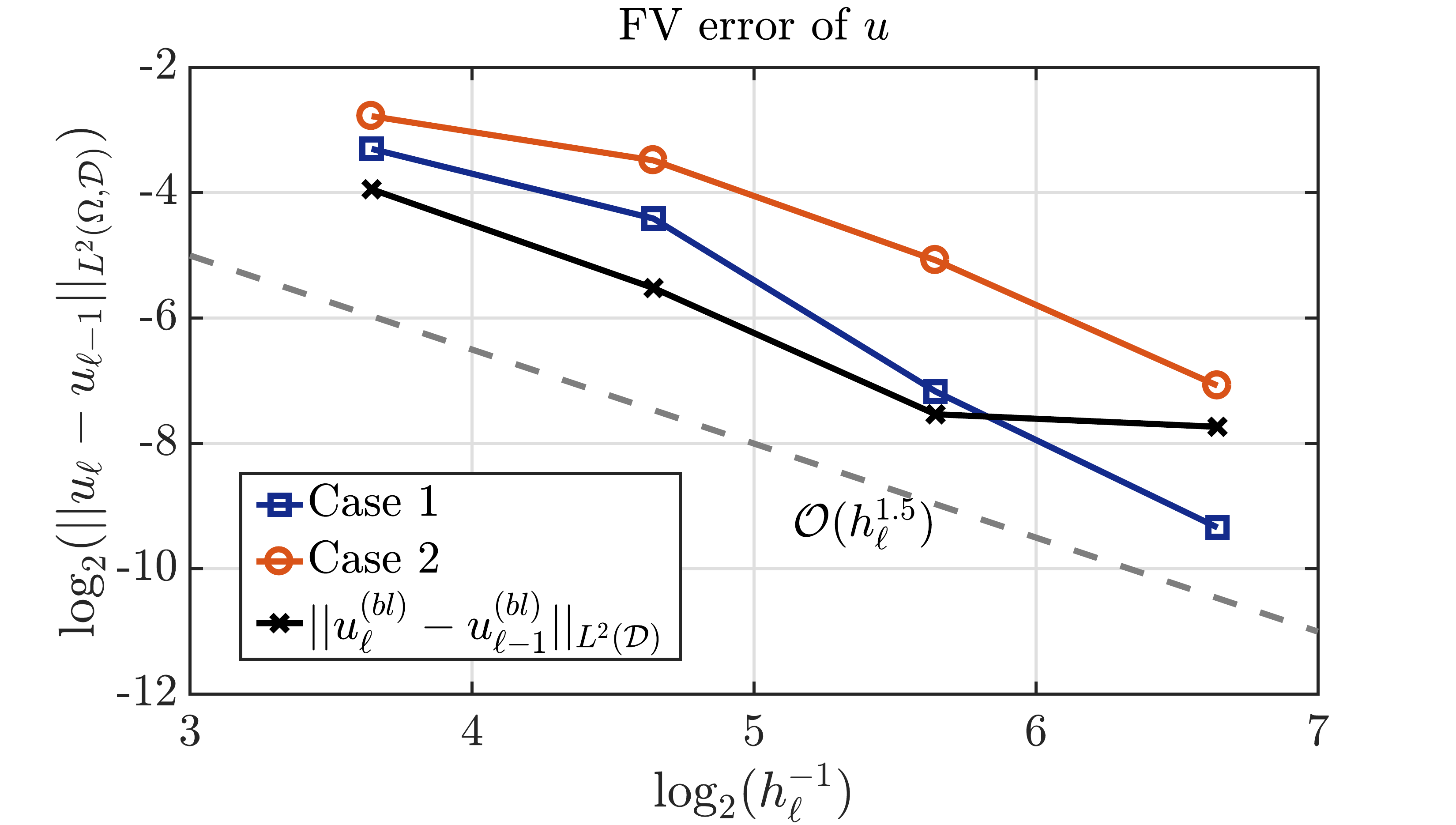}}
\end{subfigure}
\begin{subfigure}[b]{0.49\textwidth}
{\includegraphics[clip,  trim=1cm 0cm 0cm 0cm, scale=0.27]{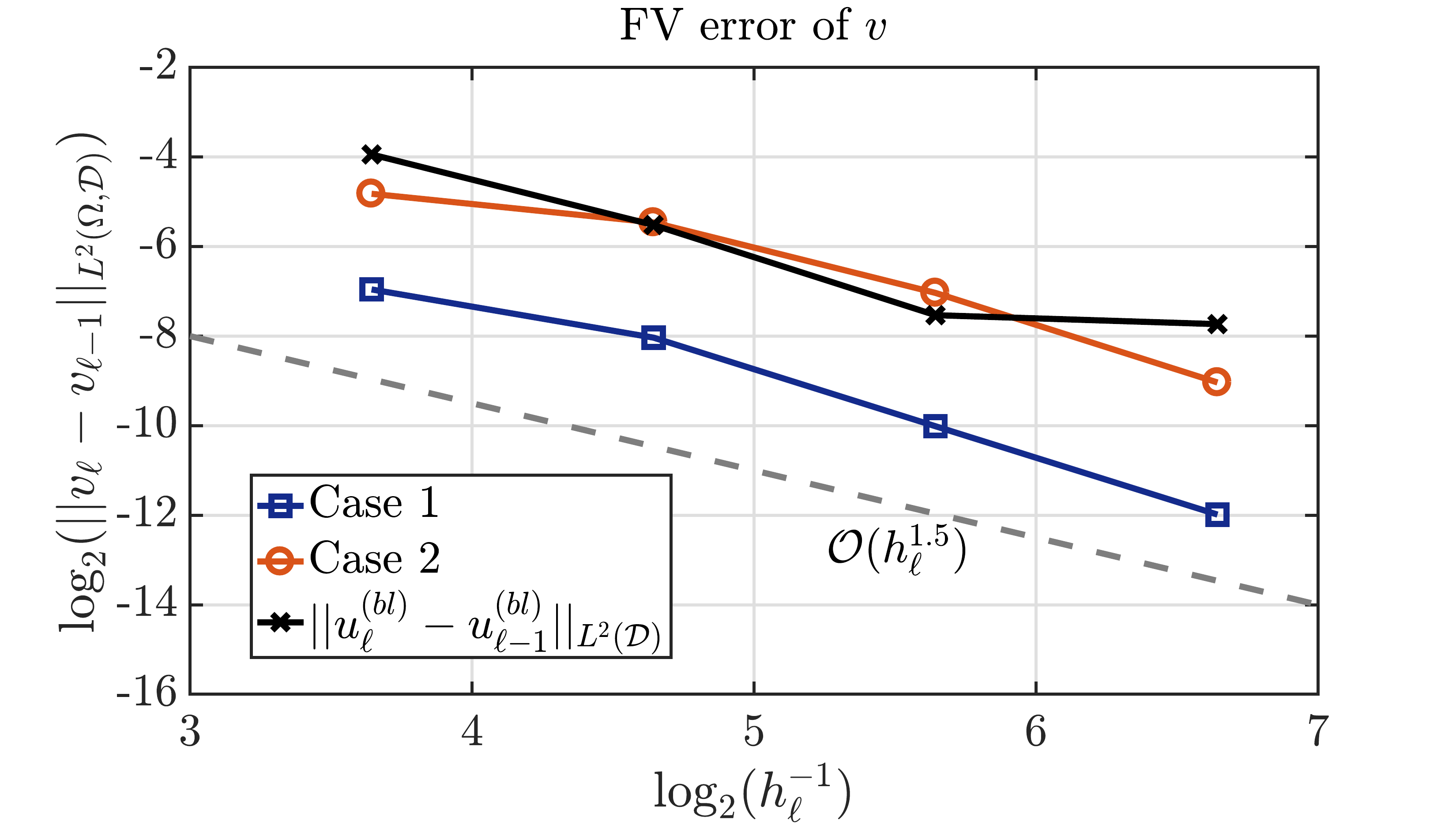}}
\end{subfigure}
\caption{Convergence of the FV error with levels along with error in baseline solution of $u$. Dotted line denotes the empirical convergence rate of baseline RANS simulations.}\label{bias_SD}
\end{figure}
\begin{figure}[H]
\begin{subfigure}[b]{0.49\textwidth}
{\includegraphics[clip, trim=1cm 0cm 0cm 0cm, scale =0.27]{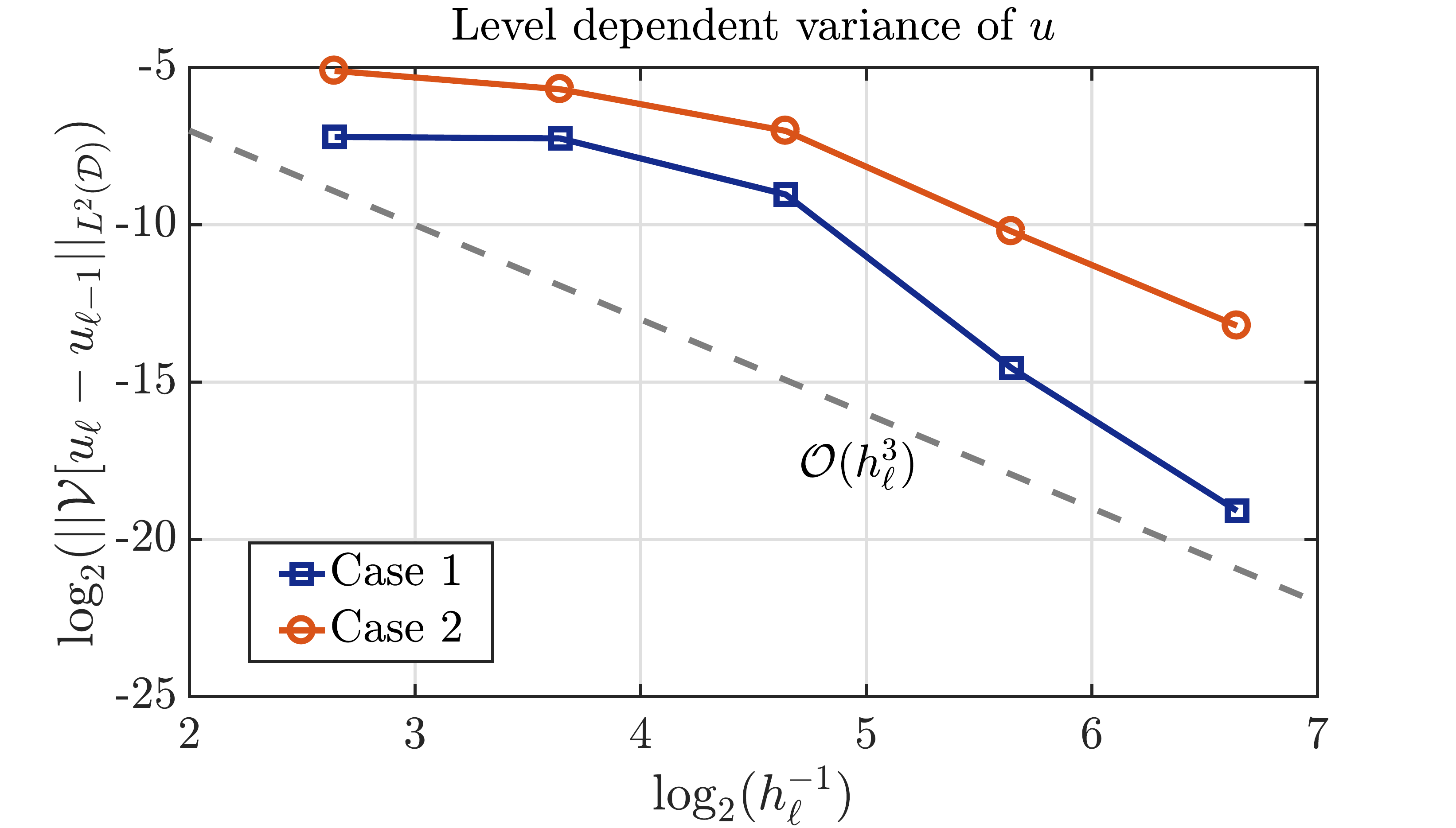}}
\end{subfigure}
\begin{subfigure}[b]{0.49\textwidth}
{\includegraphics[clip,  trim=1cm 0cm 0cm 0cm, scale=0.27]{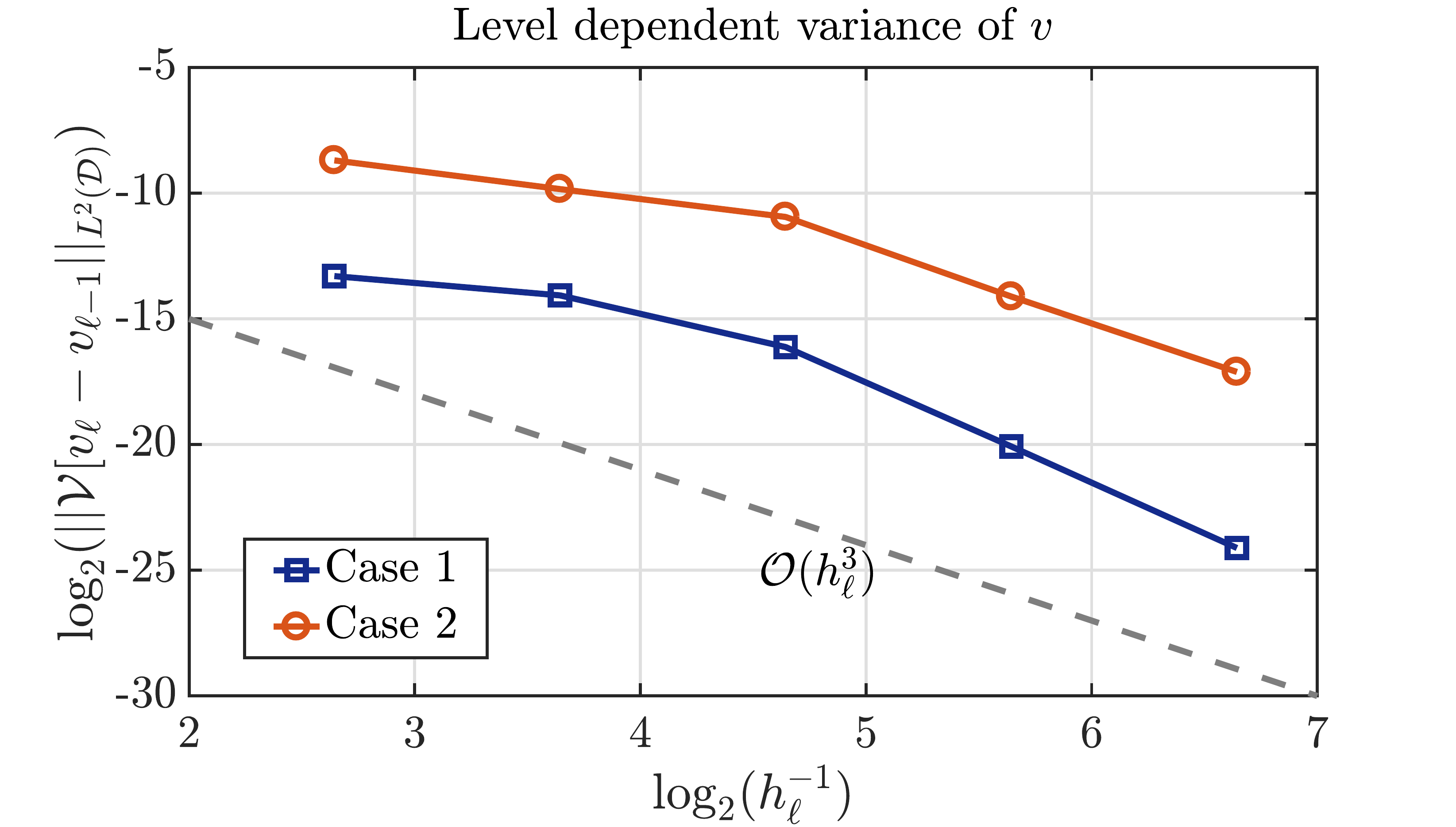}}
\end{subfigure}
\caption{Convergence of the level-dependent variance with grid refinement. The dotted line depicts an $\mathcal{O}(h_\ell^3)$ convergence.}\label{var_SD}
\end{figure}
We point out that the above convergence study can be quite expensive, as many samples over all the levels are needed to obtain accurate estimates of the MLMC rates $\alpha,\beta$. The purpose of the above analysis is to (i) demonstrate that the FV error decay rate extracted from the deterministic solves can be an accurate estimate of the rate $\alpha$ and (ii) verify that the assumption $\beta=2\alpha$ holds. With a fixed $\beta$, we can obtain the number of samples on all MLMC levels in advance and can efficiently distribute the work on a computing cluster. Alternatively, one can also implement the standard MLMC algorithm \cite{MLMC1,MLMC2} which adaptively computes and refines these parameters along with the number of samples on each level until a prescribed tolerance is achieved. Note that for such algorithms, optimal load balancing is non-trivial as the number of samples predicted on the different levels after every refinement stage varies.

Next, we compare the accuracy and computation cost of the MLMC and MC estimators to compute the mean and variance. For this analysis the in-plane velocity $v$ is chosen as the quantity of interest. To measure the accuracy, we rely on the following relative error measure \cite{mishra2012sparse,mishra2016multi}:
\begin{equation}\label{eq:MLMC_rel_err}
\varepsilon_{rel}  := \frac{||\mathpzc{E}_{ref}[v] - \mathpzc{E}^{ML}_L[v_L] ||_{L^2(\mathcal{D}_L)}}{||\mathpzc{E}_{ref}[v]||_{L^2(\mathcal{D}_L)}}.
\end{equation} 
Here, $\mathpzc{E}^{ML}_L[v_L]$ can be replaced by the standard MC estimator $\mathpzc{E}^{MC}_N[v_h]$. Analogously, the relative errors in the variance estimators $\mathcal{V}^{MC}_N$ and $\mathcal{V}^{ML}_L$ are also computed. For the MLMC estimator, we compute the mean and variance for different $h_L$ (or $h$ for the standard MC). These experiments are conducted 16 times to eliminate statistical fluctuations and the mean relative error $\overline{\varepsilon}_{rel}$ is reported. The reference solutions for the expected value $\mathpzc{E}_{ref}[v]$ and the variance $\mathcal{V}_{ref}[v]$ are computed using the 5-level MLMC estimator. Reference solutions will be discussed in detail later on.

Based on the deterministic FV error convergence study, we fix $\alpha=1.5$ and $\beta=2\alpha=3$ and $\gamma=3$ (see Table \ref{sqDuct}). Thus, for the MLMC estimator, we get a sample sequence $N_\ell = N_L2^{3{(L-\ell)}}$ based on the formula \eqref{eq:Nl}. Note that we have $\beta\approx\gamma$ and therefore we can obtain an MLMC estimator for which all levels contribute equally in terms of the cost, see \eqref{MLMC_comp}. As mentioned earlier, the number of samples on the finest level $N_L$ is a free parameter and should be set to a small value. For all experiments, we use $N_L=8$. In Table \ref{MLMC_samples_SD}, we list the number of level-wise samples for the MLMC estimators with different $L$. For the standard (or single-level) MC estimator, the number of samples is decided according to \eqref{MCsamp} resulting in $N = \mathcal{O}(h^{-3})$. This means that the number of MC samples should be  increased by a factor of eight with each grid refinement. The standard MC simulation was conducted on three grids: $16\times16, 32\times32$ and $64\times64$ with samples 8, 64 and 512, respectively. The standard MC was not performed on the grid $128\times128$ due to prohibitively large computational cost, as we would need to compute about 4096 samples on this grid. 

In Fig. \ref{erel_mean} (left), we show the mean relative errors in the expected value of $v$ computed using the MC and MLMC estimators for Case 1.  We observe that the plain MC estimator is slightly more accurate than the MLMC estimator for same finest grid $h_L$. The computational cost versus the accuracy for both methods is also shown in Fig. \ref{erel_mean} (right) and we observe that the MLMC estimator achieves same accuracy for a lower computational cost compared to the MC estimator. For reference, the predicted asymptotic cost of the MC \eqref{MC_comp} and MLMC \eqref{MLMC_comp} estimators for the considered $\alpha,\beta$ and $\gamma$ are also presented.  Similarly, the error and runtime from the two variance estimators are compared in Fig. \ref{erel_var}. Ideally, the cost of the MLMC estimator is expected to grow at half the rate of the MC estimator but this is not clearly visible for the multilevel estimator for the mean. This may very well be a pre-asymptotic effect. Nevertheless, the gains are more pronounced for the multilevel variance estimator and we clearly observe the cost scaling close to the predicted rate.
\begin{table}[H]
\begin{center}
\begin{tabular}{|c|c|c|c|c|c|}
\cline{2-6}
\multicolumn{1}{c}{} & \multicolumn{5}{|c|}{Level-wise samples $N_\ell$}\\
\cline{2-6}
\hline
No. of levels ($L+1$) & $N_0$ & $N_1$ & $N_2$ &$N_3$&$N_4$\\
\hline
1 & 8 & - & - &-&-  \\
2 & 64 & 8  & - & - &-   \\
3 & 512 & 64 & 8 & - &- \\
4 &  4096 & 512 & 64 & 8 & -   \\
5 (ref) &  32768 & 4096 & 512 & 64 & 8 \\
\hline
\end{tabular}
\end{center}
\vspace{0.2cm}
\caption{Number of samples used for the MLMC estimators with different $L$ for the square duct flow. The 5-level MLMC estimator was utilized as the reference solution.}\label{MLMC_samples_SD}
\end{table} 

\begin{figure}[H]
\begin{subfigure}[b]{0.49\textwidth}   
            \centering 
      {\includegraphics[clip,  trim=1cm 0cm 0cm 0cm,scale=0.27]{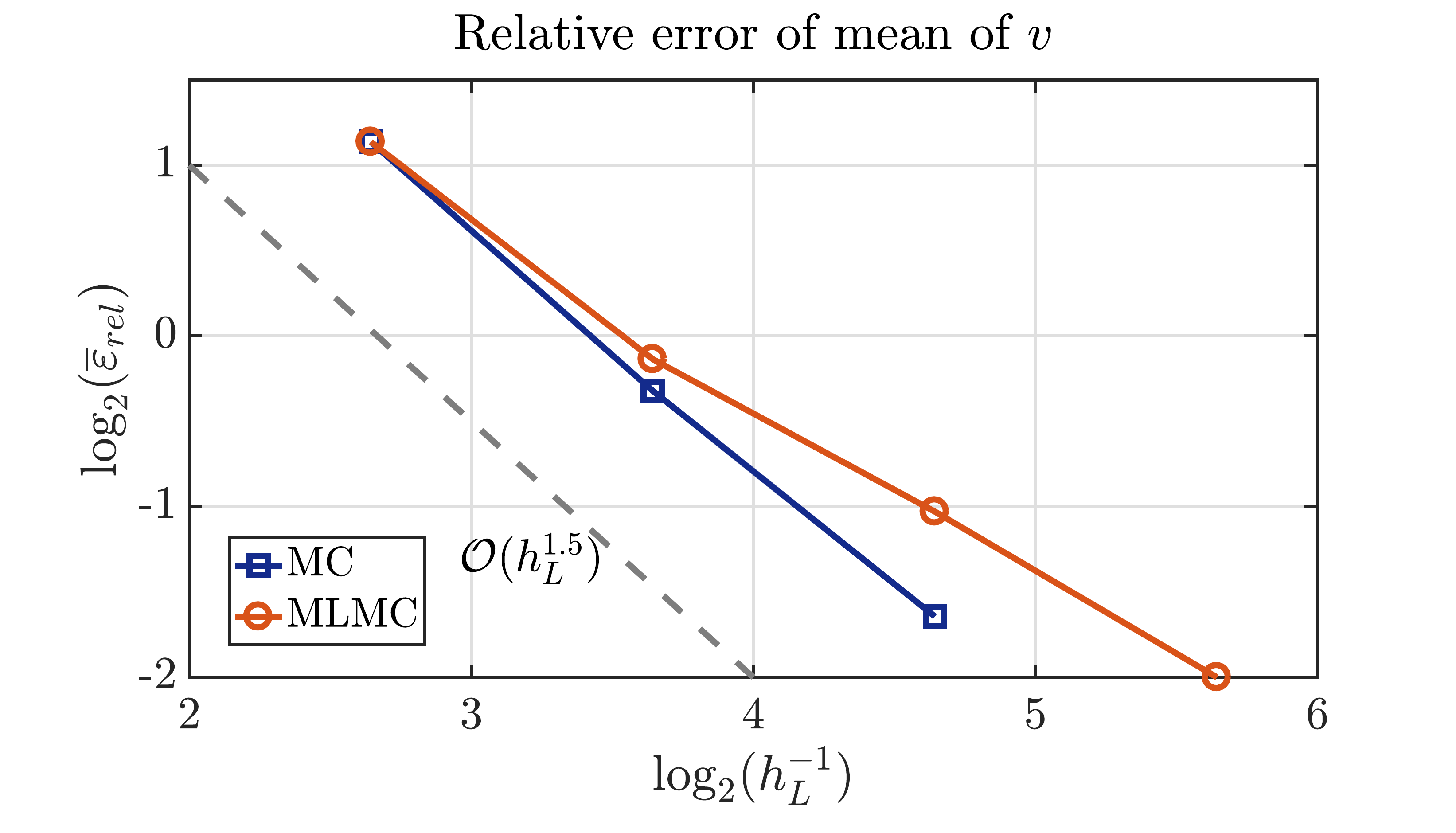}}
        \end{subfigure}
        \begin{subfigure}[b]{0.49\textwidth}   
            \centering 
 {\includegraphics[clip,  trim=1cm 0cm 0cm 0cm,scale=0.27]{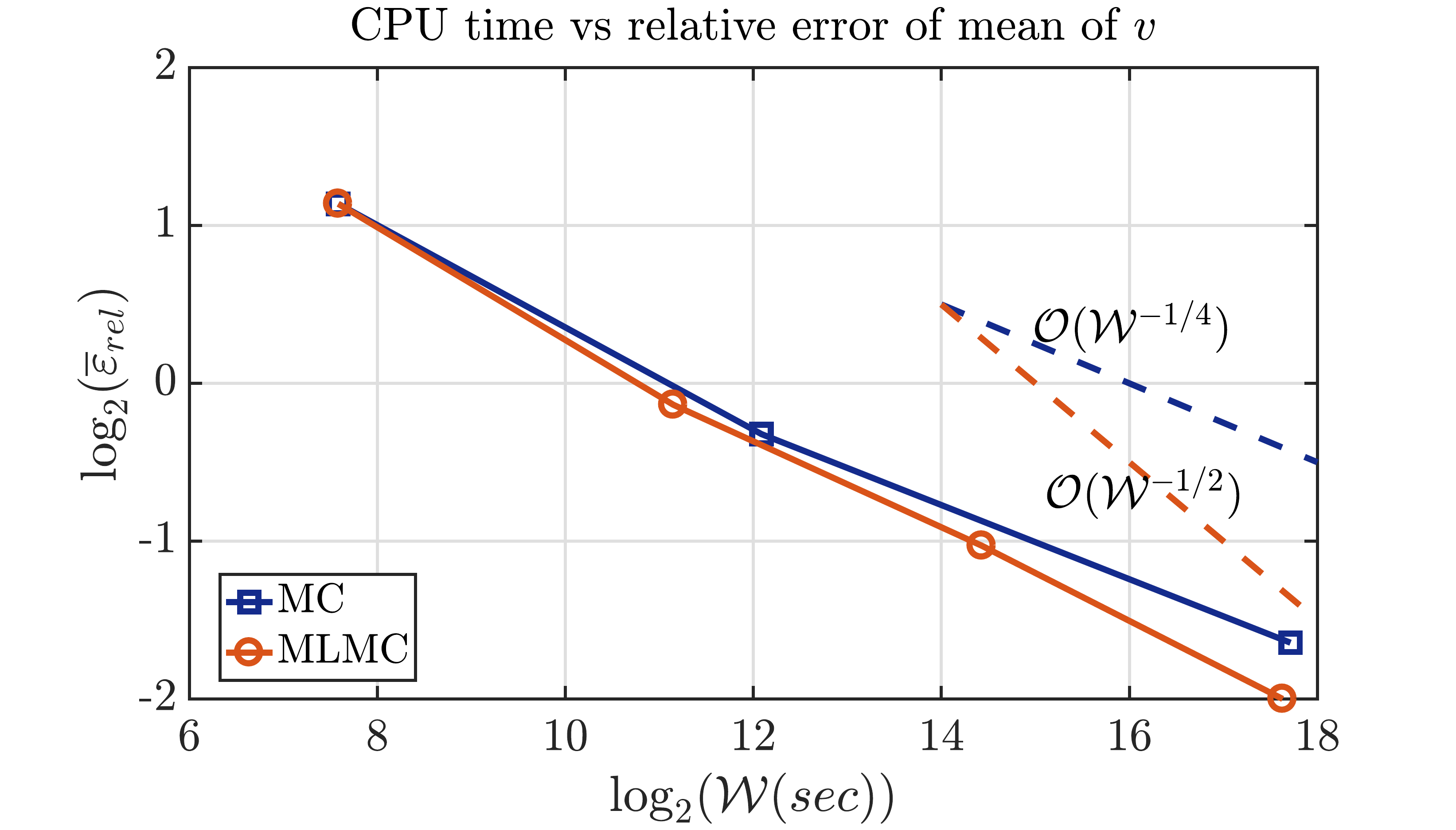}}
        \end{subfigure}
\caption{(Left) Comparison of the mean relative error $\overline{\varepsilon}_{rel}$ in the expected value of $v$ for different  meshes for Case 1. (Right) Computational work versus  accuracy for the MC and MLMC estimators. Dotted lines show the predicted asymptotic cost for the MC (blue) and MLMC (red) estimators.}\label{erel_mean}
\end{figure}

\begin{figure}[H]
\begin{subfigure}[b]{0.49\textwidth}   
            \centering 
      {\includegraphics[clip,  trim=1cm 0cm 0cm 0cm,scale=0.27]{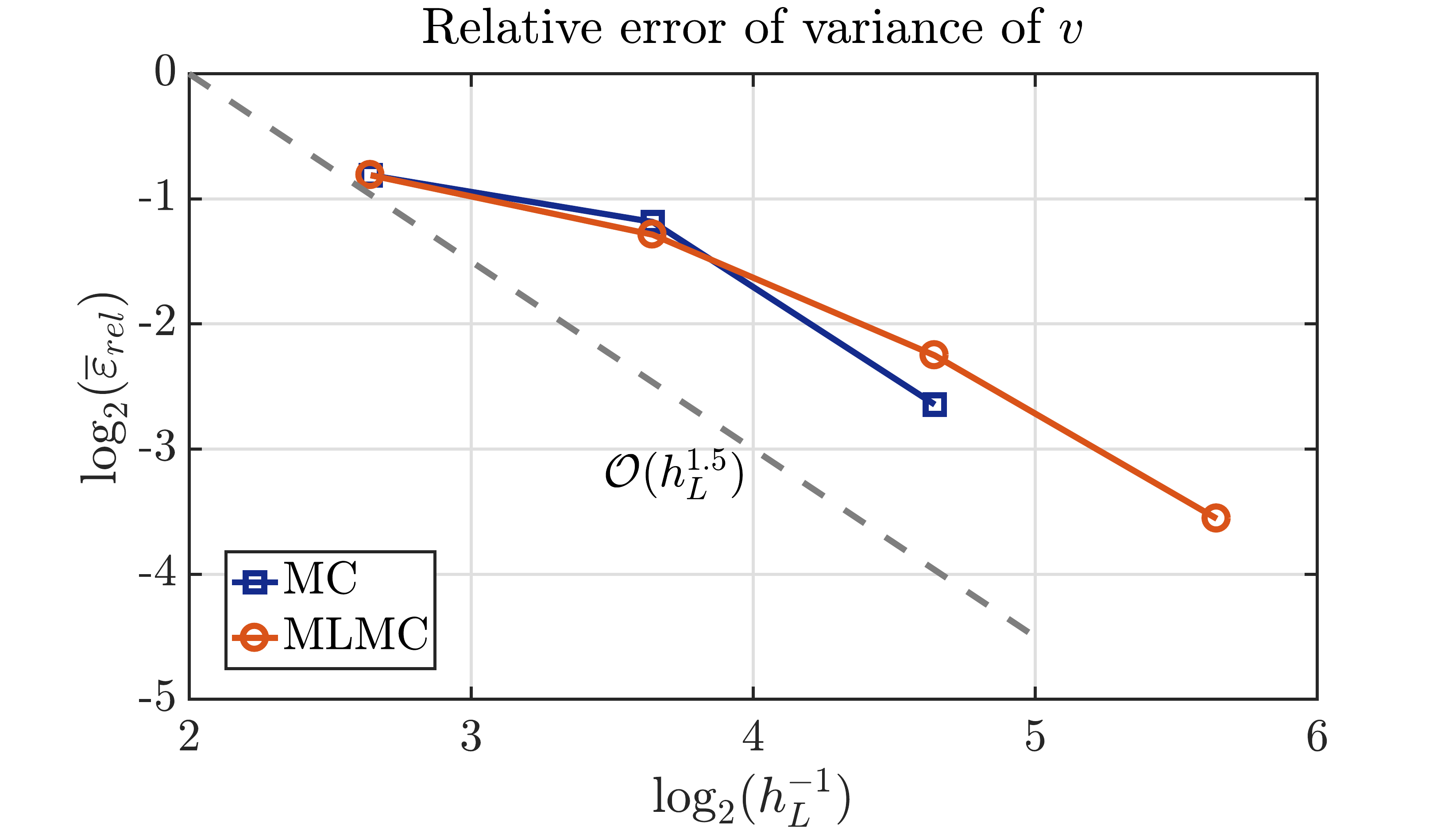}}
        \end{subfigure}
        \begin{subfigure}[b]{0.49\textwidth}   
            \centering 
 {\includegraphics[clip,  trim=1cm 0cm 0cm 0cm,scale=0.27]{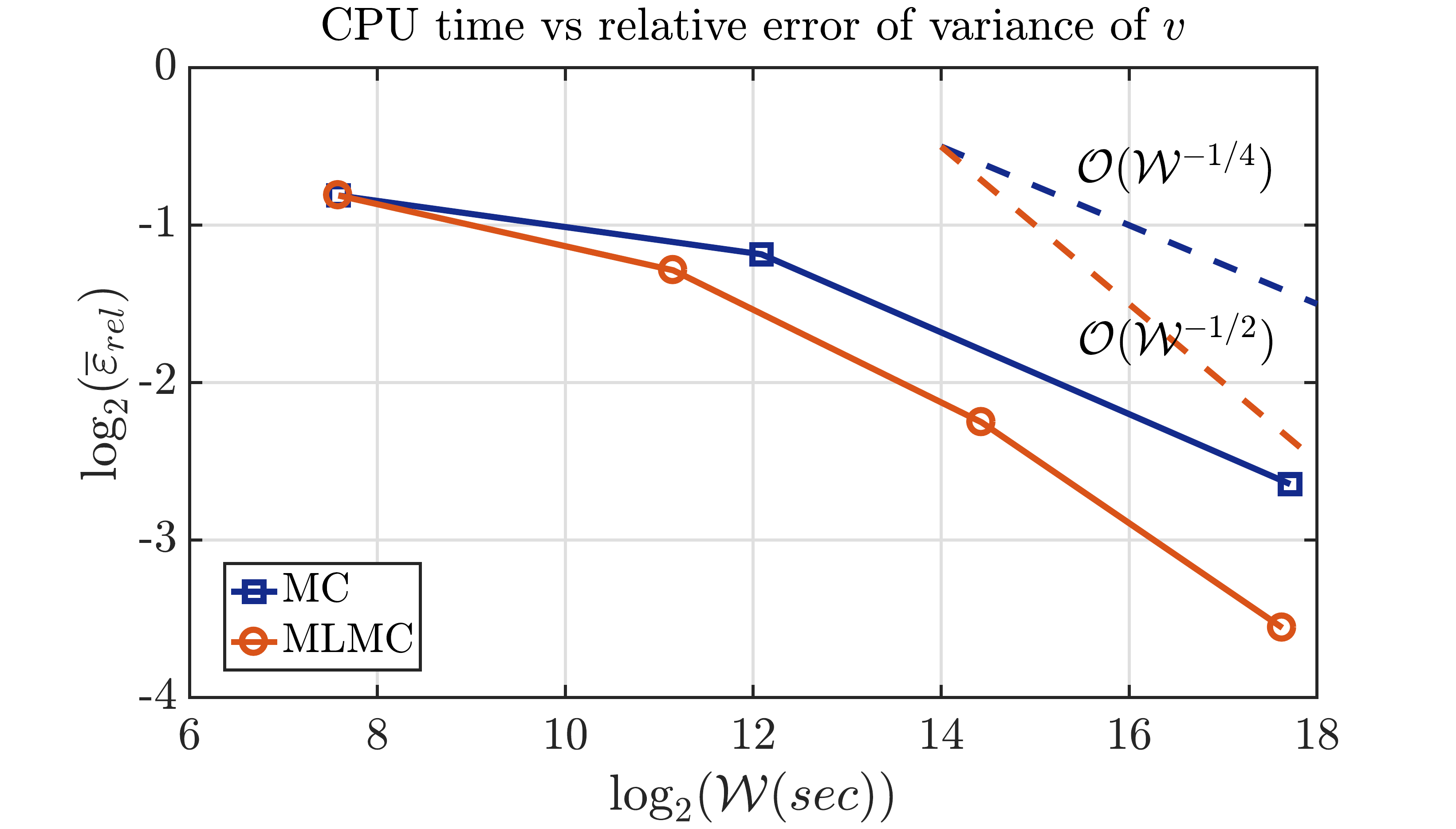}}
        \end{subfigure}
        \caption{(Left) Comparison of the mean relative error $\overline{\varepsilon}_{rel}$ in the variance of $v$ for different meshes for Case 1. (Right) Computational work versus accuracy for the MC and MLMC estimators.}\label{erel_var}
\end{figure}

We now compare the stochastic solutions (mean and variance) for the RRST model computed using the 5-level estimator for Case 1 and 2 with the DNS data. The goal in this setting is to establish that the stochastic model is sufficiently general to (reliably) envelope DNS data at high probability as required for a good prior. 

For the MLMC estimator, an appropriate spatial interpolation method is required to combine all expectations from the telescopic sum \eqref{MLMCestimator}. To interpolate scalar fields from grid $\mathcal{D}_{\ell-1}$ to $\mathcal{D}_\ell$, a second-order spatial interpolation is employed. For instance, when using the multilevel estimator to compute $\mathpzc{E}^{ML}_L[v_L]$, we proceed as follows. We begin by computing  $\mathpzc{E}_{N_0}^{MC}[v_0]$ on the coarsest grid $\mathcal{D}_0$. This is then interpolated to the next finer grid $\mathcal{D}_1$ and is added to the correction term $\mathpzc{E}^{MC}_{N_1}[v_1-v_0]$ resulting in a two-level estimate (a scalar field) $\mathpzc{E}_1^{ML}[v_1]$. Similarly, this scalar field is further interpolated to the next grid and summed with the next correction term $\mathpzc{E}^{MC}_{N_2}[v_2-v_1]$. This process is repeated until the finest level is reached. Another possibility is to interpolate all expectations to the finest level and then add them together. Based on our experience, this may lead to interpolation artifacts in the final outcome.  

In Fig. \ref{stream_SD} streamlines and magnitude of the in-plane velocities from the two cases are compared with the DNS data. We have observed that the size and the number of vortices are sensitive to the correlation length; shorter lengths leading to more vortices. The secondary motions are entirely driven by the RRST model with magnitude of the velocities dependent on the value of the dispersion parameter. The mean $\pm$ standard deviation for the $v$ velocity component at three locations is shown in Fig. \ref{mean_std_SD}. We see that the two standard deviations envelopes the entire DNS velocity well. It is also pointed out that for Case 2, an even larger enveloping region is obtained. As mentioned earlier, we do not take into the account any available data and the hyper-parameters considered to generate the random Reynolds stresses were chosen arbitrarily. This high sensitivity of mean velocities with respect to change in Reynolds stresses is also demonstrated in \cite{THOMPSON20161} where an error of $1\%$ in Reynolds stresses resulted in about $30\%$ error in the mean velocity profile for the plain channel flow.
\begin{figure}[H]
\begin{subfigure}[b]{0.34\textwidth}
{\includegraphics[clip, trim=0cm 0cm 0cm 0cm,scale =0.2]{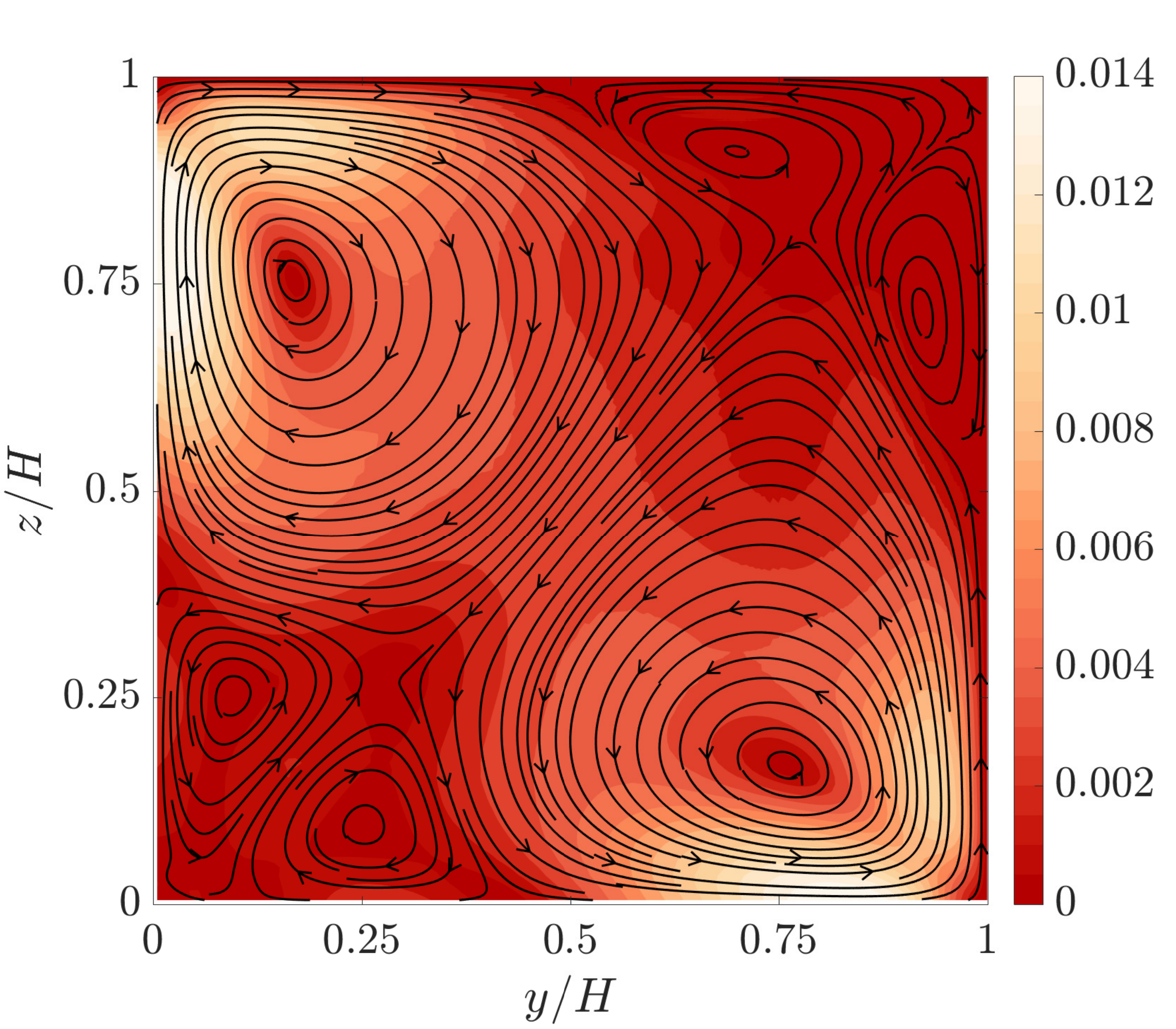}}
\caption{Case 1}
\end{subfigure}
\begin{subfigure}[b]{0.33\textwidth}
{\includegraphics[clip,  trim=0cm 0cm 0cm 0cm,scale=0.2]{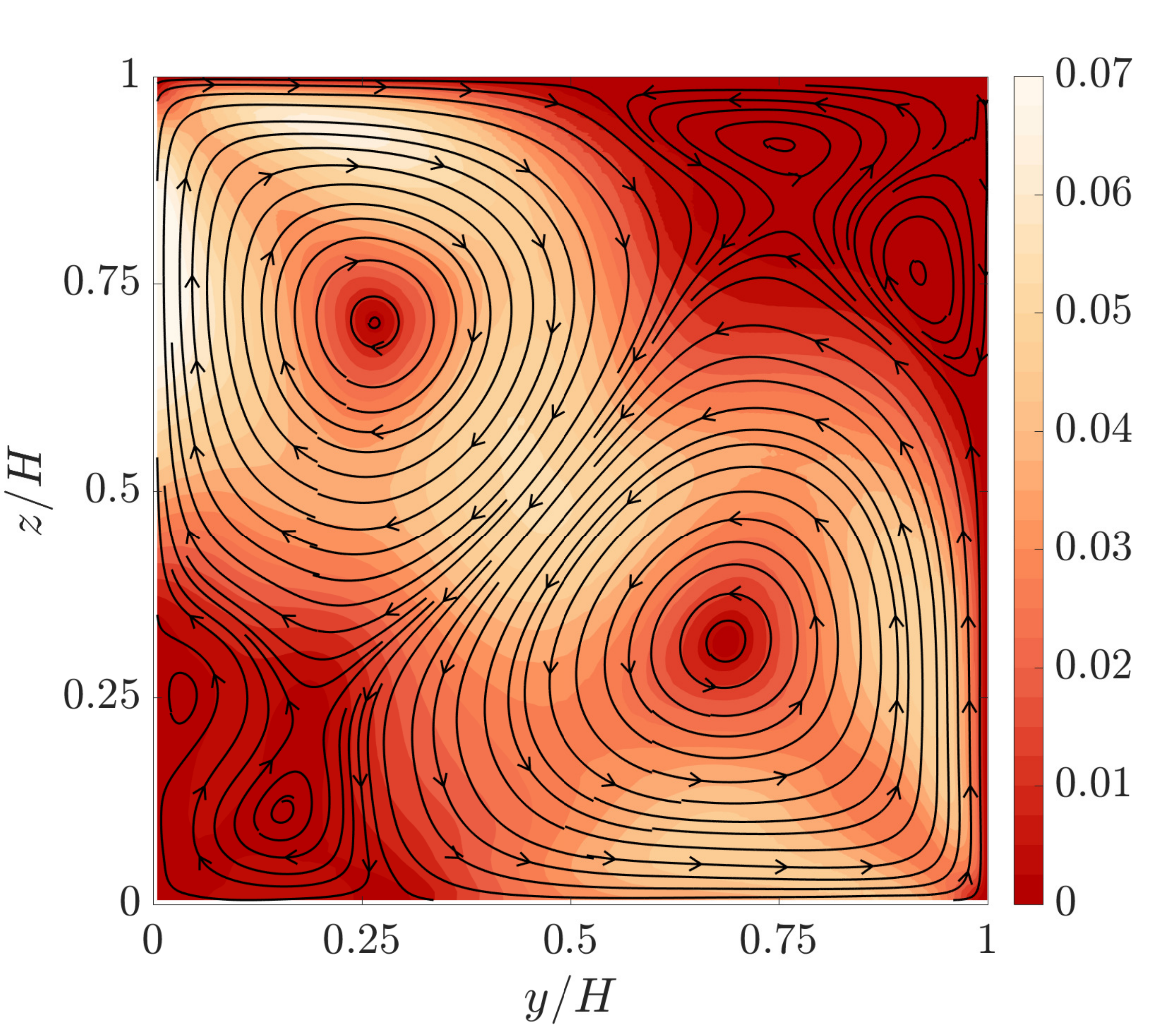}}
\caption{Case 2}
\end{subfigure}
\begin{subfigure}[b]{0.31\textwidth}
{\includegraphics[clip,  trim=0cm 0cm 0cm 0cm,scale=0.2]{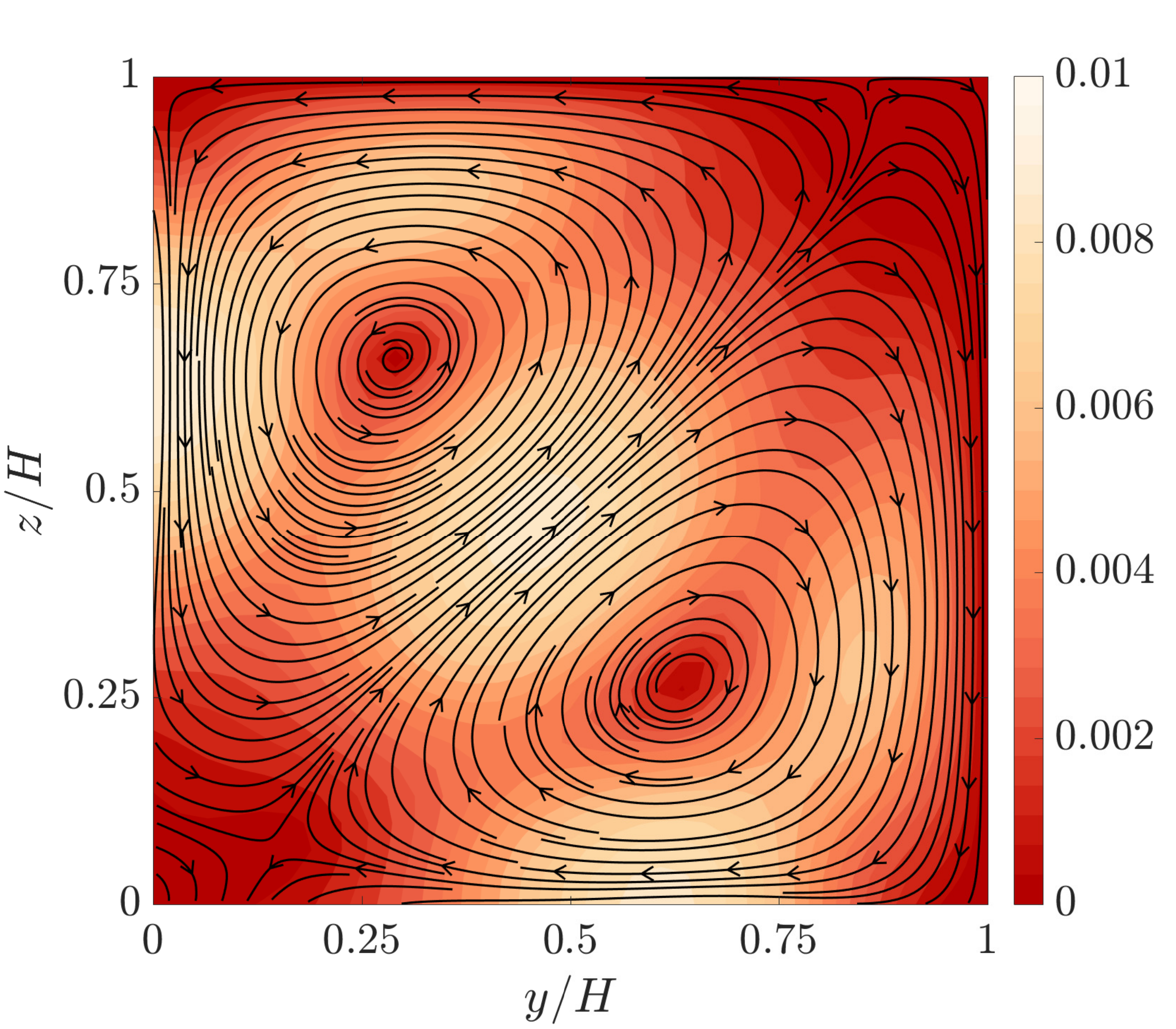}}
\caption{DNS}
\end{subfigure}
\caption{Comparison between the 5-level MLMC solution and the benchmark DNS data of the in-plane velocities $v$ and $w$. Streamlines are constructed using $\mathpzc{E}^{ML}_L[v_L]$ and $\mathpzc{E}^{ML}_L[w_L]$ ($L=4$) and contour indicates the magnitude of in-plane velocity vector $(\mathpzc{E}^{ML}_L[v_L], \mathpzc{E}^{ML}_L[w_L])$. Notice that with increase in dispersion $\delta$ an increase in the magnitude is observed.}\label{stream_SD}
\end{figure}

\begin{figure}[H]
\begin{subfigure}[b]{0.33\textwidth}
{\includegraphics[clip, trim=0cm 0cm 0cm 0cm,scale =0.18]{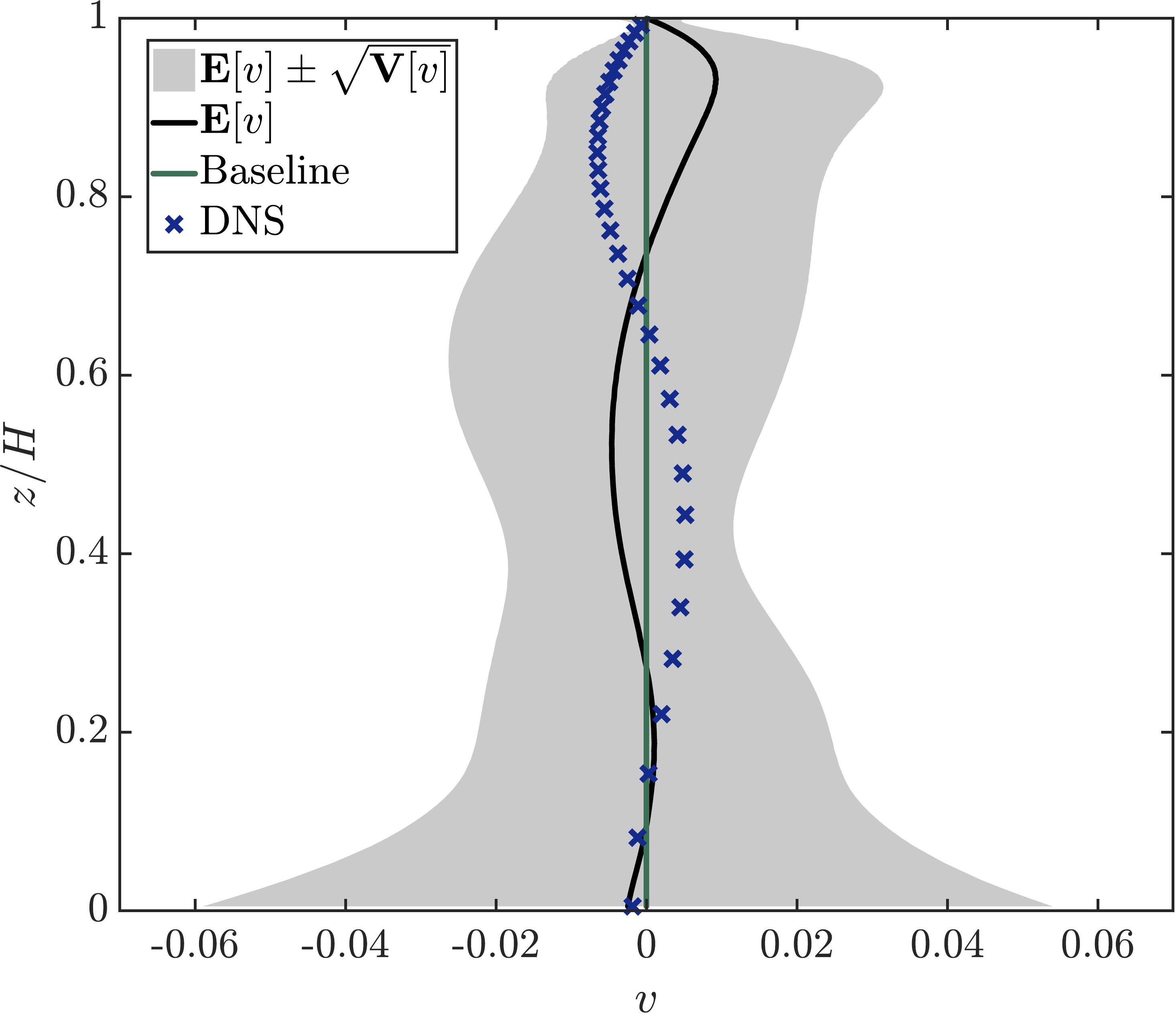}}
\caption{$y/H =0.25$}
\end{subfigure}
\begin{subfigure}[b]{0.33\textwidth}
{\includegraphics[clip,  trim=0cm 0cm 0cm 0cm,scale=0.18]{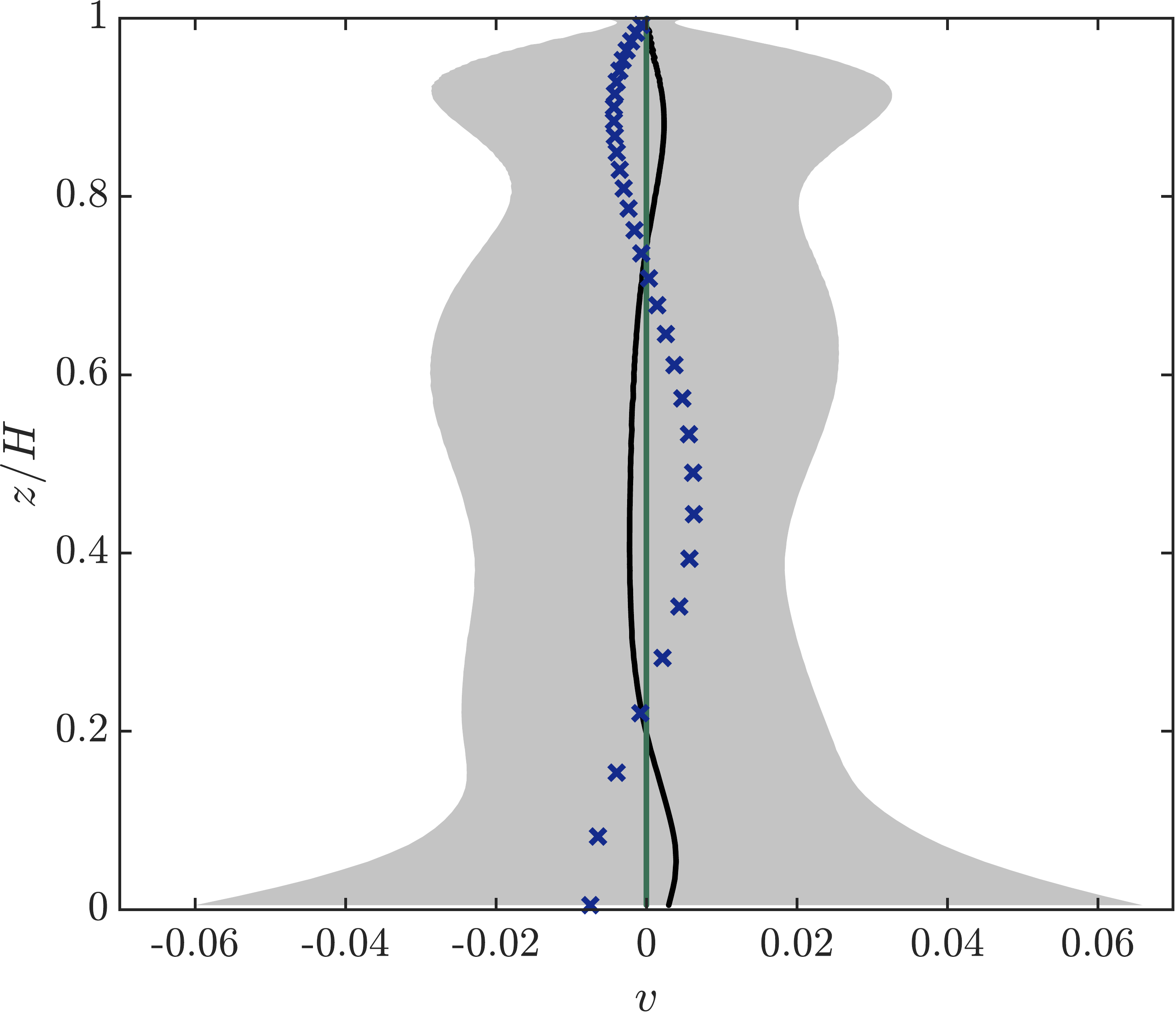}}
\caption{$y/H =0.50$}
\end{subfigure}
\begin{subfigure}[b]{0.31\textwidth}
{\includegraphics[clip,  trim=0cm 0cm 0cm 0cm,scale=0.18]{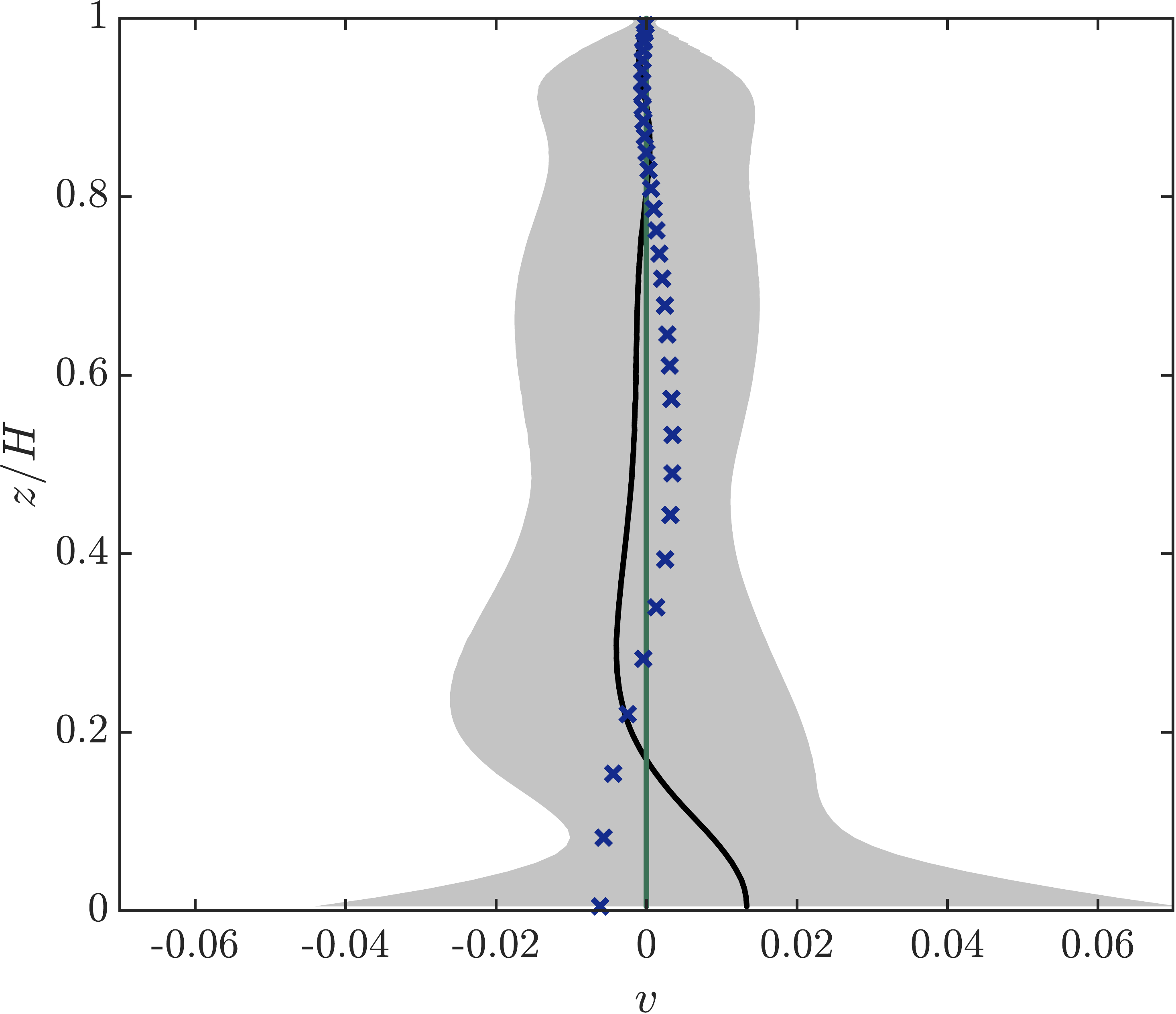}}
\caption{$y/H =0.75$}
\end{subfigure}
\caption{Mean $\mathpzc{E}^{ML}_L[v_L]$ and variance $\mathcal{V}^{ML}_L[v_L]$ of the $v$-component of the velocity computed using the 5-level estimator for Case 1 at three spanwise locations $y/H =0.25,0.50\text{ and }0.75$ with comparison to the baseline and DNS data.}\label{mean_std_SD}
\end{figure}
We have propagated the uncertainty with about $10^5$ degrees of freedom on the finest level due to six Reynolds stress components each sampled on a $128\times128$ grid. Note that there is a negligible change in the computational cost with an increase in dimensionality.  Although, the uncertain dimension can be reduced by using the KL expansion one might still have to deal with a relatively large number of uncertainties rendering any deterministic sampling method impractical (for instance, the stochastic collocation method).
\subsection{Flow over periodic hills}
The specification of the periodic hill geometry is adopted from \cite{breuer}. The time-averaged flow from the DNS data is shown in Fig. \ref{DNSflow}. The size of the computational domain is $\mathcal{D}_x = 9H$ and $\mathcal{D}_y = 3.036H$ along the streamwise and wall-normal direction, respectively, with $H=1$ denoting the hill height. The hill crest is situated at $(x/H,y/H) =  (0,1)$. Periodic boundary conditions are applied along inlet and outlet boundaries and a solid stationary wall at the top and the bottom. The Reynolds number of the flow is given by $Re = u_bH/\nu =2800$  where $u_b$ is the average velocity above the hill crest and $\nu$ is the molecular viscosity. The numerical solutions are obtained on a curvilinear block-structured grid with two blocks of size $[0,9]\times[0,2]$ and $[0,9]\times[2,3.036]$, and refinement near the lower and upper walls. 

Similar to the square duct case here also we use a pre-defined hierarchy of nested grids $\mathcal{D}_\ell$ such that we have $h_\ell \approx 0.5h_{\ell-1}$. In Fig. \ref{nestedPhill2}, the two coarsest meshes are plotted. Also, the distribution of the  $y^{+1}_\ell$ values (from the Launder-Sharma $k-\epsilon$ model) along the lower wall for the five grids levels is depicted in Fig. \ref{yplus}.  All grid levels except the coarsest satisfy the criterion $y^{+1}<1$. A separate grid hierarchy is used for the generation of the random fields. For a given grid level, these random fields are first sampled on a uniform rectangular mesh in a domain-sized $[0,9]\times[0,3]$ and are then interpolated to the cell-centres of the RANS simulation mesh. In Table \ref{pHillLevel}, we list the specification for the different levels and the CPU times needed to obtain one sample on each level. For the considered combination of numerical schemes, we again observe a cost scaling roughly as $\mathcal{O}(h_\ell^{-3})$ or $\gamma=3$.
\begin{table}[H]
\begin{center}
\begin{tabularx}{\textwidth}{@{}l|YYYY@{}}\toprule[1pt]
Level $(\ell)$ & OF mesh &$h_{\ell}$  & RF mesh  & CPU time (sec)\\\midrule\midrule
$0$ &$16\times24$ &0.5625  &$24\times8$ &$0.26\times10^2$\\
$1$ &$32\times48$ &0.2812  &$48\times16$&$0.69\times10^2$ \\
$2$ &$64\times96$ &0.1406  &$96\times32$ &$6.82\times10^2$\\
$3$ &$128\times192$ &0.0703  &$192\times64$&$5.01\times10^3$ \\
$4$ &$256\times384$ &0.0352   &$384\times128$&$4.70\times10^4$\\
\bottomrule[1pt]
\end{tabularx}
\end{center}
\caption{Specification of the MLMC grid hierarchy for the periodic hill case with $Re=2800$. ``OF mesh'' denotes the simulation mesh in OpenFOAM and ``RF mesh'' denotes the grid used for the generation of the random eddy viscosity field. CPU time is the total time for one sample.}\label{pHillLevel}
\end{table}
\begin{figure}[H]
\begin{subfigure}[b]{0.49\textwidth}
\hbox{\hspace{-0.5cm}\includegraphics[clip, trim=0cm 0cm 0cm 0cm,scale=0.19]{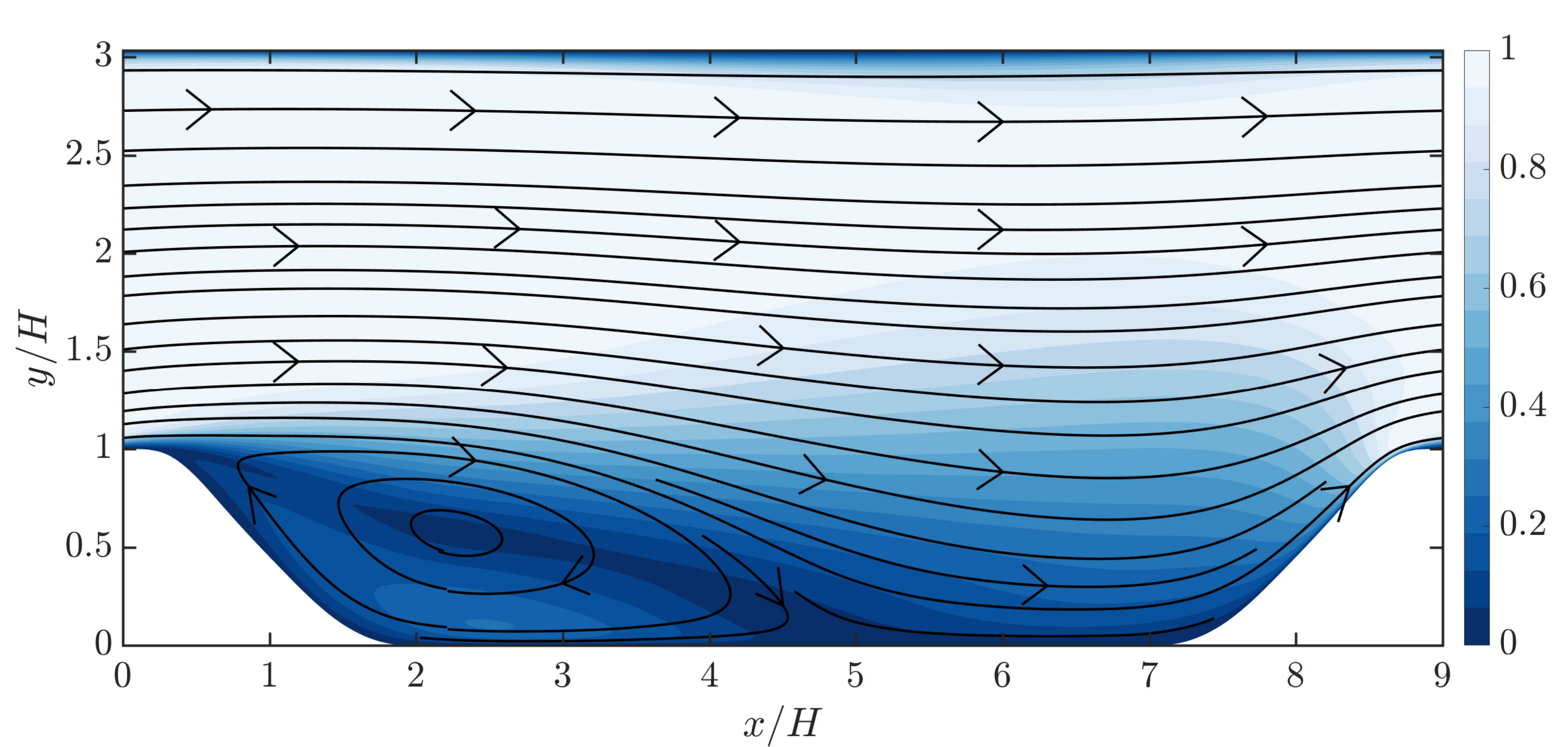}}
\caption{}\label{DNSflow}
\end{subfigure}
\begin{subfigure}[b]{0.49\textwidth}
{\includegraphics[clip, trim=0cm 0cm 0cm 0cm,scale=0.19]{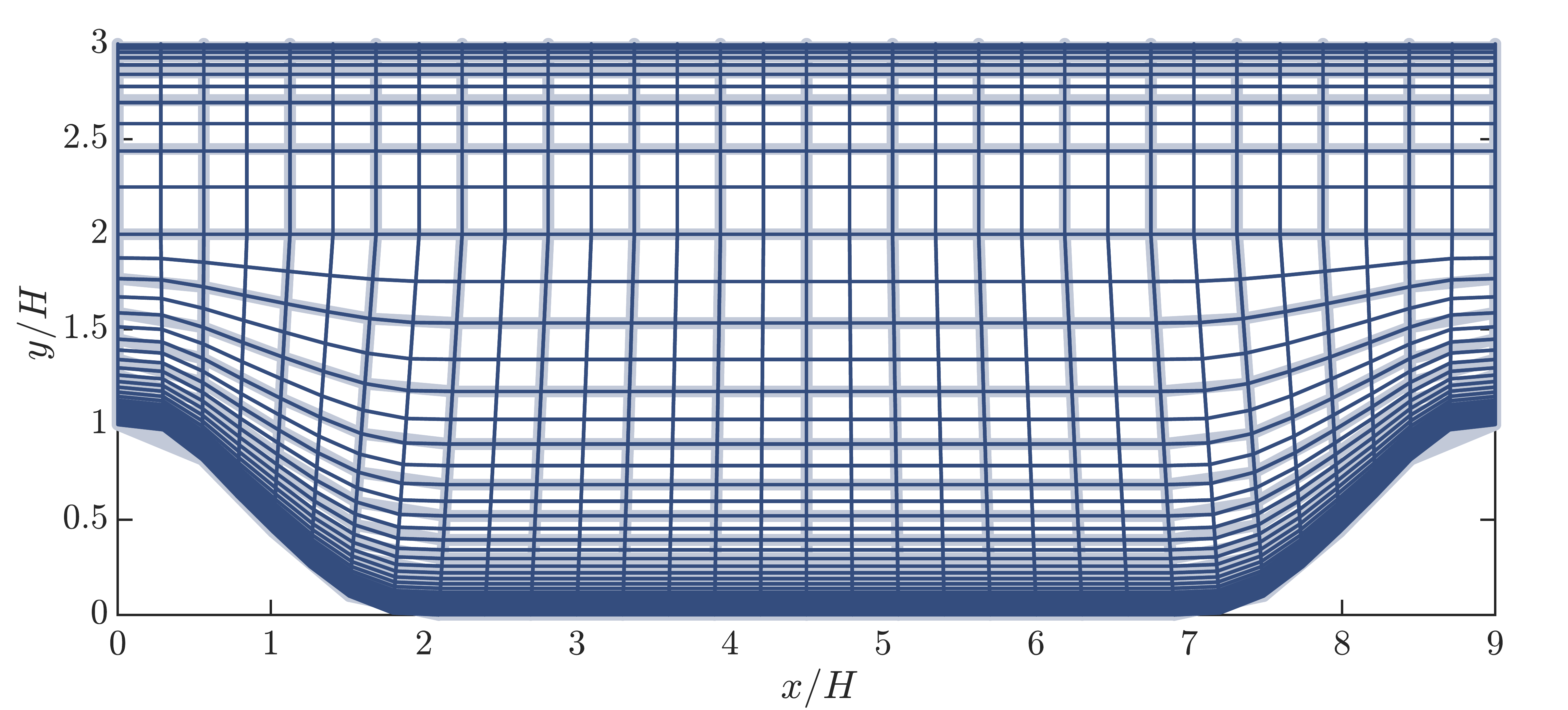}}
\caption{}\label{nestedPhill2}
\end{subfigure}
\caption{(a) Time-averaged flow on a periodic hill with Re = 2800 obtained from DNS data (Breuer et al. 2009) and (b) Nested curvilinear grids, the light blue lines depict 16x24 grid corresponding to $\ell=0$ and dark blue lines depict 32x48 grid corresponding to $\ell=1$.}
\end{figure}
\begin{figure}[H]
\centering
\includegraphics[clip, trim=2cm 0cm 2cm 0cm,width=0.6\textwidth]{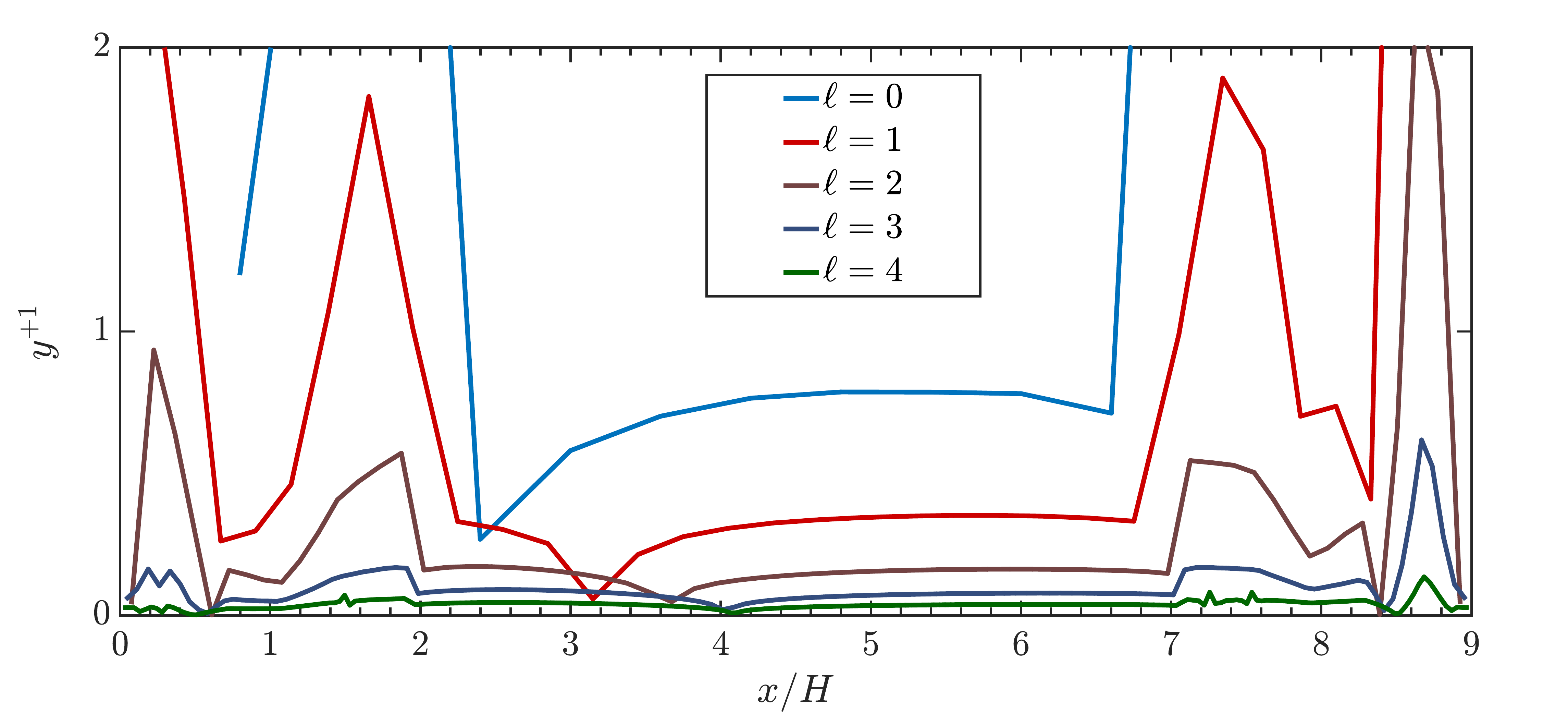}
\caption{The $y^{+1}_\ell$ values along the lower wall computed from the baseline simulations.}\label{yplus}
\end{figure}
\subsubsection{MLMC with the REV model}
We now analyze the performance of the MLMC method for the periodic hill flow using the random eddy viscosity (REV) stochastic model. To generate the samples of the random eddy viscosity two set of parameters are utilized, denoted by Case 1 and Case 2, see Table \ref{PH_params}. The two cases differ only in terms of the correlation length along the x- and y-directions. In Fig. \ref{ex_REV}, we show an example of a REV field for each case along with the baseline field $\nu_t^{(bl)}(\mathbf{x})$ obtained from the converged solution from a $k-\epsilon$ model at the finest level $\mathcal{D}_4$ with $256\times384$ cells. Due to small correlation lengths for Case 2, we observe more peaks in the random eddy viscosity field with a relatively large magnitude. Here, for the sake of generality we do not consider a periodic random eddy viscosity field, but, can be easily implemented as the circulant embedding method naturally yields a periodic random field. 

\begin{table}[H]
\begin{center}
\begin{tabularx}{.8\textwidth}{>{\hsize=.5\hsize}X>{\hsize=2\hsize}X>{\hsize=.4\hsize}X>{\hsize=.4\hsize}X}\toprule[1pt]
Parameter & Description & Case 1 & Case 2  \\\midrule\midrule
$l_x/H$ & Correlation length along x-direction & 1.5&0.6\\
$l_y/H$ & Correlation length along y-direction & 0.5&0.2\\
$\sigma^2_c$ & Marginal variance of the random field&0.5&0.5\\
\bottomrule[1pt]
\end{tabularx}
\end{center}
\caption{Parameter sets to generate random eddy viscosity field for the periodic hill flow.}\label{PH_params}
\end{table}

\begin{figure}[H]
\centering
{\includegraphics[clip, trim=0cm 0cm 0cm 0cm,scale=0.26]{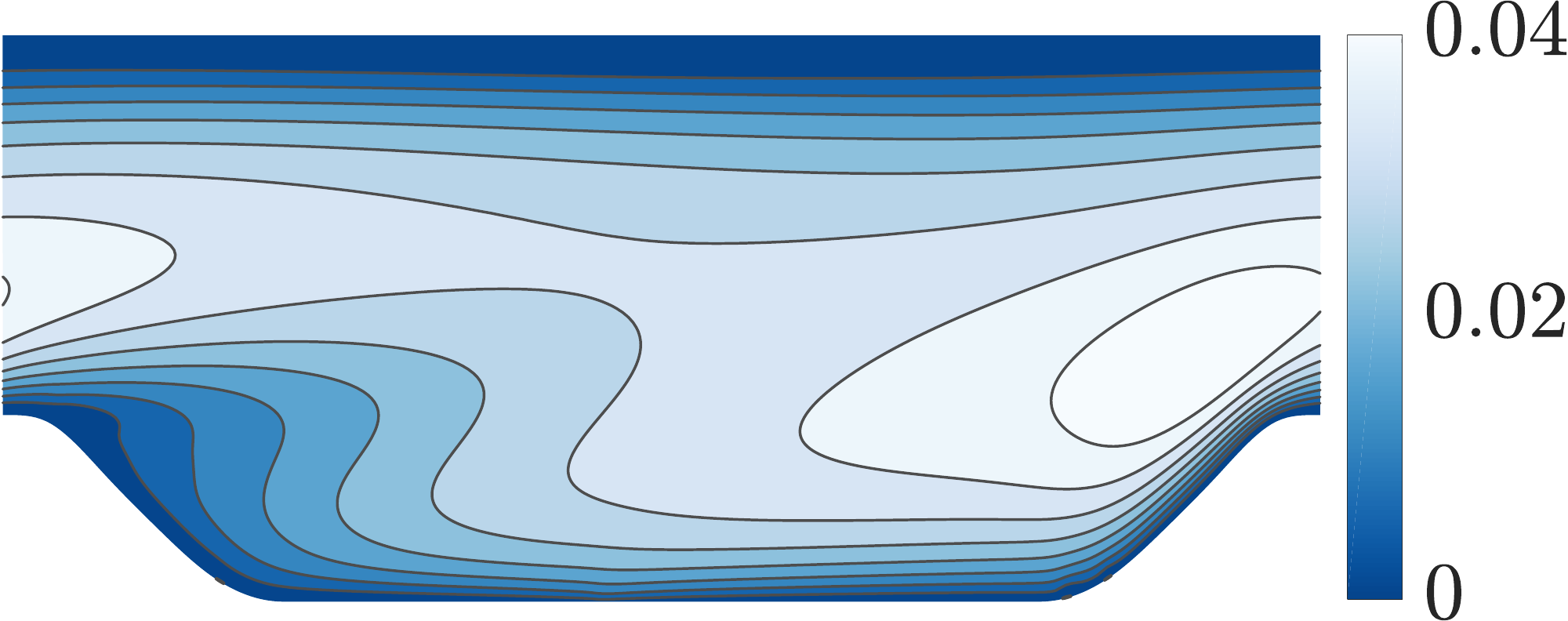}}
{\includegraphics[clip, trim=0cm 0cm 0cm 0cm,scale=0.26]{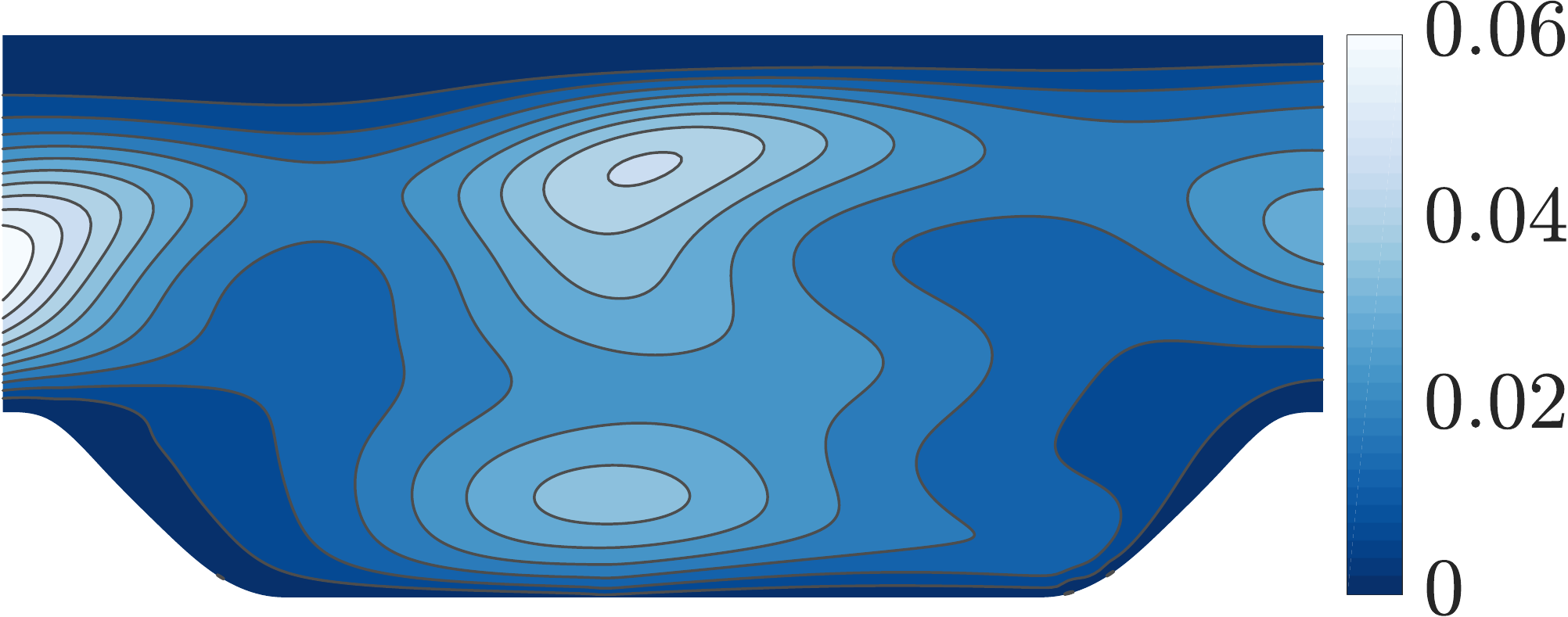}}
{\includegraphics[clip, trim=0cm 0cm 0cm 0cm,scale=0.26]{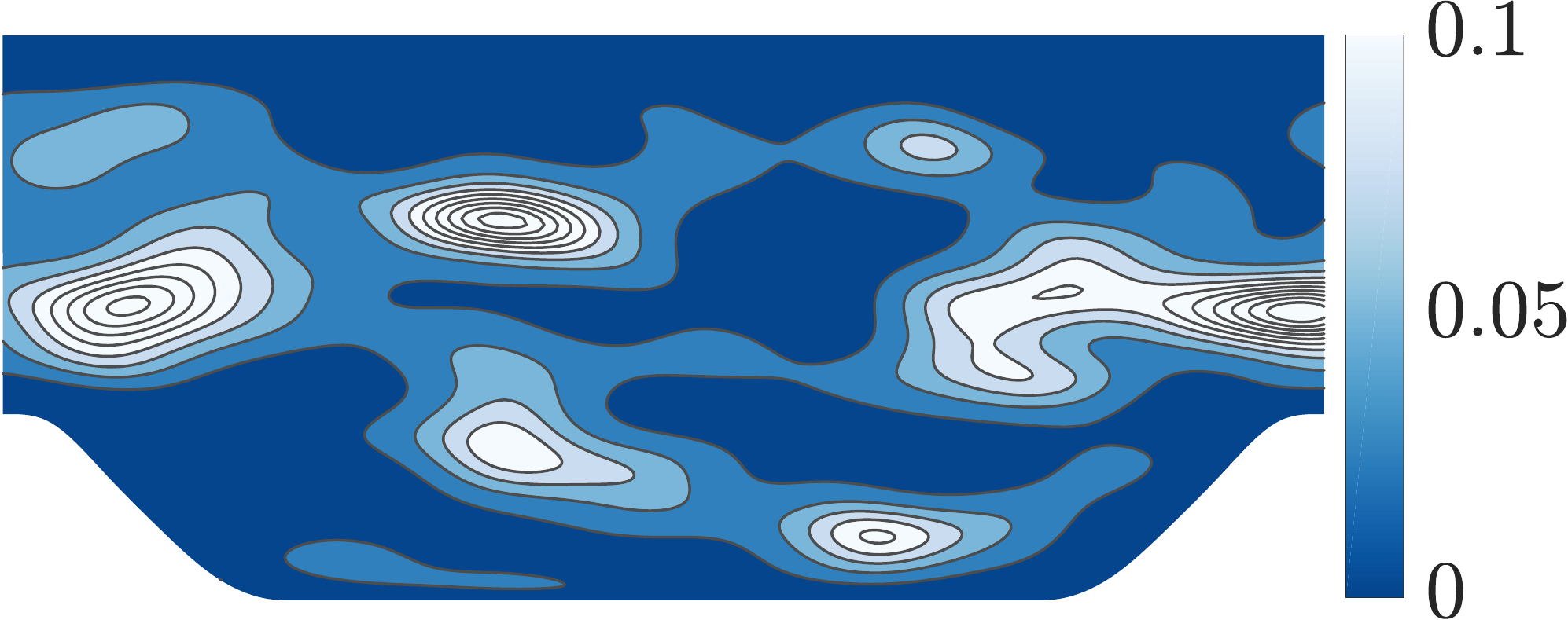}}
\caption{Baseline EV field from the Launder-Sharma $k-\epsilon$ model (left) and typical realizations of REV fields generated using the parameter set from Case 1 (middle) and Case 2 (right).}\label{ex_REV}
\end{figure}
%\subsection{Numerical convergence analysis}
We perform a similar analysis as the square duct flow to obtain the MLMC parameters. We begin by comparing the FV error in the deterministic and stochastic version of the problem for the streamwise velocity $u$ and the wall shear stress $\tau^w$ in Fig. \ref{bias_PH}. The error in the baseline converges as $\mathcal{O}(h_\ell)$ for both quantities of interest. The error in the random variables also decays at roughly the same rate. Here too the slower convergence rate can be primarily attributed to complex curvilinear meshes. Also, note that the relative errors in Case 1 and 2 are very close, indicating that they result in similar mean solutions. The sampling variance on different levels is depicted in Fig. \ref{var_PH}. As expected the variance decays at a rate twice of the discretization error coinciding with observations made in case of the square duct flow.

\begin{figure}[H]
\begin{subfigure}[b]{0.49\textwidth}
\centering
\hbox{\hspace{0cm}\includegraphics[clip, trim=1cm 0cm 0cm 0cm,scale =0.27]{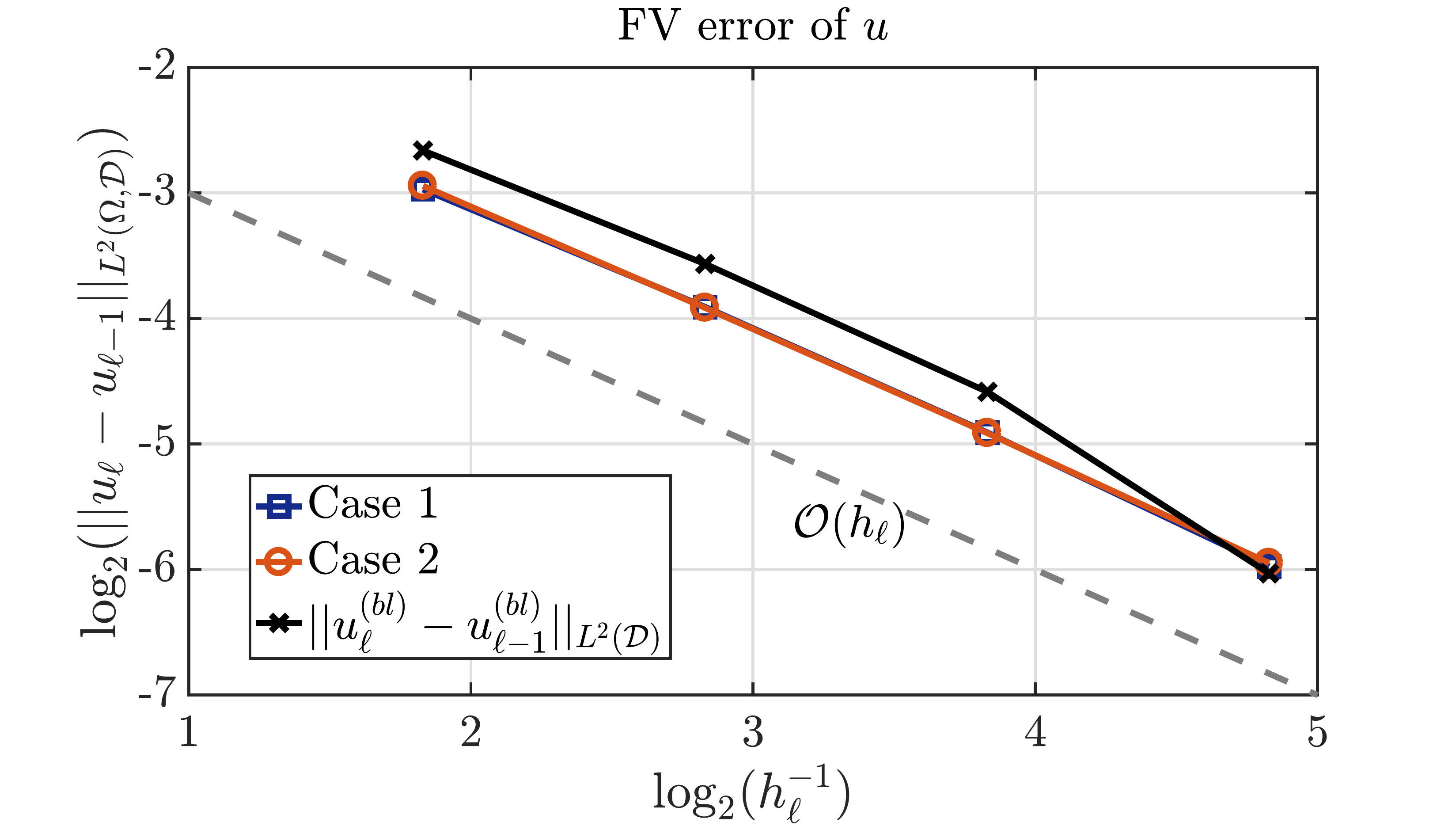}}
\end{subfigure}
\begin{subfigure}[b]{0.49\textwidth}
\centering
\hbox{\hspace{0cm}\includegraphics[clip,  trim=1cm 0cm 0cm 0cm,scale=0.27]{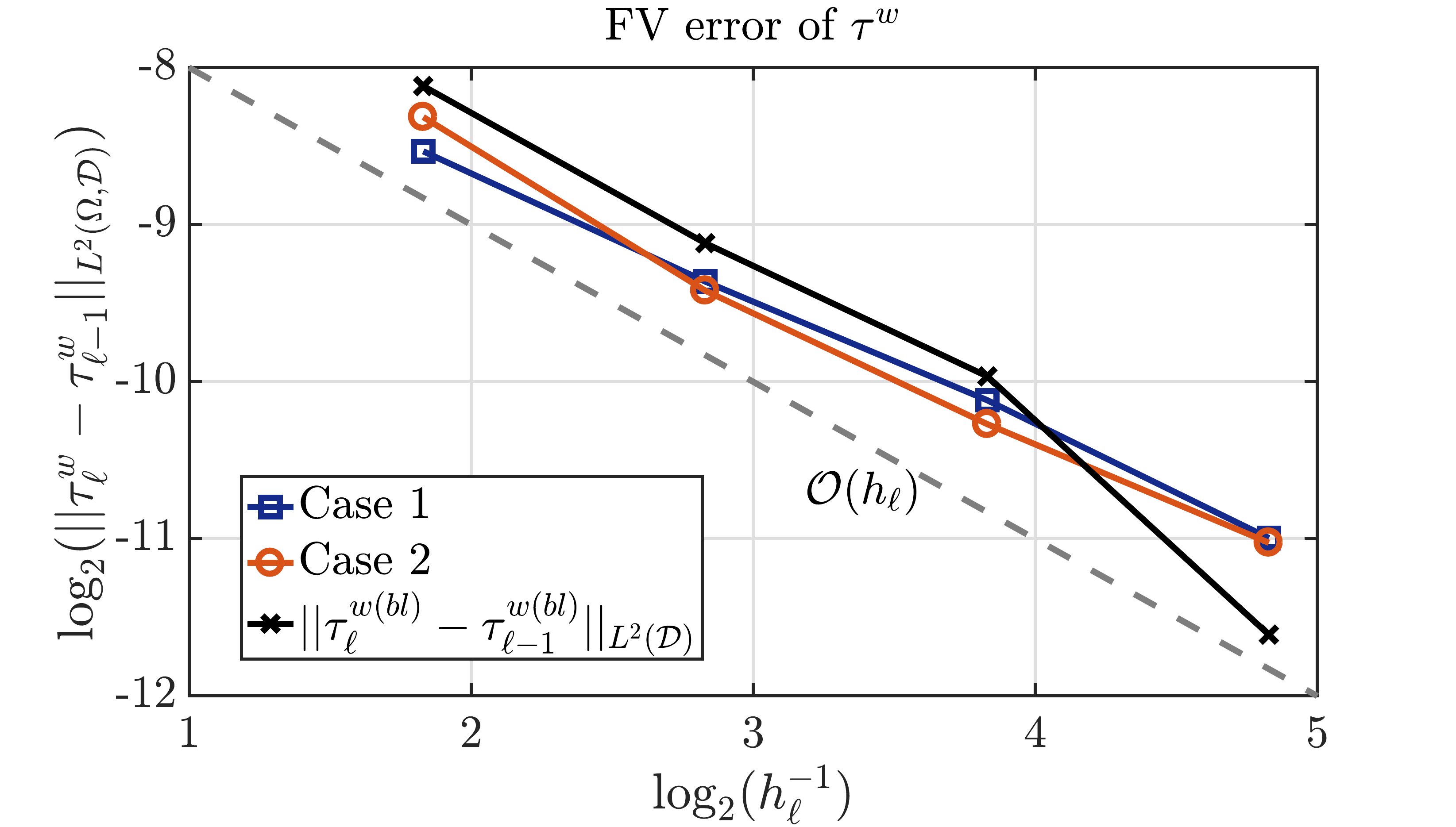}}
\end{subfigure}
\caption{Convergence of the FV error with levels along with error in baseline solution. The dotted line denote the empirical convergence rate of baseline RANS simulations.}\label{bias_PH}
\end{figure}
\begin{figure}[H]
\begin{subfigure}[b]{0.49\textwidth}
\centering
\hbox{\hspace{0cm}\includegraphics[clip, trim=1cm 0cm 0cm 0cm,scale =0.27]{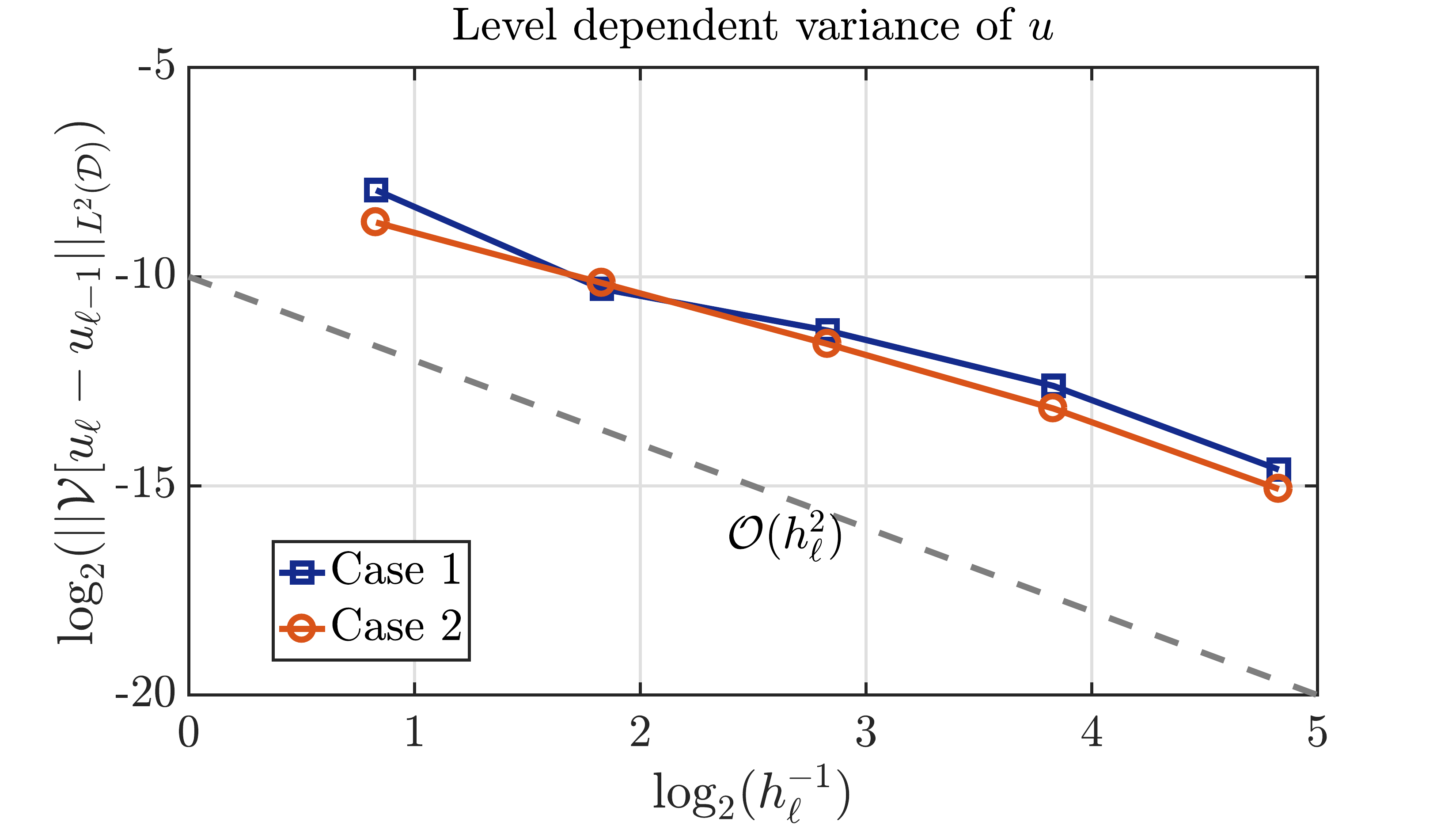}}
\end{subfigure}
\begin{subfigure}[b]{0.49\textwidth}
\centering
\hbox{\hspace{0cm}\includegraphics[clip,  trim=1cm 0cm 0cm 0cm,scale=0.27]{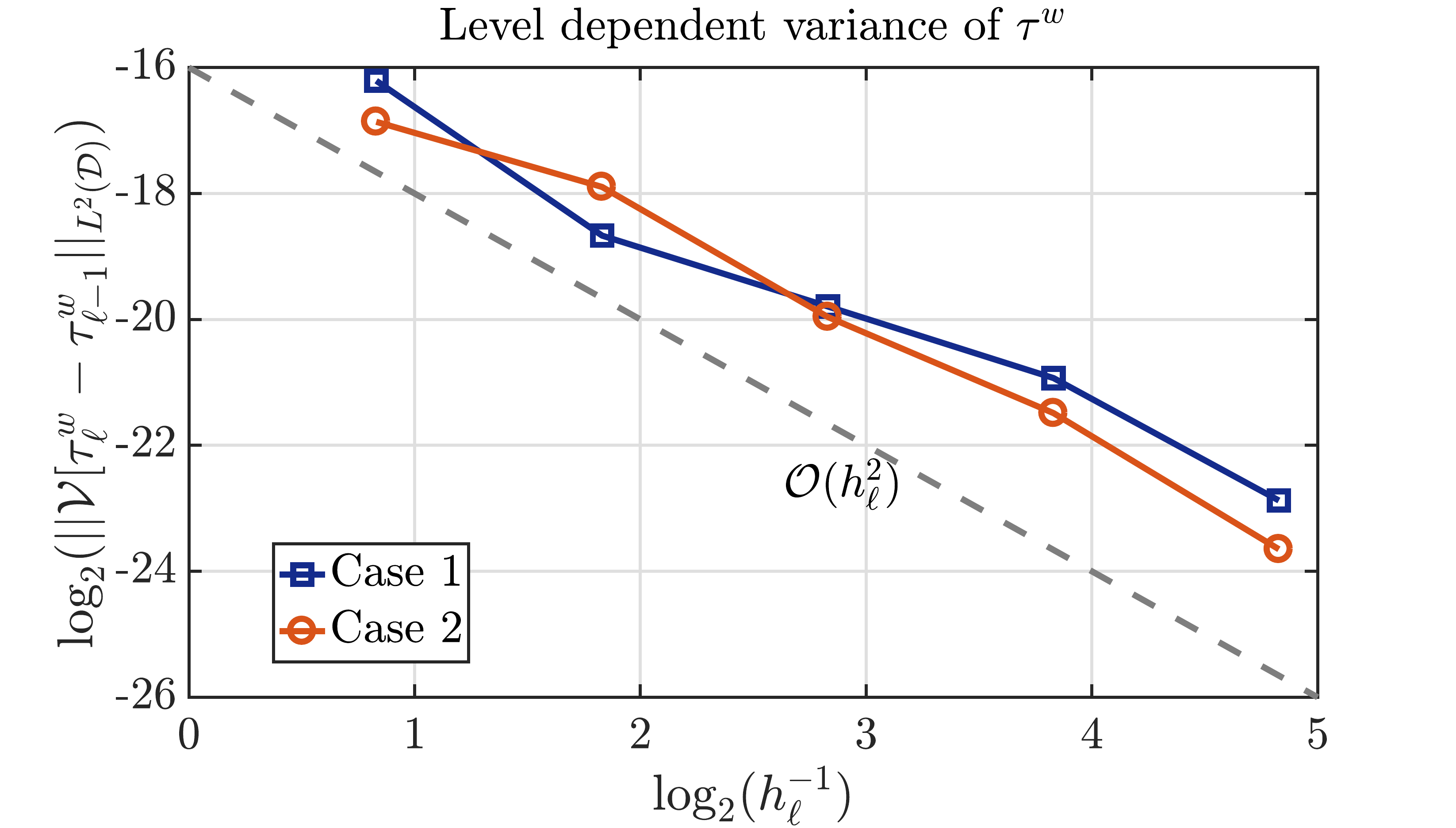}}
\end{subfigure}
\caption{Convergence of the level-dependent variance for different grids. The dotted line depicts an $\mathcal{O}(h_\ell^2)$ convergence.}\label{var_PH}
\end{figure}

From the above study, we again illustrate that the rate from the baseline solution can provide a good estimate for the MLMC simulation parameters. Next, we analyze the relative errors in the MC and MLMC estimators for the streamwise velocity $u$ in a similar fashion as for the square duct flow. As the quantity of interest we consider the streamwise velocity $u$ and set $\alpha=1$ and $\beta=2\alpha=2$ and $\gamma=3$. Recall that, with these rates we end up in the third scenario $\gamma>\beta$ in \eqref{MLMC_comp}, resulting in an asymptotically optimal MLMC estimator. The level-wise samples for the MLMC estimator is given by $N_\ell = N_L2^{2(L-\ell)}$ with $N_L=8$. The number of samples for the MLMC estimator with different $L$ is given in  Table \ref{MLMC_samples_PH}. The reference solutions for the mean and variance $\mathpzc{E}_{ref}[u]$ and $\mathcal{V}_{ref}[u]$, respectively are again based on the 5-level estimator. In case of the standard MC estimator, we follow $N = \mathcal{O}(h^{-2})$, thus the number of MC samples is increased by a factor of four with grid refinements. The standard MC simulation is conducted on four grids: $16\times24, 32\times48$, $64\times96$ and $128\times192$ with samples 8, 32,128 and 512, respectively.

\begin{table}[H]
\begin{center}
\begin{tabular}{|c|c|c|c|c|c|}
\cline{2-6}
\multicolumn{1}{c}{} & \multicolumn{5}{|c|}{Level-wise samples $N_\ell$}\\
\cline{2-6}
\hline
No. of levels ($L+1$) & $N_0$ & $N_1$ & $N_2$ &$N_3$&$N_4$\\
\hline
1 & 8 & - & - &-&-  \\
2 & 32 & 8  & - & - &-   \\
3 & 128 & 32 & 8 & - &- \\
4 &  512 & 128 & 32 & 8 & -   \\
5 (ref) &  2048 & 512 & 128 & 32 & 8 \\
\hline
\end{tabular}
\end{center}
\vspace{0.2cm}
\caption{Number of samples used for the MLMC estimators with different $L$ for the flow over periodic hills. The 5-level MLMC estimator was utilized as the reference solution.}\label{MLMC_samples_PH}
\end{table} 

The mean relative error in the expectation of $u$ approximated using the MC and MLMC methods is shown in Fig. \ref{erel_mean_PH}. The random eddy viscosity is based on Case 1. Both estimators are able to achieve similar accuracies, of order $\mathcal{O}(h_L)$. Also, the cost for both estimators scales similarly to the theoretical predictions in \eqref{MLMC_comp}. For $L=3$, we see a speedup of up to 30 times using the MLMC estimator.  In the case of the variance estimator in Fig. \ref{erel_var_PH}, we observe slightly slower rates and the MLMC method appears to be a bit more accurate for the same grid.  In terms of computational cost, similar gains are observed  as for the expected value of $u$. We point out that for the MLMC estimator, the dominant cost comes from the finest level and as the number of samples $N_L$ is a constant, we obtain a computational complexity of $\mathcal{O}(h_L^{-3})$. This is, up to a constant term, the same as solving one deterministic problem on the finest level, thus the MLMC estimator for this problem can be regarded as optimal.

 \begin{figure}[H]
\begin{subfigure}[b]{0.49\textwidth}   
            \centering 
      {\includegraphics[clip,  trim=1cm 0cm 0cm 0cm,scale=0.27]{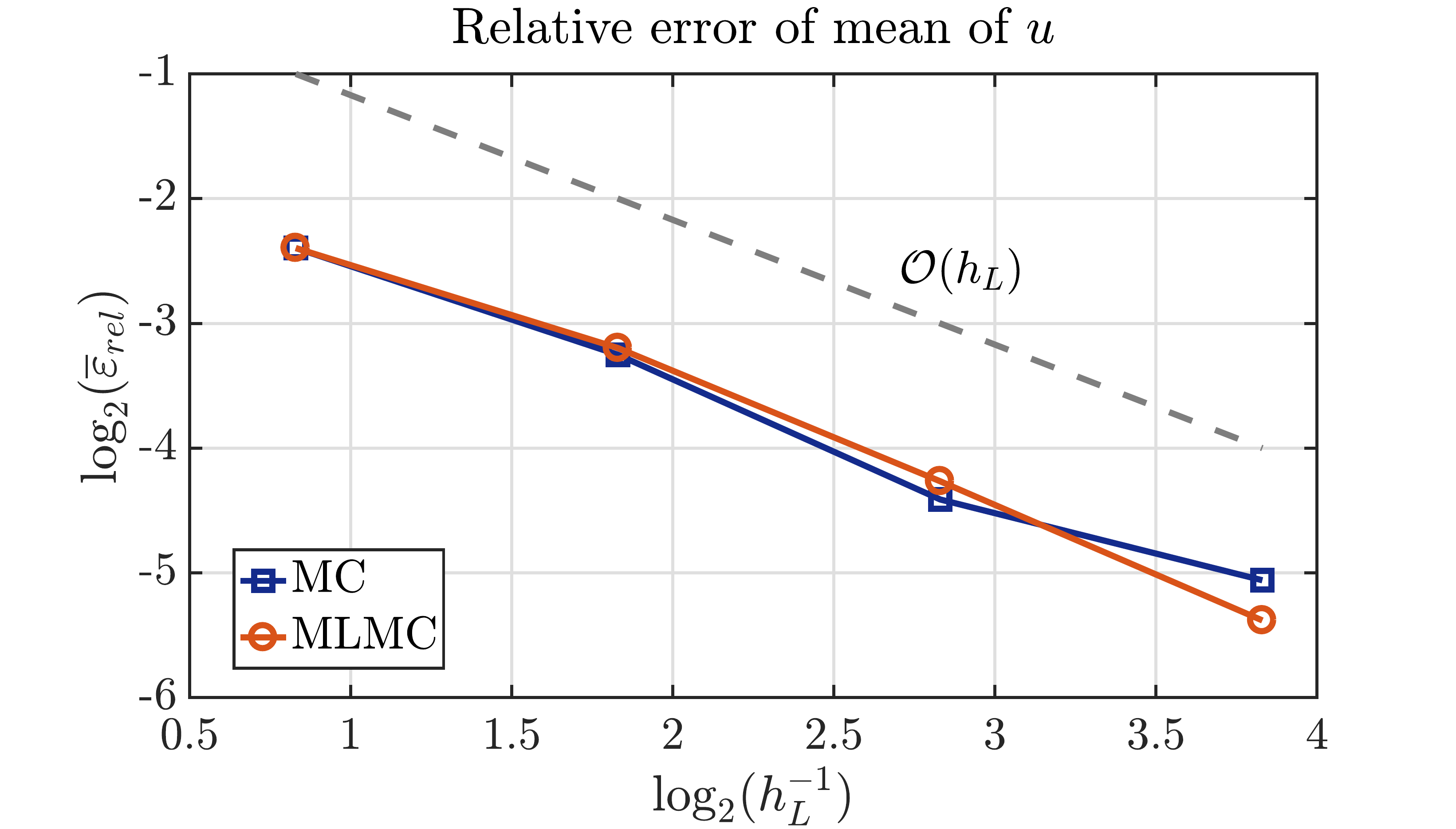}}
        \end{subfigure}
        \begin{subfigure}[b]{0.49\textwidth}   
            \centering 
 {\includegraphics[clip,  trim=1cm 0cm 0cm 0cm,scale=0.27]{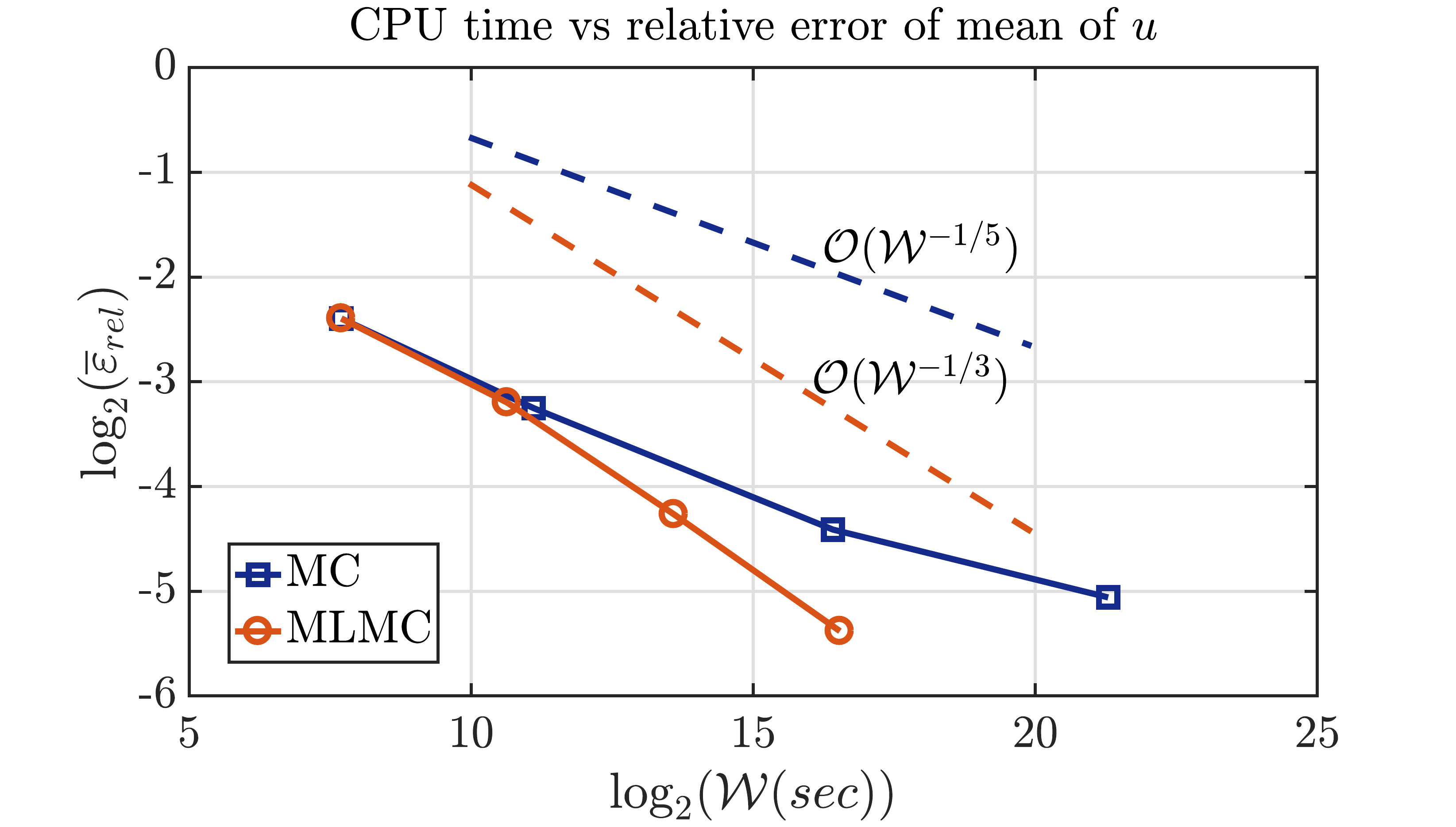}}
        \end{subfigure}
\caption{(Left) Comparison of the mean relative error $\overline{\varepsilon}_{rel}$ in the expected value of $u$ for different  meshes for Case 1. (Right) Computational work versus accuracy for the MC and MLMC estimators. Dotted lines show the predicted asymptotic cost for the MC (blue) and MLMC (red) estimators.}\label{erel_mean_PH}
\end{figure}

\begin{figure}[H]
\begin{subfigure}[b]{0.49\textwidth}   
            \centering 
      {\includegraphics[clip,  trim=1cm 0cm 0cm 0cm,scale=0.27]{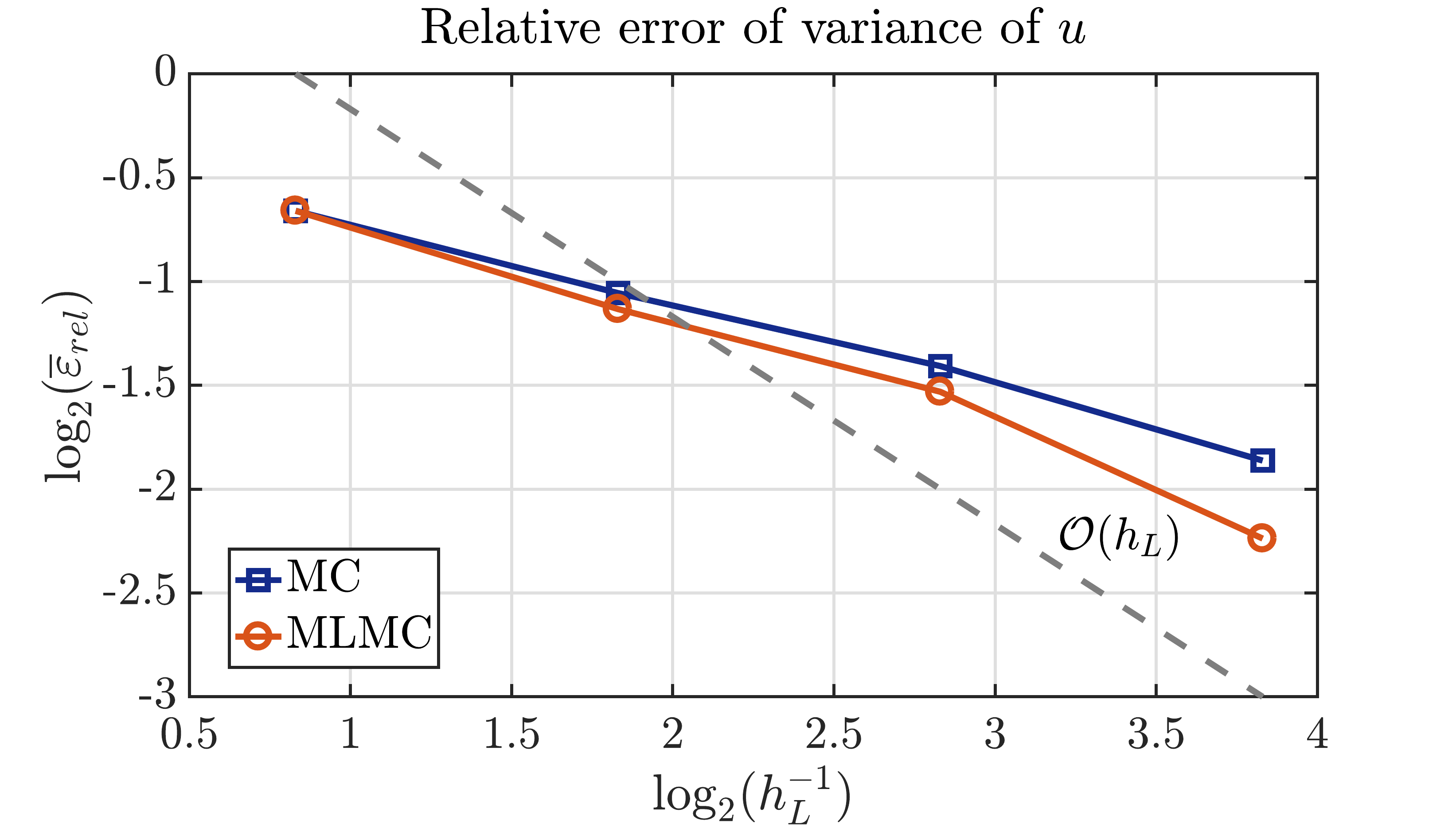}}
        \end{subfigure}
        \begin{subfigure}[b]{0.49\textwidth}   
            \centering 
 {\includegraphics[clip,  trim=1cm 0cm 0cm 0cm,scale=0.27]{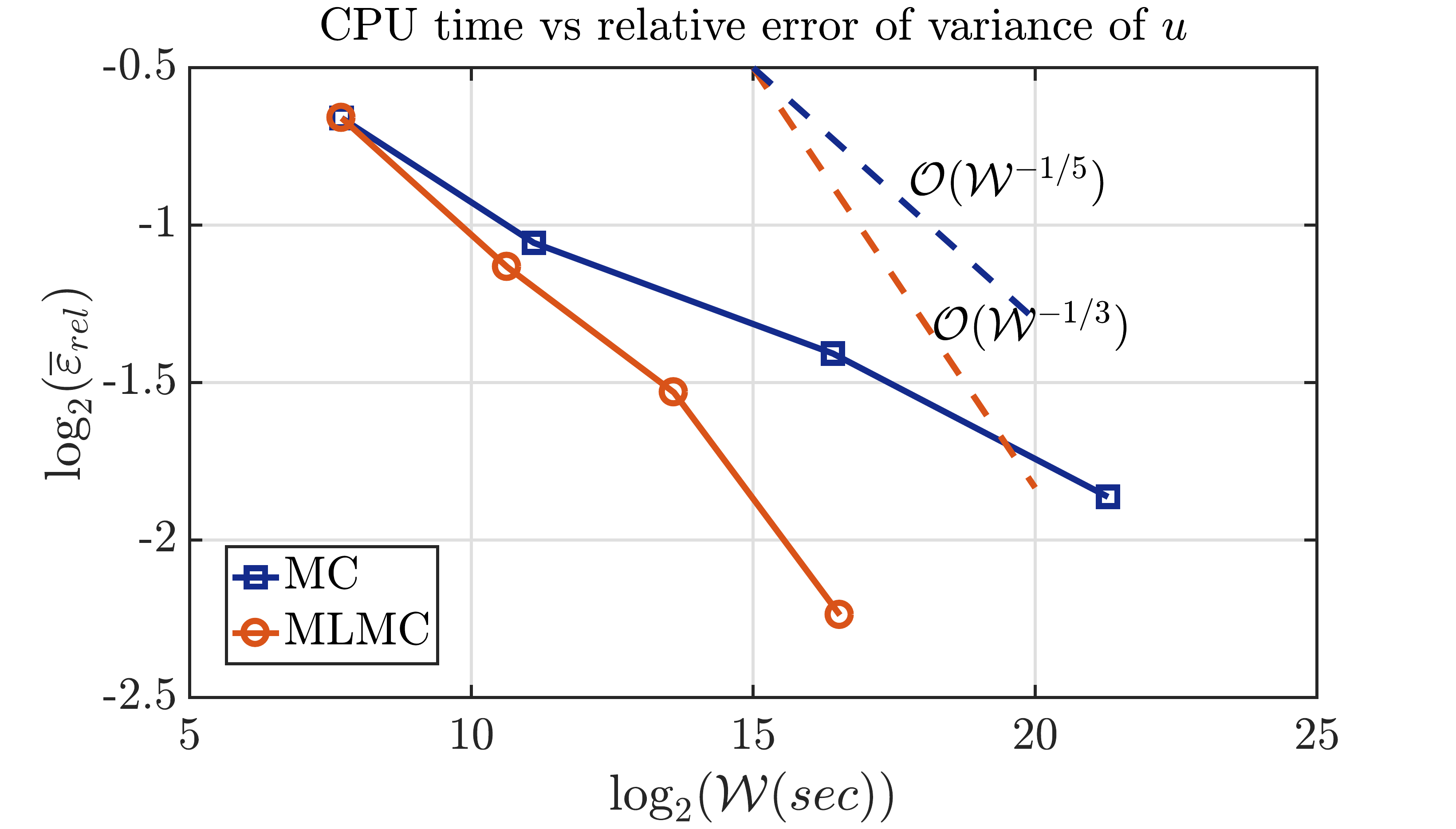}}
        \end{subfigure}
        \caption{(Left) Comparison of the mean relative error $\overline{\varepsilon}_{rel}$ in the variance of $u$ for different meshes for Case 1. (Right) Computational work  versus accuracy for the MC and MLMC estimators.}\label{erel_var_PH}
\end{figure}

Next we compare the reference solutions computed using the 5-level MLMC estimator with the REV model and DNS data. The main motivation of using the REV model was to obtain an uncertainty bound of the QoIs due to uncertainties arising from the transport equations or the closure parameters. Therefore, we are interested in the computations of the variance field using the MLMC method. In Fig. \ref{var_eddy}, the variance field for the streamwise wise velocity $u$ for the two cases based on the 5-level estimator is shown. A relatively high variance is observed near boundary layers and near the recirculation zone around $0.5<x/H<4.5$. Case 1 is visibly able to generate a larger variance than Case 2 indicating larger length scales can produce larger variation. The mean $\pm$ standard deviation is compared with the DNS data at various locations in Fig. \ref{mean_std_u}. It can be seen that the MLMC velocity profiles are very close to the baseline RANS solution for both cases. Further, we observe that this stochastic model is less sensitive in the free shear layer as it fails to capture the DNS data very well. However, we have tested that a combination of larger marginal variance $\sigma_c^2$ and length scales $l_x/H, l_y/H$ can result in a larger uncertainty bound around the free shear layer. Despite randomly chosen turbulence models, interesting regions such as flow separation and reattachment can be detected from the variance field. Lastly, the mean and standard deviation obtained for the wall shear stress $\tau^w$ are also compared with the DNS data in Fig. \ref{mean_std_wss}. Largest variances appear near the baseline reattachment point $x_{re}^{(bl)}$ near $x/H\approx4$ for both the cases. For comparison, the DNS data is also plotted which falls within $\pm$ one standard deviation bound of $\tau^w$ for both cases. 

The reference solution presented above is based on $5\times10^4$ degrees of freedom because the random eddy viscosity field on the finest mesh was sampled on a $384\times128$ grid. Here too, the KL expansion based dimension reduction can be employed and may still result in a large number of random inputs, especially when the size of the domain is much larger than the correlation lengths.
\begin{figure}[H]
        \begin{subfigure}[b]{0.51\textwidth}   
            \centering 
            \includegraphics[clip, trim=0cm 0cm 0cm 0cm,scale=0.24]{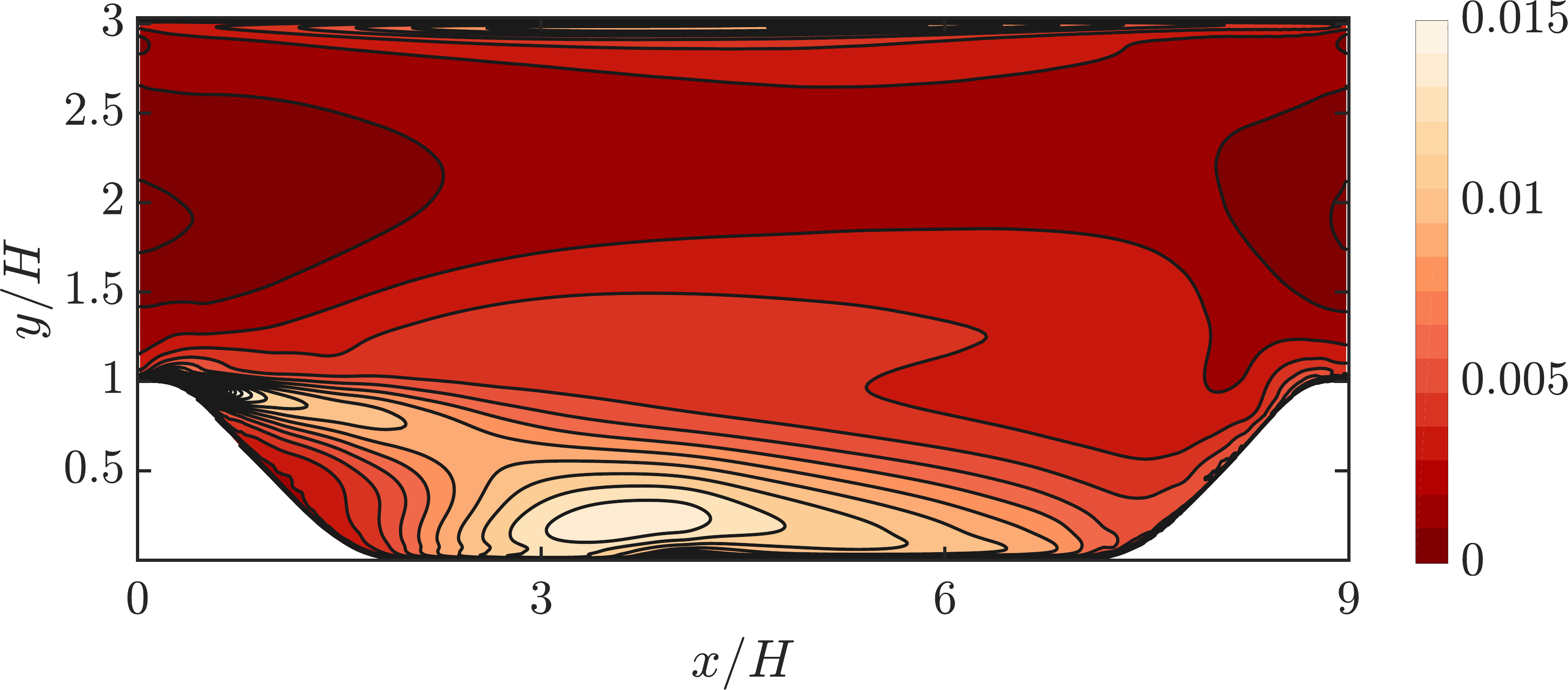}
         \caption{Case 1}            
        \end{subfigure}
        \begin{subfigure}[b]{0.47\textwidth}   
            \centering 
            \includegraphics[clip, trim=1cm 0cm 0cm 0cm,scale=0.24]{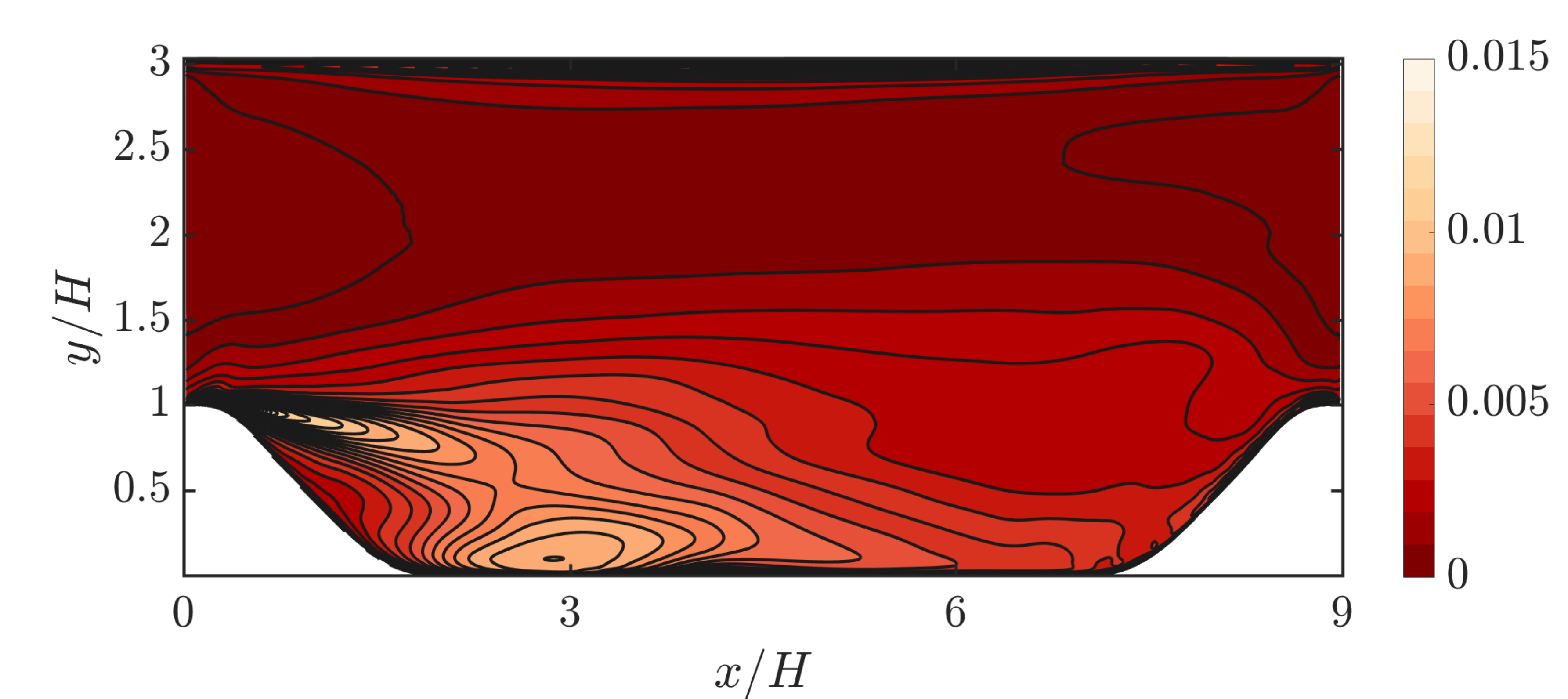}
                        \caption{Case 2}    
        \end{subfigure}
\caption{Variance field $\mathcal{V}^{ML}_L[u_L]$ for the streamwise velocity $u$ computed using the 5-level estimator. Variance is large near top and bottom boundary layers.}\label{var_eddy}
\end{figure}
\begin{figure}[H]
        \begin{subfigure}[b]{0.98\textwidth}   
            \centering 
            {\includegraphics[clip, trim=2.8cm 0cm 4cm 0cm,width=.75\textwidth]{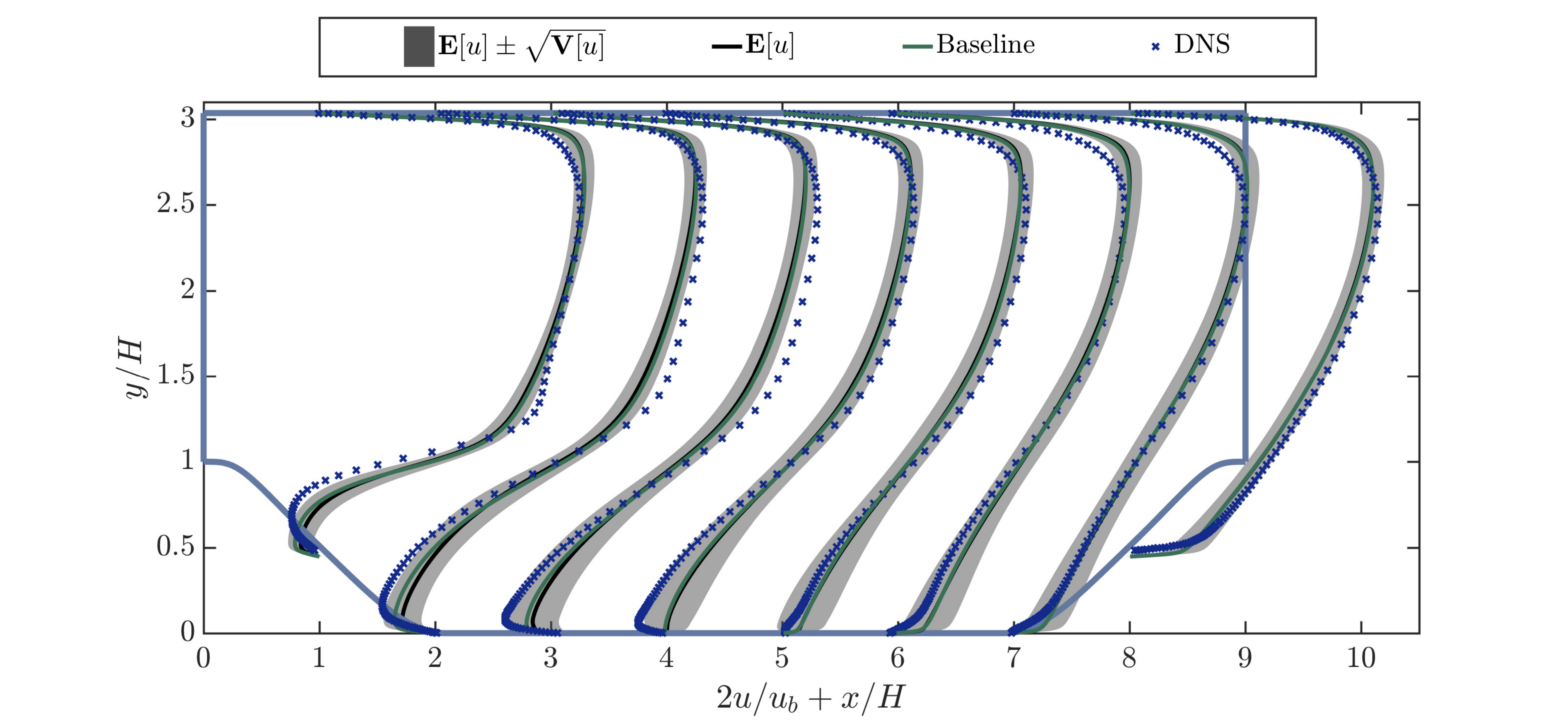}}
            \caption{Case 1}
        \end{subfigure}
                
        \vspace{1cm}
        \begin{subfigure}[b]{0.98\textwidth}   
            \centering 
            \includegraphics[clip, trim=2.8cm 0cm 4cm 0cm,width=.75\textwidth]{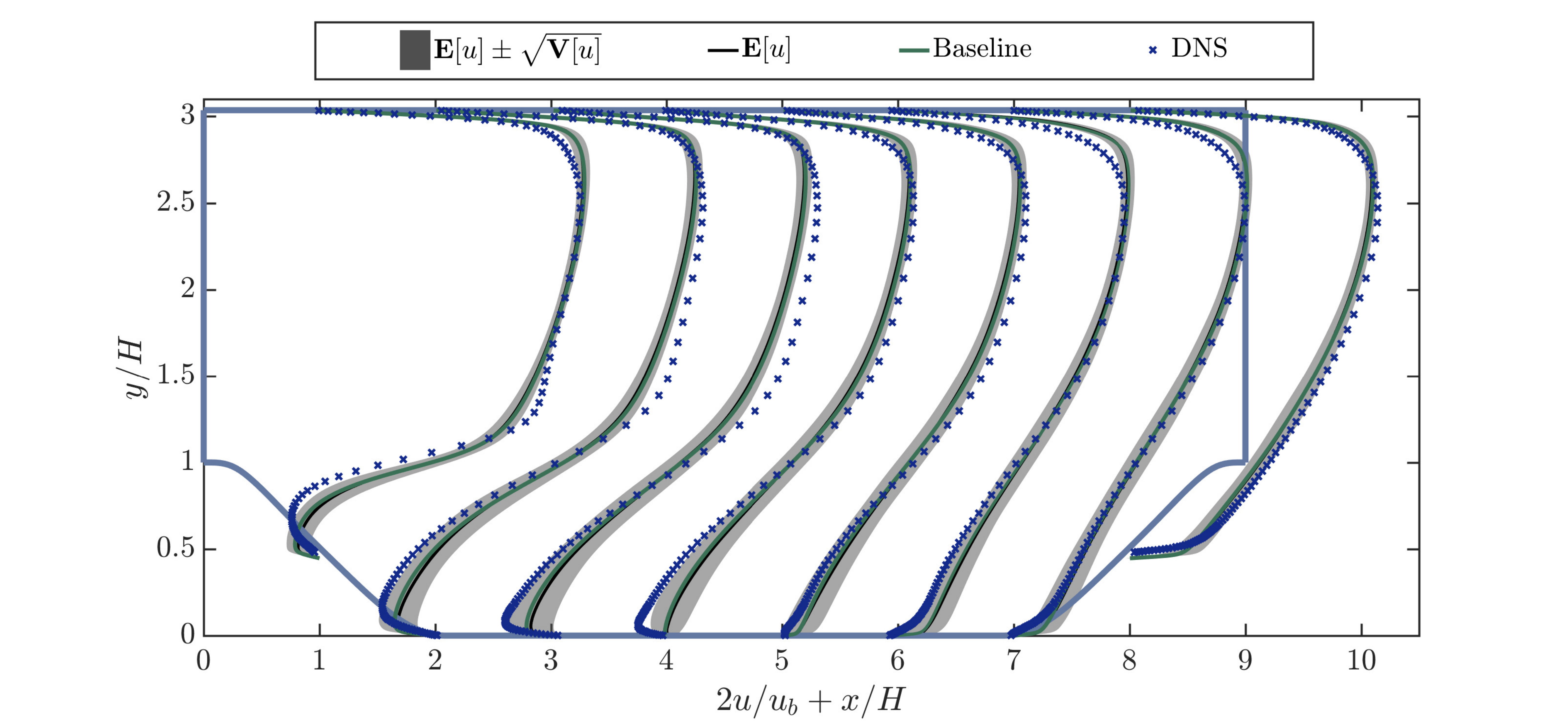} 
            \caption{Case 2}
        \end{subfigure}
\caption{Mean and variance of  the streamwise velocity computed using the 5-level estimator and comparison with DNS data at locations $x/H=1,2,3,...,8$. Velocities are scaled by a factor of two to facilitate visualization.}\label{mean_std_u}
\end{figure}
\begin{figure}[H]
        \begin{subfigure}[b]{0.98\textwidth}   
            \centering 
            {\includegraphics[clip, trim=2cm 0cm 2cm 0cm,width=0.9\textwidth]{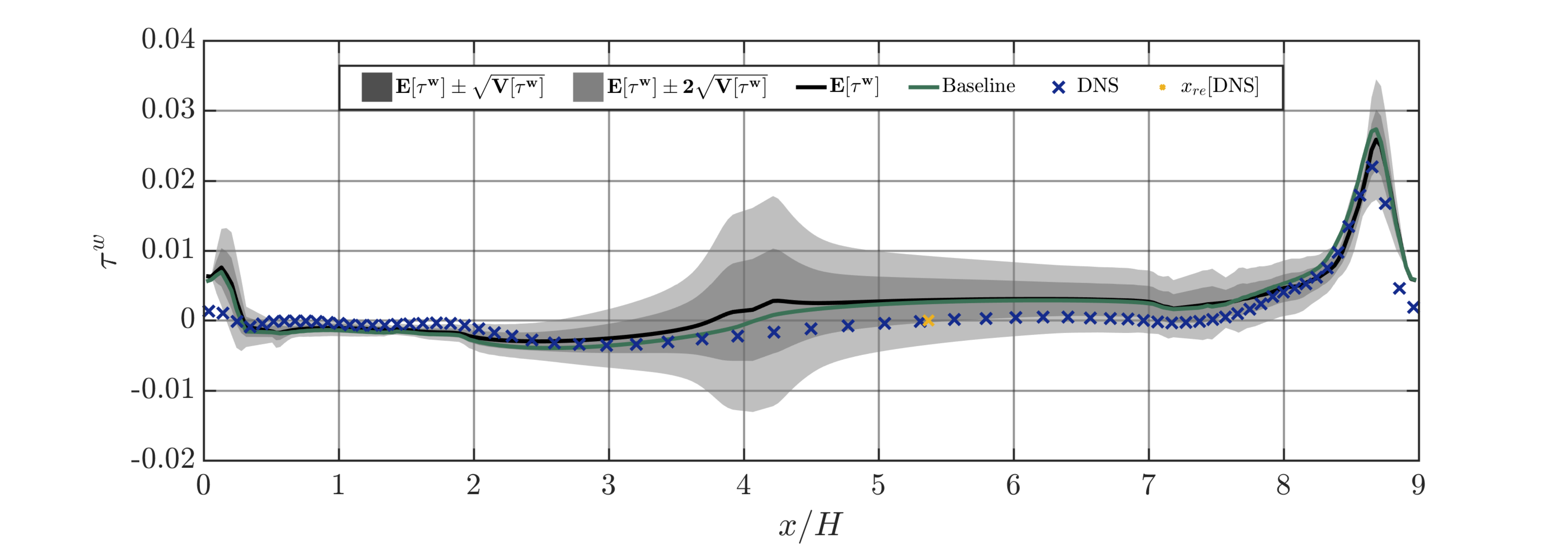}} 
            \caption{Case 1}
        \end{subfigure}
                
        \vspace{1cm}
        \begin{subfigure}[b]{0.98\textwidth}   
            \centering 
            {\includegraphics[clip, trim=2cm 0cm 2cm 0cm,width=0.9\textwidth]{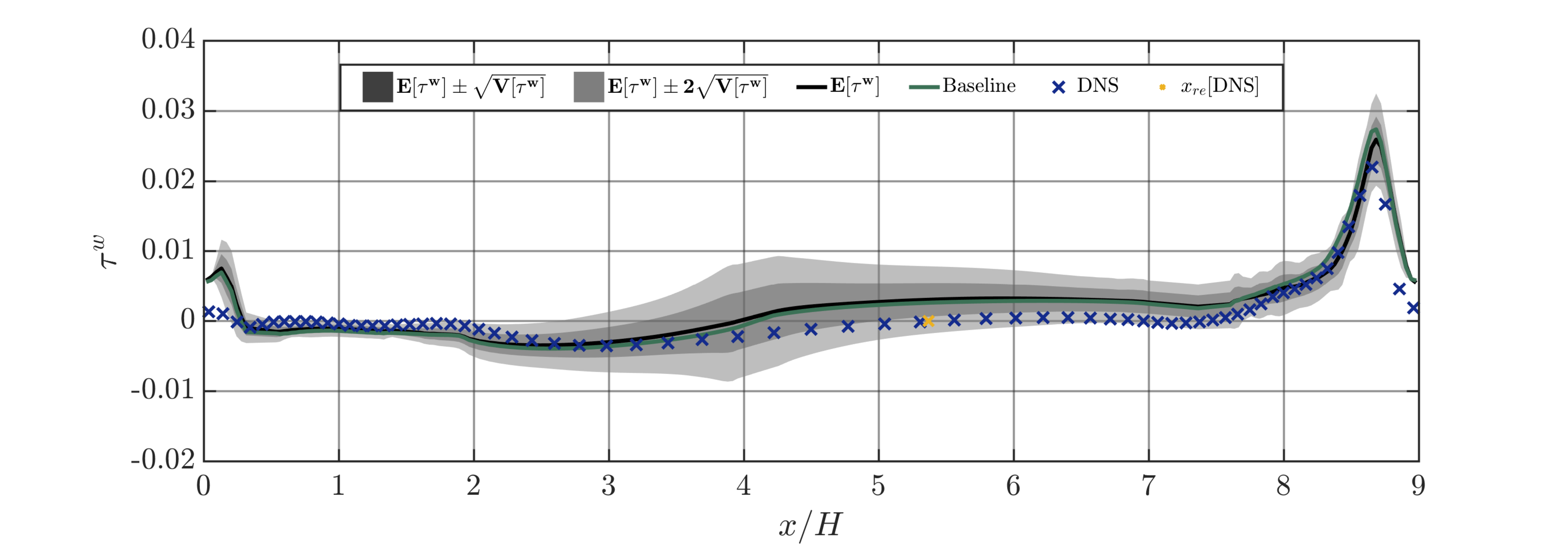}}
            \caption{Case 2}
        \end{subfigure}
\caption{Mean and variance of the wall shear stress $\tau^w$ computed using the 5-level estimator and comparison with DNS data.}\label{mean_std_wss}
\end{figure}

\subsubsection{MLMC with the RRST model}
In the final numerical experiment, we test the performance of the MLMC method with the RRST model applied to the periodic hill test case. We use the same 5-grid hierarchy as was considered for the random eddy viscosity experiments to study the convergence of the bias and sampling error with respect to the levels. Also, the same number of OpenFOAM iterations was used to propagate the random stress tensor as was used to propagate the random eddy viscosity, thus, we have same the CPU time per sample as was given in Table \ref{pHillLevel} (neglecting the cost for sampling a single random tensor field). The two parameter sets for generating the random tensor fields are listed in Table \ref{PH_params2}. For a fair comparison, we fix the blending parameter to $\xi=0.6$ for both cases, although a higher blending is possible for the easier Case 1. Sample profiles of $R_{12}$ for the two cases are compared in Fig. \ref{samples_RRST_PH} along with the baseline profile $R_{12}^{(bl)}$ (from the $k-\epsilon$ model). The effect of a larger dispersion and small correlation lengths is clearly visible for Case 2.
\begin{table}[H]
\begin{center}
\begin{tabularx}{.8\textwidth}{>{\hsize=.5\hsize}X>{\hsize=2\hsize}X>{\hsize=.4\hsize}X>{\hsize=.4\hsize}X}\toprule[1pt]
Parameter & Description & Case 1 & Case 2  \\\midrule\midrule
$l_x/H$ & Correlation length along $x$-direction & 1.5&0.6\\
$l_y/H$ & Correlation length along $y$-direction & 0.5&0.2\\
$\sigma^2_c$ & Variance of log-normal random field&1&1\\
$\delta(\mathbf{x})$ & Dispersion parameter &0.2 & 0.4\\
$N_{PC}$ & Order of polynomial chaos expansion & 5&5\\
$\xi$& Blending factor&0.6&0.6\\
\bottomrule[1pt]
\end{tabularx}
\end{center}
\caption{Parameter sets to generate random Reynolds stress tensor for the flow over periodic hills.}\label{PH_params2}
\end{table}
\begin{figure}[H]
\begin{subfigure}[b]{0.49\textwidth}
\centering
\hbox{\hspace{-0.2cm}\includegraphics[clip, trim=0cm 0cm 0cm 0cm,scale = 0.21]{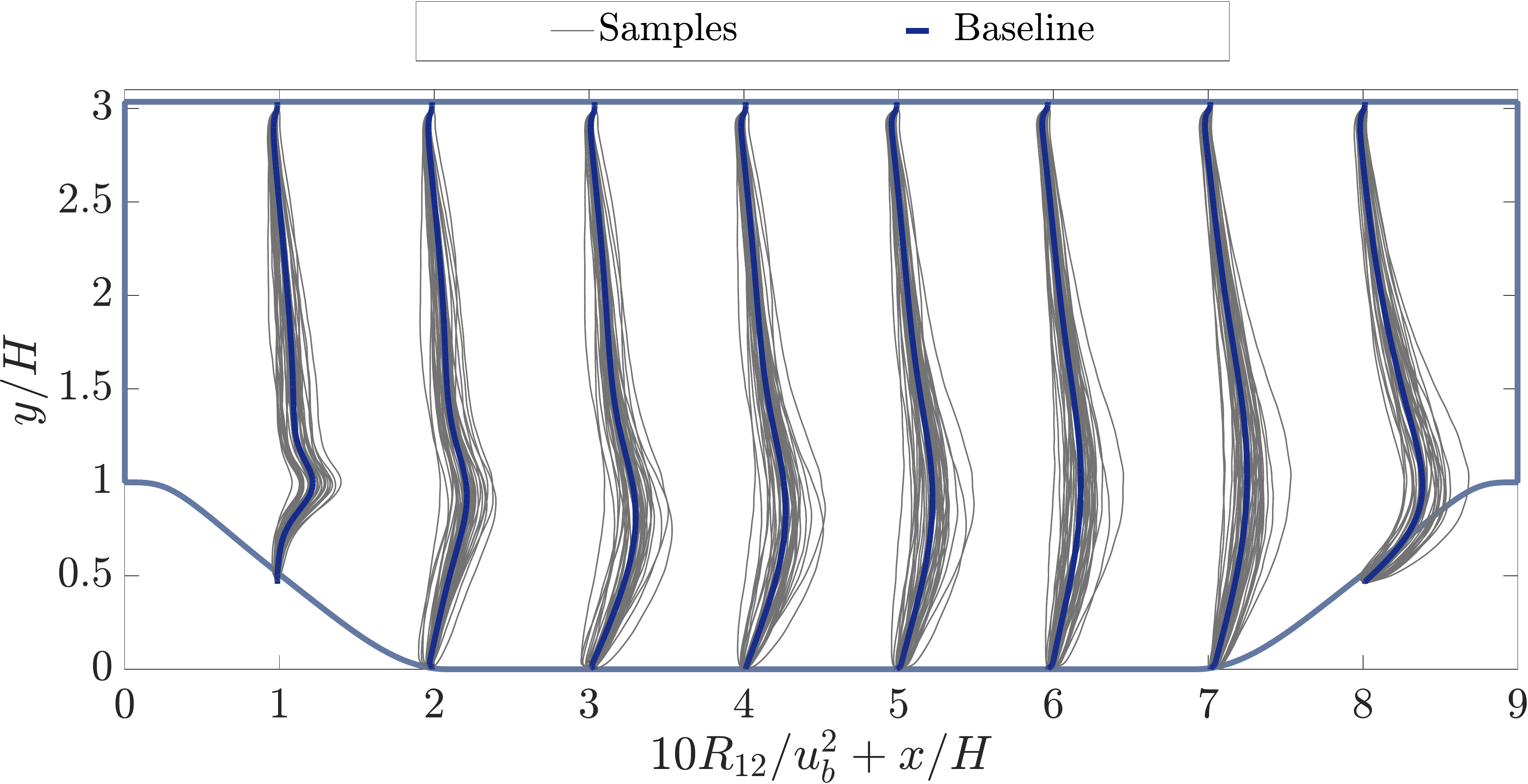}}
\caption{Case 1}
\end{subfigure}
\begin{subfigure}[b]{0.49\textwidth}
\centering
\hbox{\hspace{-0.2cm}\includegraphics[clip,  trim=0cm 0cm 0cm 0cm,scale= 0.21]{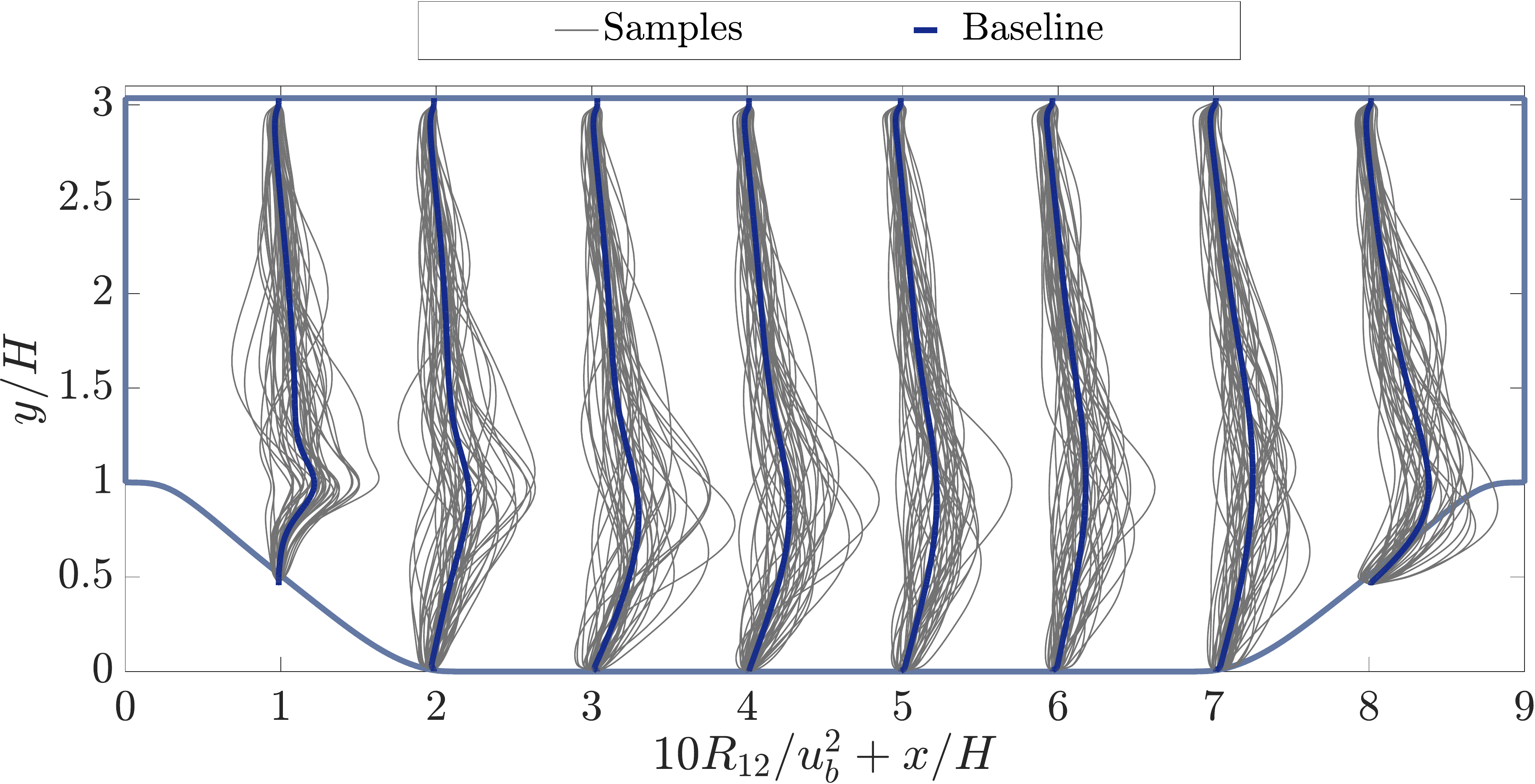}}
\caption{Case 2}
\end{subfigure}
\caption{Comparison of sample profiles of $R_{12}$ at different locations along with baseline values.}\label{samples_RRST_PH}
\end{figure}
We begin by analyzing the convergence of the FV bias with grid refinements in Fig. \ref{bias_PH_SM2} for the streamwise velocity (left) and the wall shear stress (right). A first-order convergence is seen for the first four levels, similar to the REV model. But, for both cases the error is not reduced up to the discretization accuracy on the finest $256\times384$ grid. Similar behaviour is observed for the level-dependent variance in Fig. \ref{var_PH_SM2}, where the fifth level exhibits a larger variance compared to the fourth level. 
\begin{figure}[H]
\begin{subfigure}[b]{0.49\textwidth}
\centering
\hbox{\hspace{0cm}\includegraphics[clip, trim=1cm 0cm 0cm 0cm,scale = 0.27]{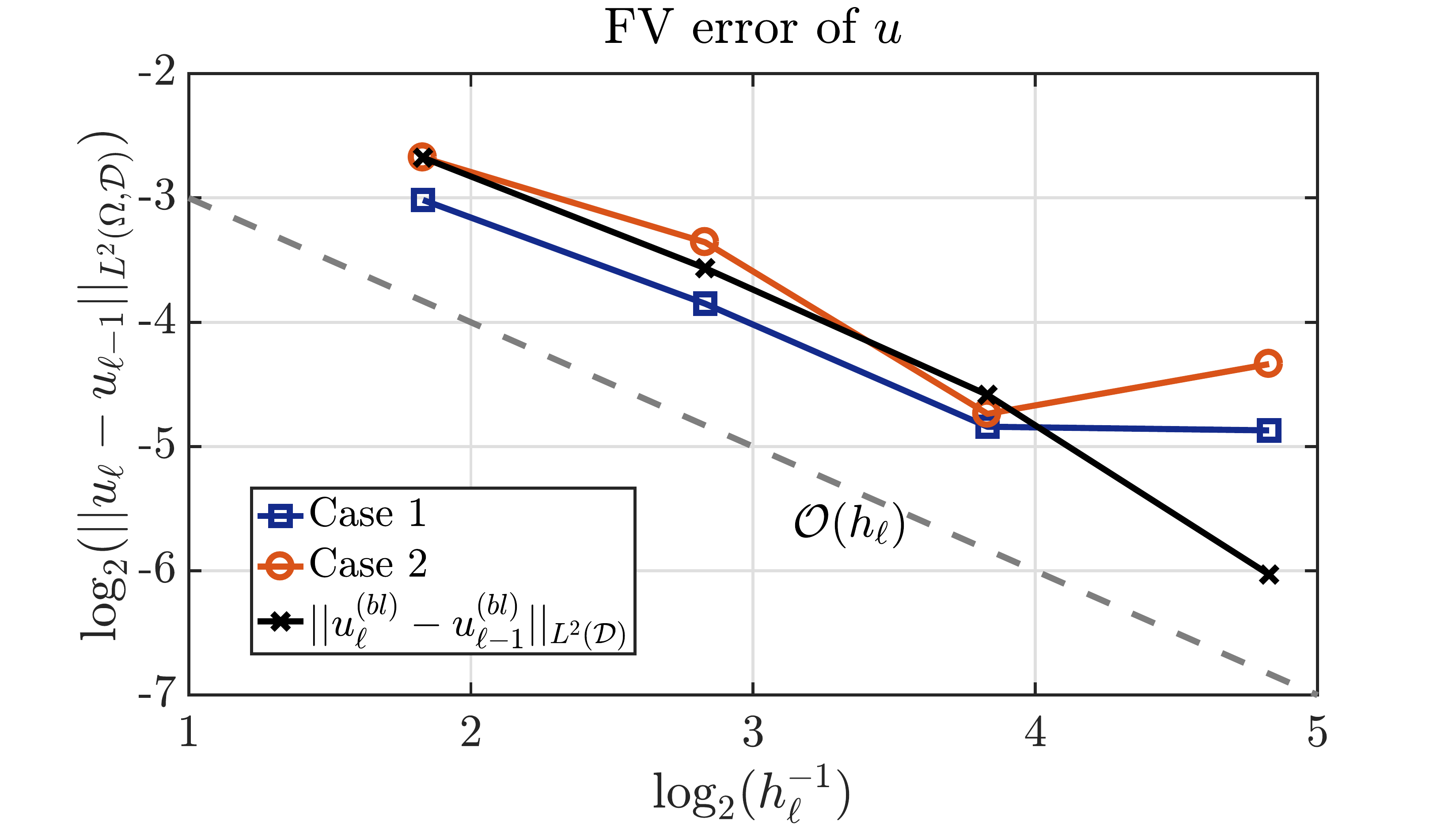}}
\end{subfigure}
\begin{subfigure}[b]{0.49\textwidth}
\centering
\hbox{\hspace{0cm}\includegraphics[clip,  trim=1cm 0cm 0cm 0cm,scale= 0.27]{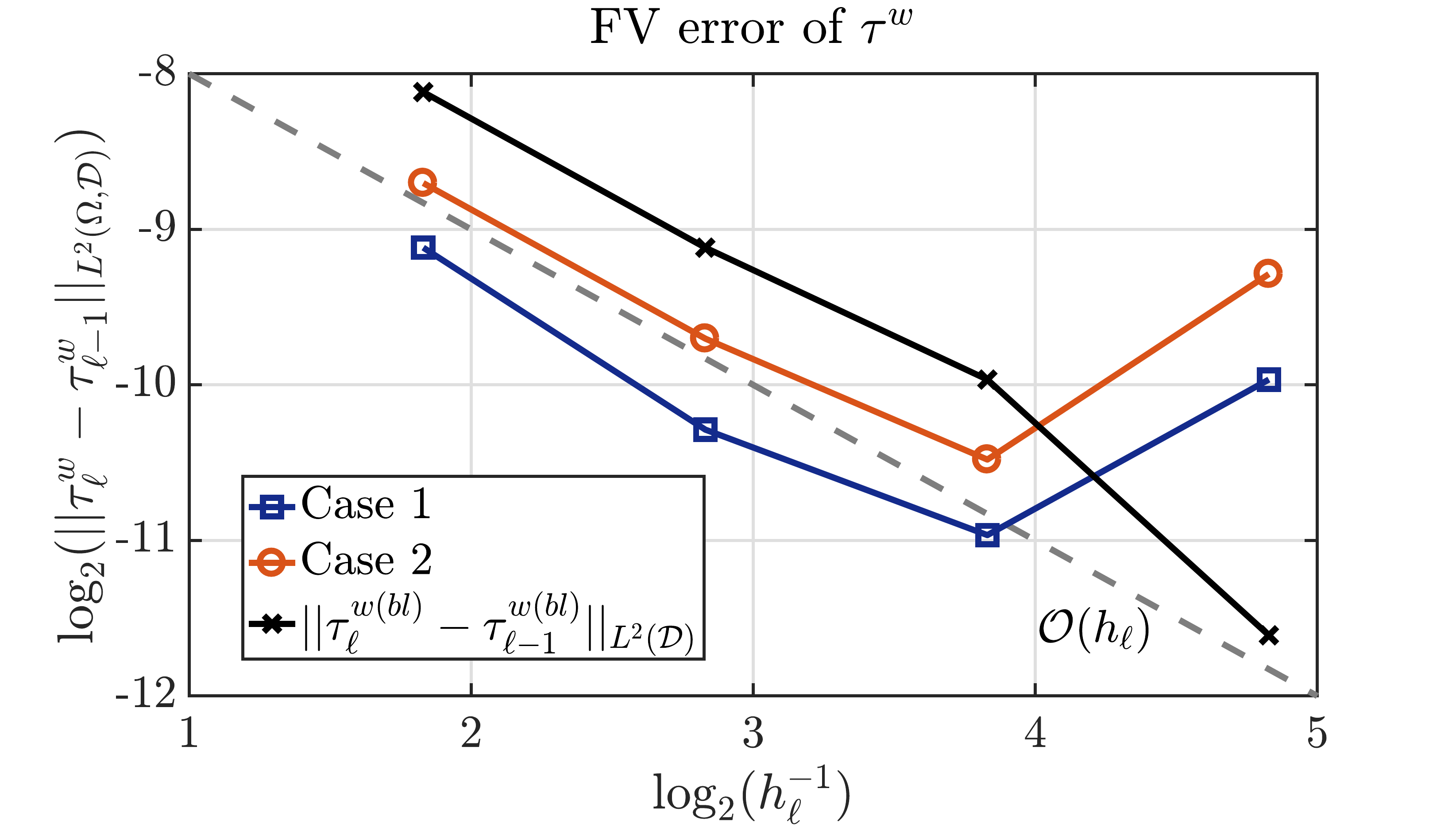}}
\end{subfigure}
\caption{Convergence of the FV error with levels for the RRST model along with the error in baseline solution. The dotted line depicts $\mathcal{O}(h_\ell)$ convergence.}\label{bias_PH_SM2}
\end{figure}
\begin{figure}[H]
\begin{subfigure}[b]{0.49\textwidth}
\centering
\hbox{\hspace{0cm}\includegraphics[clip, trim=1cm 0cm 0cm 0cm,scale = 0.27]{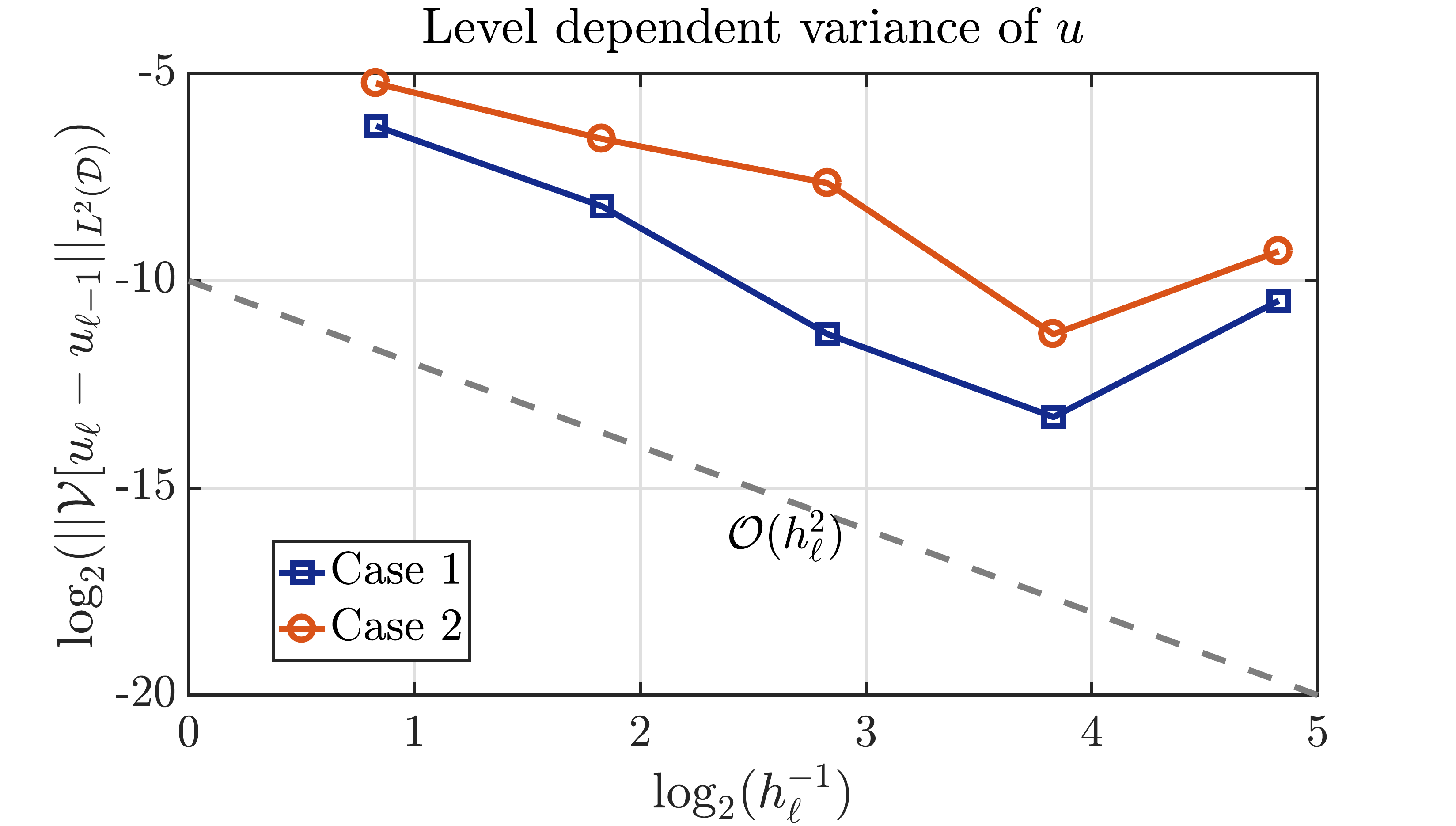}}
\end{subfigure}
\begin{subfigure}[b]{0.49\textwidth}
\centering
\hbox{\hspace{0cm}\includegraphics[clip,  trim=1cm 0cm 0cm 0cm,scale = 0.27]{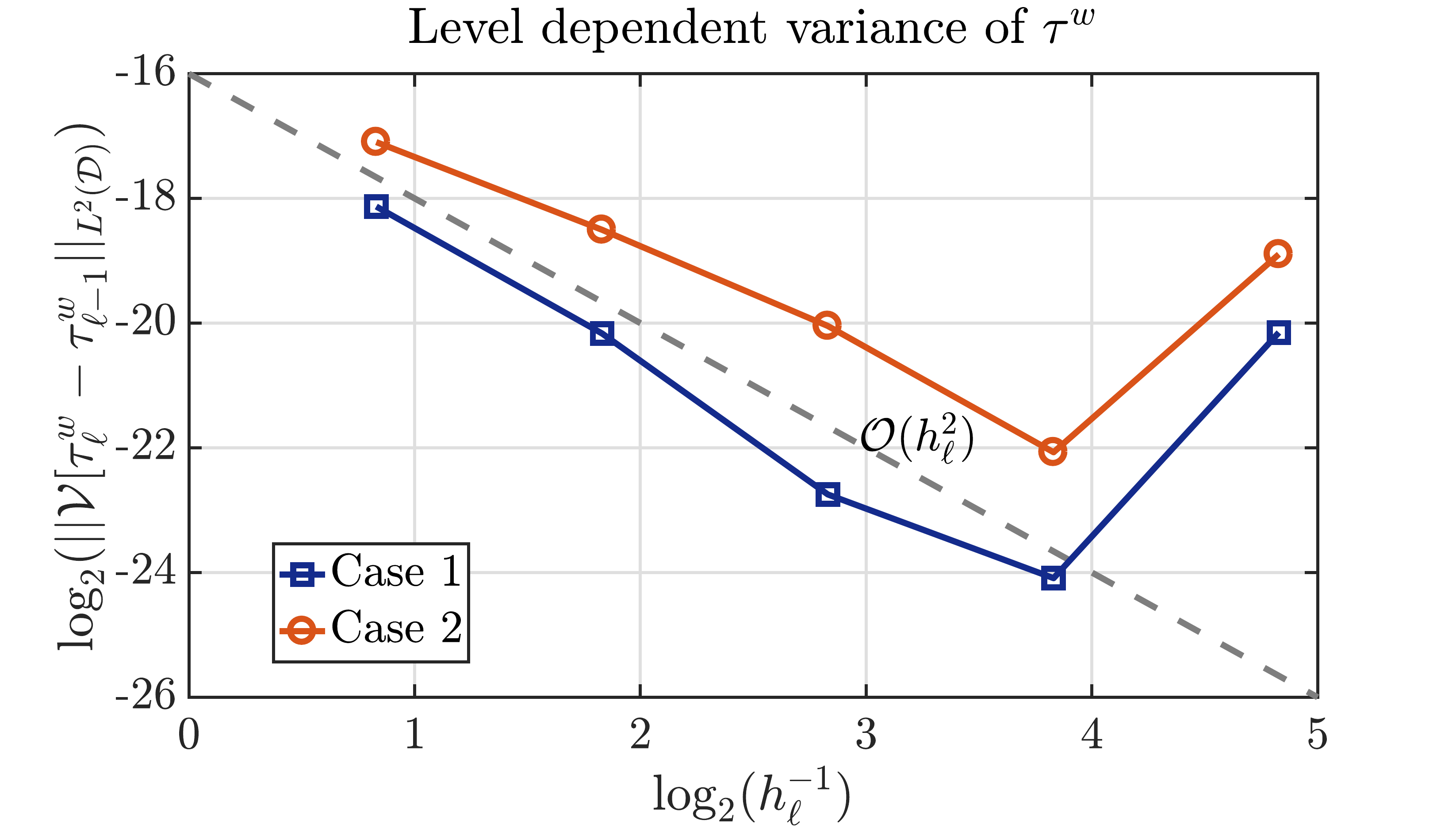}}
\end{subfigure}
\caption{Convergence of the level-dependent variance for the RRST model. The dotted line depicts an $\mathcal{O}(h_\ell^2)$ convergence.}\label{var_PH_SM2}
\end{figure}
As we have similar rates as for the $\alpha,\beta,\gamma$ as the REV model, we use the same number of MLMC samples, from Table \ref{MLMC_samples_PH}. Similarly, for the plain MC method $8,32,128$ samples are used for the $16\times24$,$32\times48$ and $64\times96$ grids, respectively. The reference solution for the mean and variance, $\mathpzc{E}_{ref}[u]$ and $\mathcal{V}_{ref}[u]$, are based on a 4-level Monte Carlo estimator as the fifth level does not provide any improvement in the accuracy (for the considered solver). In Figs. \ref{erel_mean_PH_SM2} and \ref{erel_var_PH_SM2}, we show the mean relative errors and cost scaling for the mean and variance for Case 1.  The speedup is similar to the REV model and close to the theoretically predicted rates. 
 \begin{figure}[H]
\begin{subfigure}[b]{0.49\textwidth}   
            \centering 
      {\includegraphics[clip,  trim=1cm 0cm 0cm 0cm,scale=0.27]{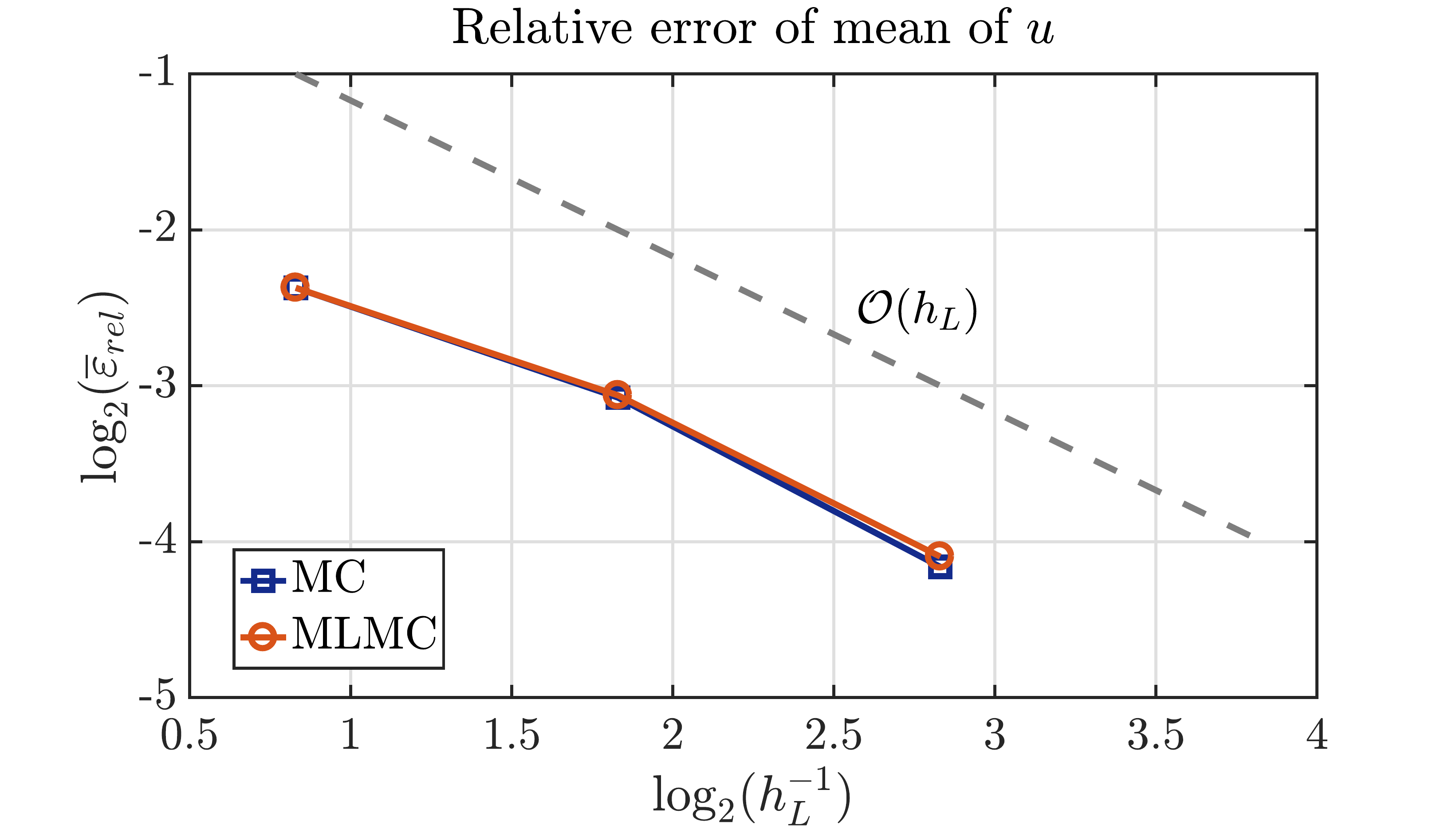}}
        \end{subfigure}
        \begin{subfigure}[b]{0.49\textwidth}   
            \centering 
 {\includegraphics[clip,  trim=1cm 0cm 0cm 0cm,scale=0.27]{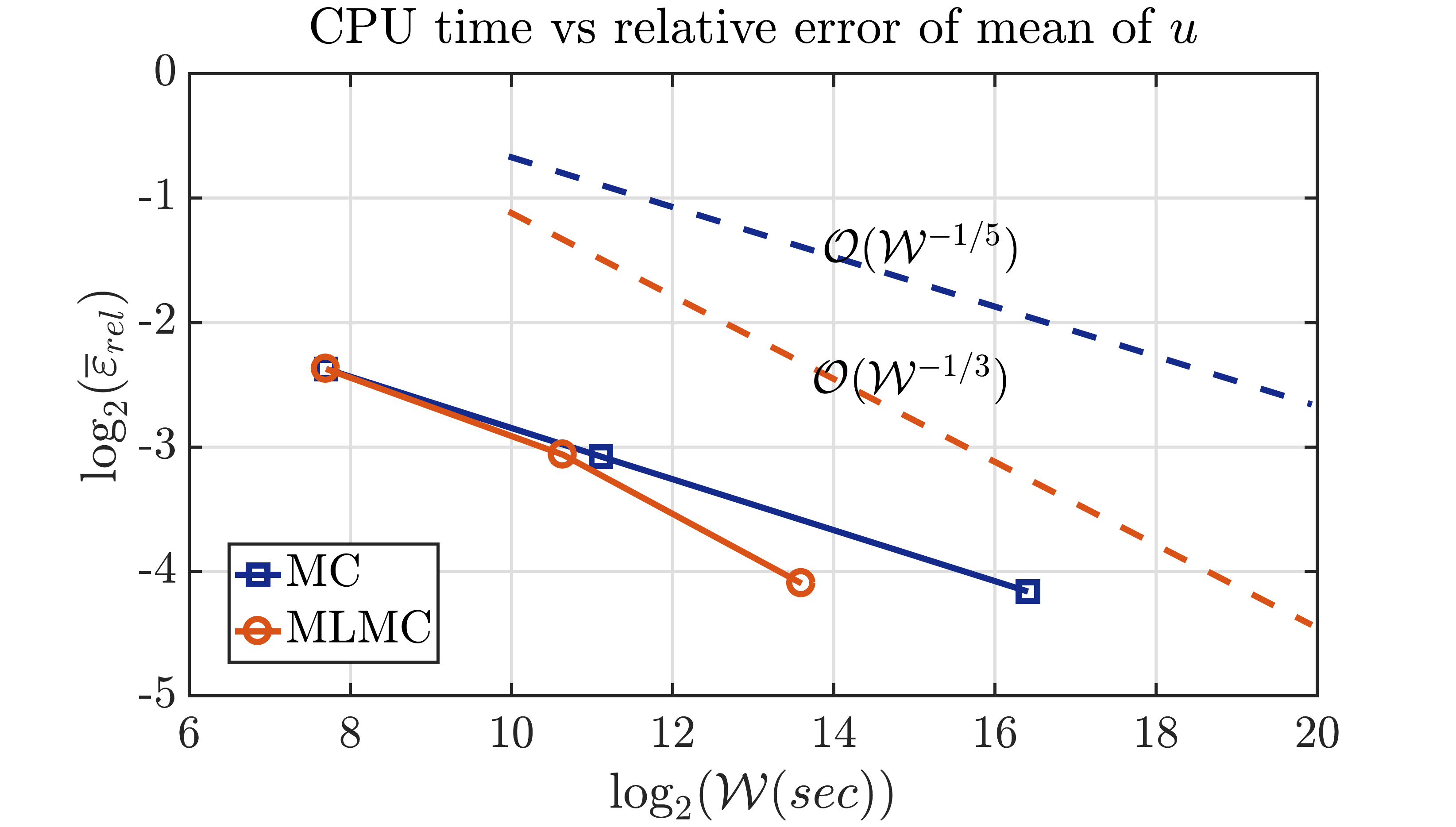}}
        \end{subfigure}
\caption{(Left) Comparison of the mean relative error $\overline{\varepsilon}_{rel}$ in the expected value of $u$ for different  meshes for Case 1. (Right) Computational work versus accuracy for the MC and MLMC estimators. Dotted lines show the predicted asymptotic cost for the MC (blue) and MLMC (red) estimators.}\label{erel_mean_PH_SM2}
\end{figure}

\begin{figure}[H]
\begin{subfigure}[b]{0.49\textwidth}   
            \centering 
      {\includegraphics[clip,  trim=1cm 0cm 0cm 0cm,scale=0.27]{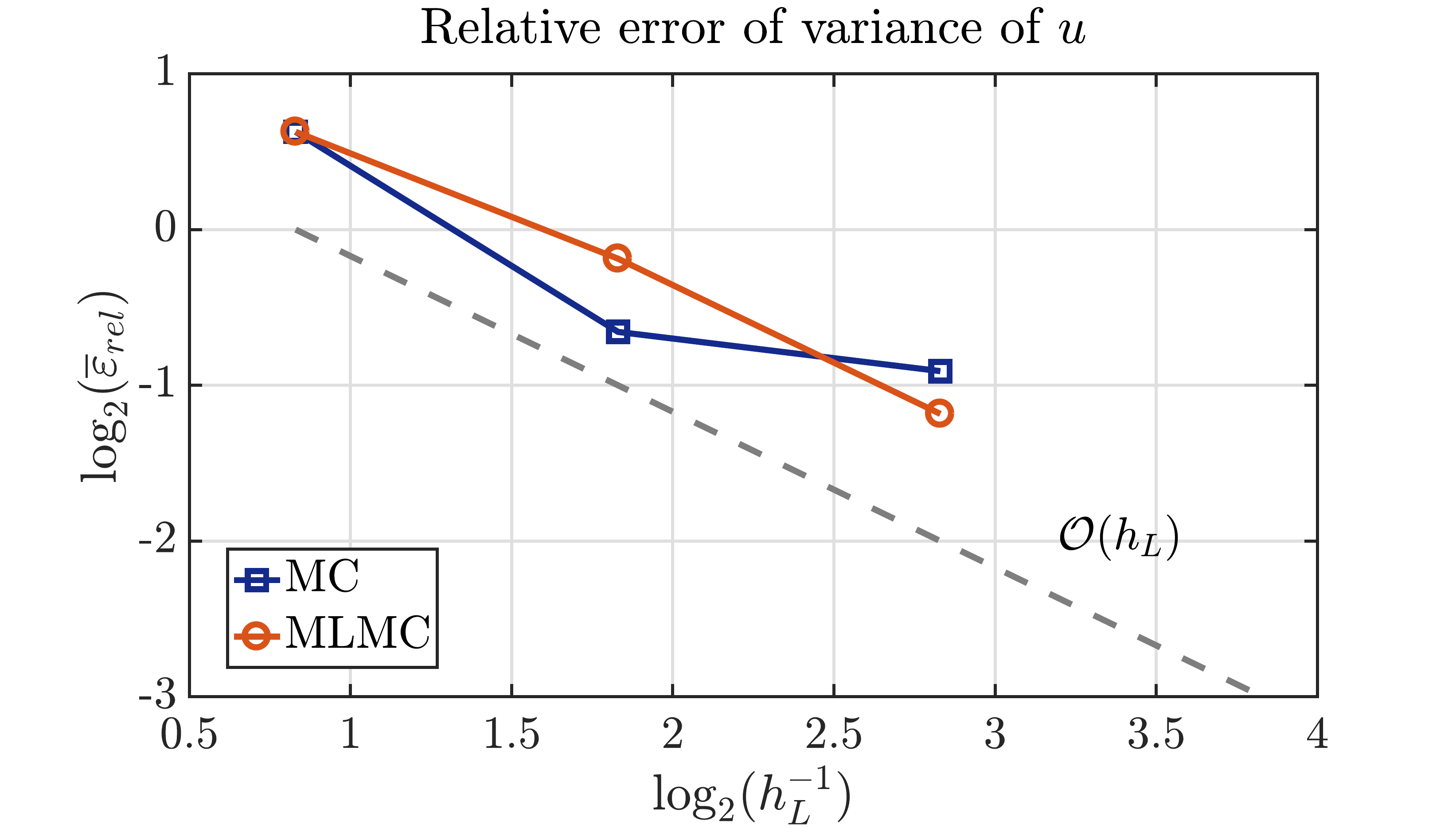}}
        \end{subfigure}
        \begin{subfigure}[b]{0.49\textwidth}   
            \centering 
 {\includegraphics[clip,  trim=1cm 0cm 0cm 0cm,scale=0.27]{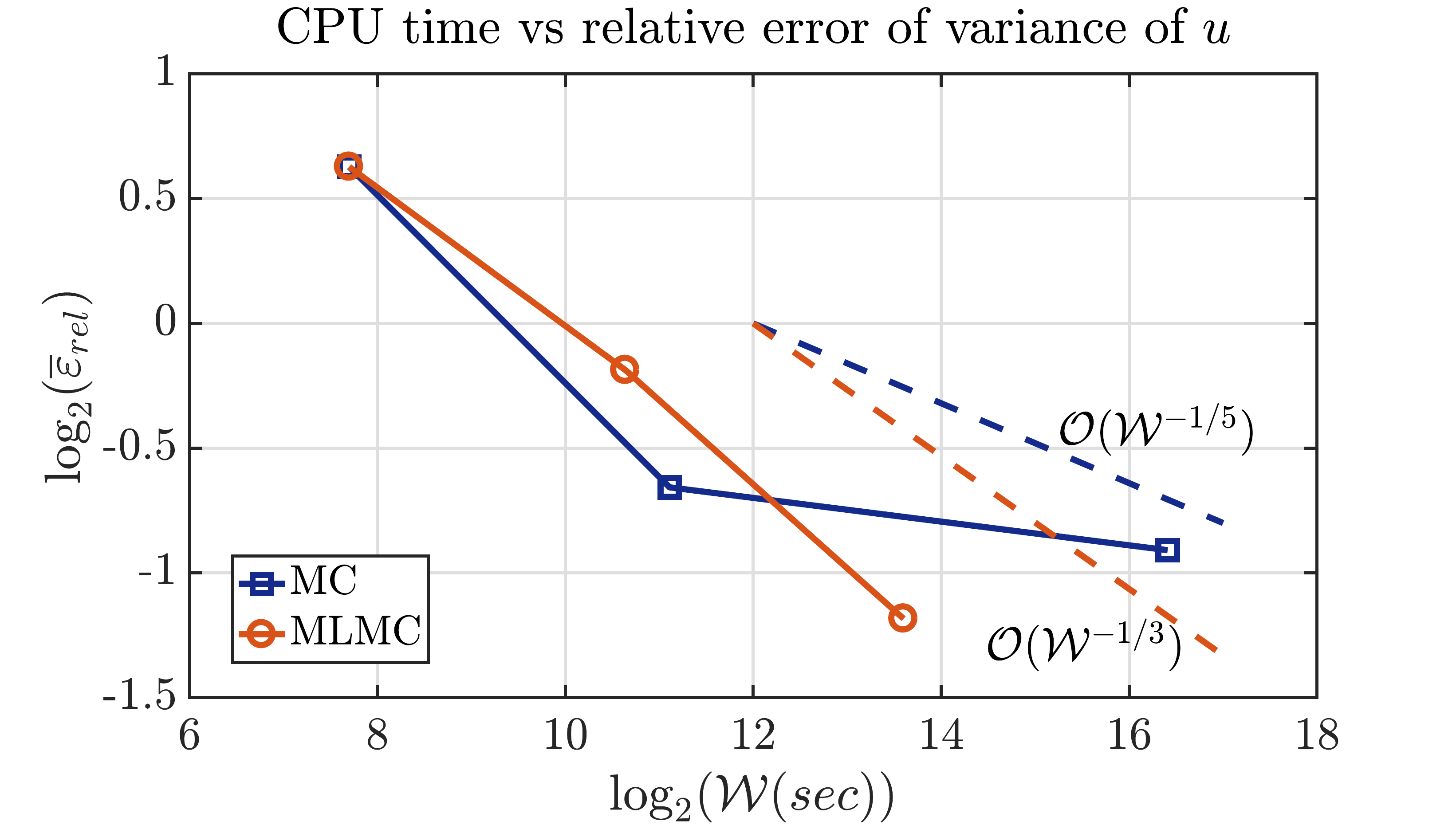}}
        \end{subfigure}
        \caption{(Left) Comparison of the mean relative error $\overline{\varepsilon}_{rel}$ in the variance of $u$ for different meshes for Case 1. (Right) Computational work  versus accuracy for the MC and MLMC estimators.}\label{erel_var_PH_SM2}
\end{figure}
The variance fields computed using the 4-level Monte Carlo for the two cases are presented in Fig. \ref{var_RRST}. Larger variances are observed at locations where the effect of the turbulence is high, for example, near boundary layers and around locations where the flow starts to separate. The mean $\pm$ standard deviation of $u$ at different locations is compared with the baseline and DNS data in Fig. \ref{mean_std_u_SM2}. As expected, a larger enveloping region is obtained for larger dispersion $\delta$. The mean $\pm$ two standard deviations for the wall shear stress is also plotted in Fig. \ref{mean_std_wss_SM2}. Again, a high variation is observed near the reattachment points obtained from the RANS simulation. We see that the DNS data falls within 2 standard deviations for both cases. We remind readers that the standard deviation observed are underestimated as the random tensor only contributes $60\%$ of the propagated Reynolds stress tensor. For both quantities of interest, the observed means are very close to the baseline RANS solution, possibly indicating approximately linear dependence of $u$ on the randomized RST.
\begin{figure}[H]
        \begin{subfigure}[b]{0.51\textwidth}   
            \centering 
           \hbox{\hspace{-0.5cm}\includegraphics[clip, trim=1cm 0cm 0cm 0cm,scale=0.24]{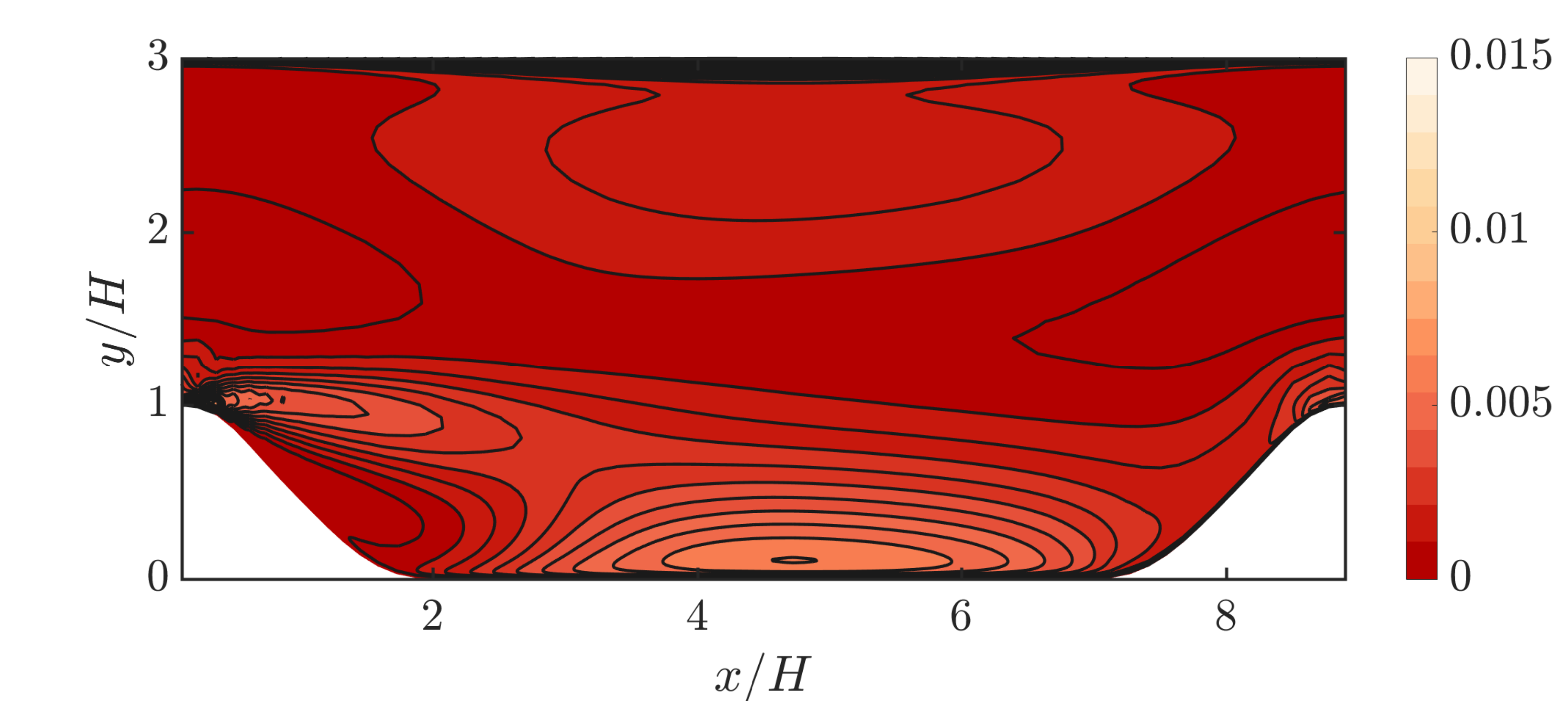}}
         \caption{Case 1}            
        \end{subfigure}
        \begin{subfigure}[b]{0.47\textwidth}   
            \centering 
             \hbox{\hspace{-0.5cm}\includegraphics[clip, trim=1cm 0cm 0cm 0cm,scale=0.24]{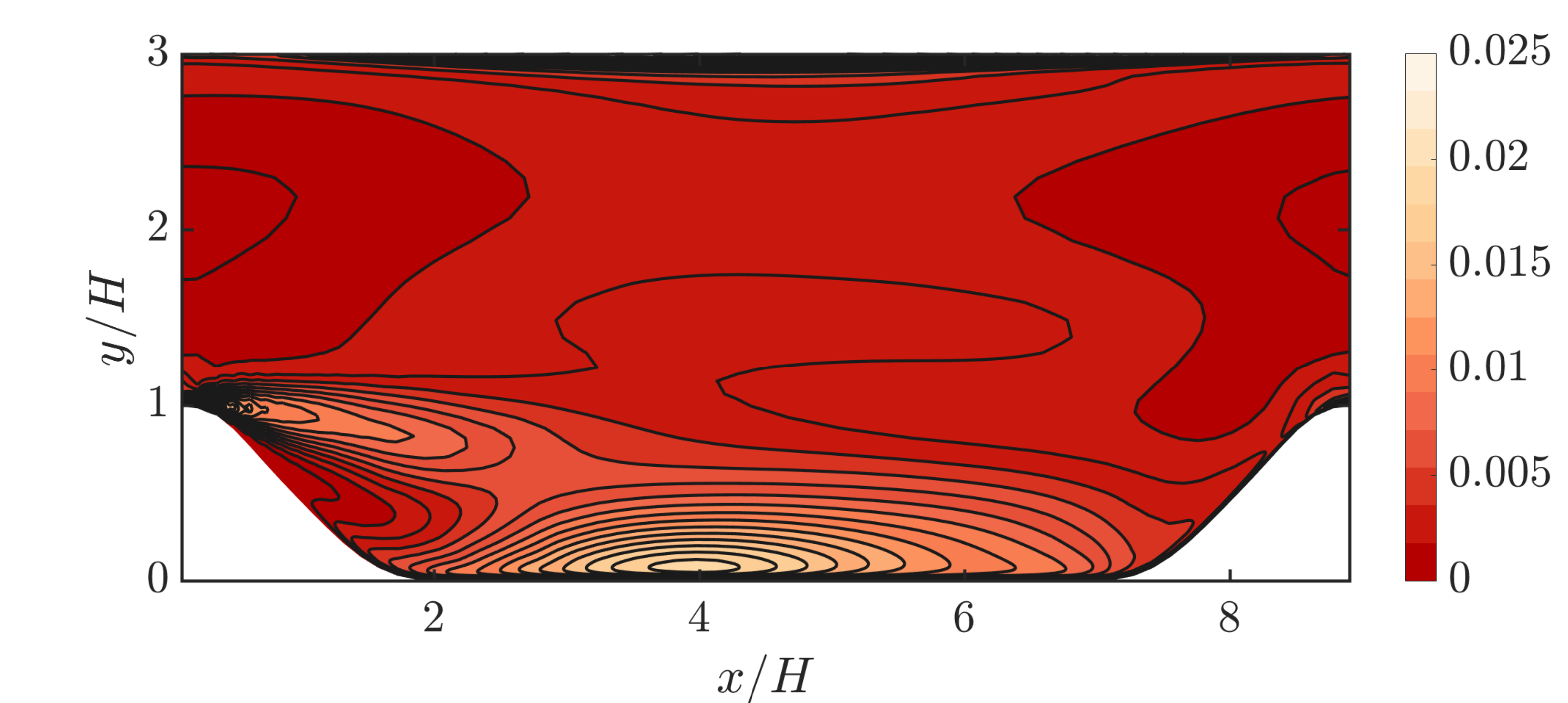}}
                        \caption{Case 2}    
        \end{subfigure}
\caption{Variance field $\mathcal{V}^{ML}_L[u_L]$ for the streamwise velocity $u$ computed using the 4-level estimator.}\label{var_RRST}
\end{figure}
\begin{figure}[H]
        \begin{subfigure}[b]{0.98\textwidth}   
            \centering 
            {\includegraphics[clip, trim=2.8cm 0cm 4cm 0cm,width=.75\textwidth]{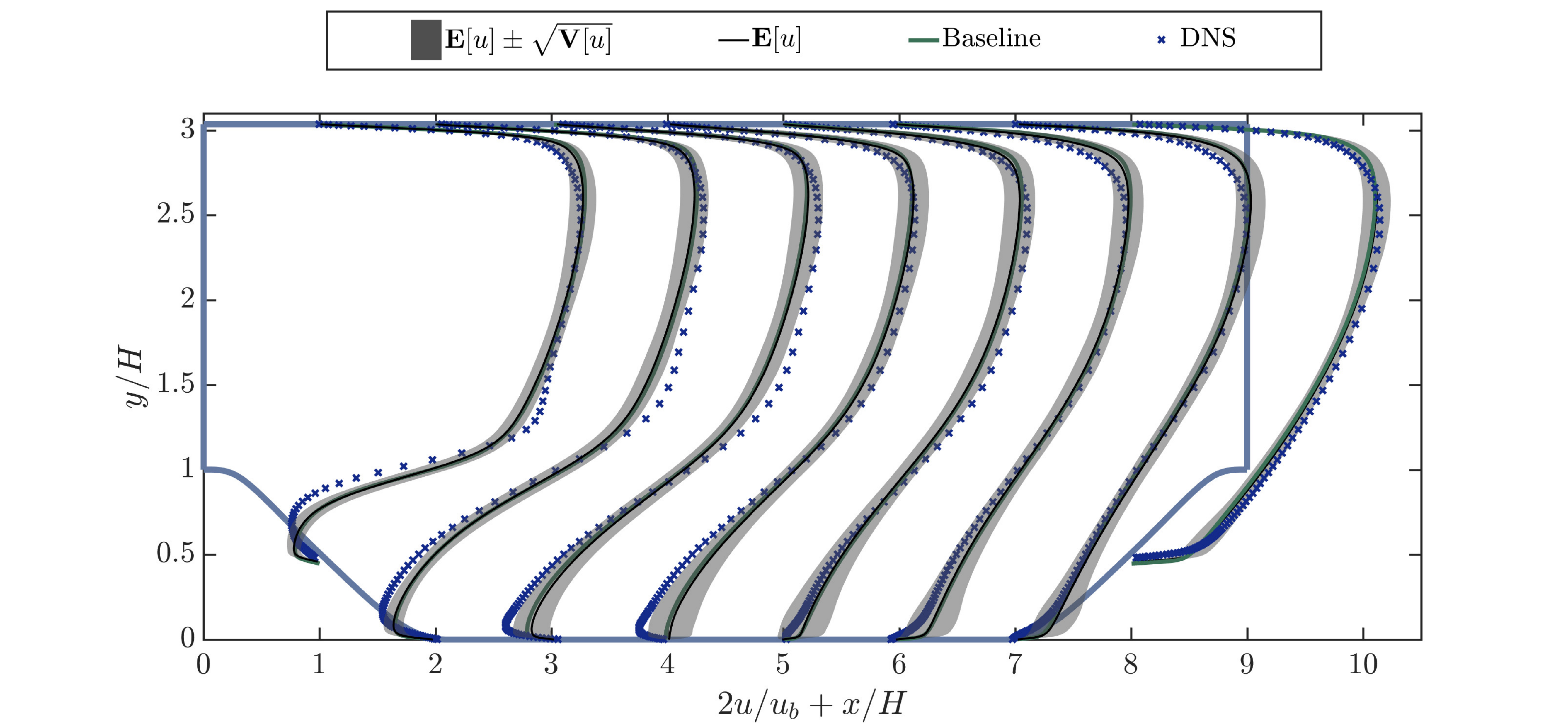}}
            \caption{Case 1}
        \end{subfigure}
        
        \vspace{1cm}
        \begin{subfigure}[b]{0.98\textwidth}   
            \centering 
            \includegraphics[clip, trim=2.8cm 0cm 4cm 0cm,width=.75\textwidth]{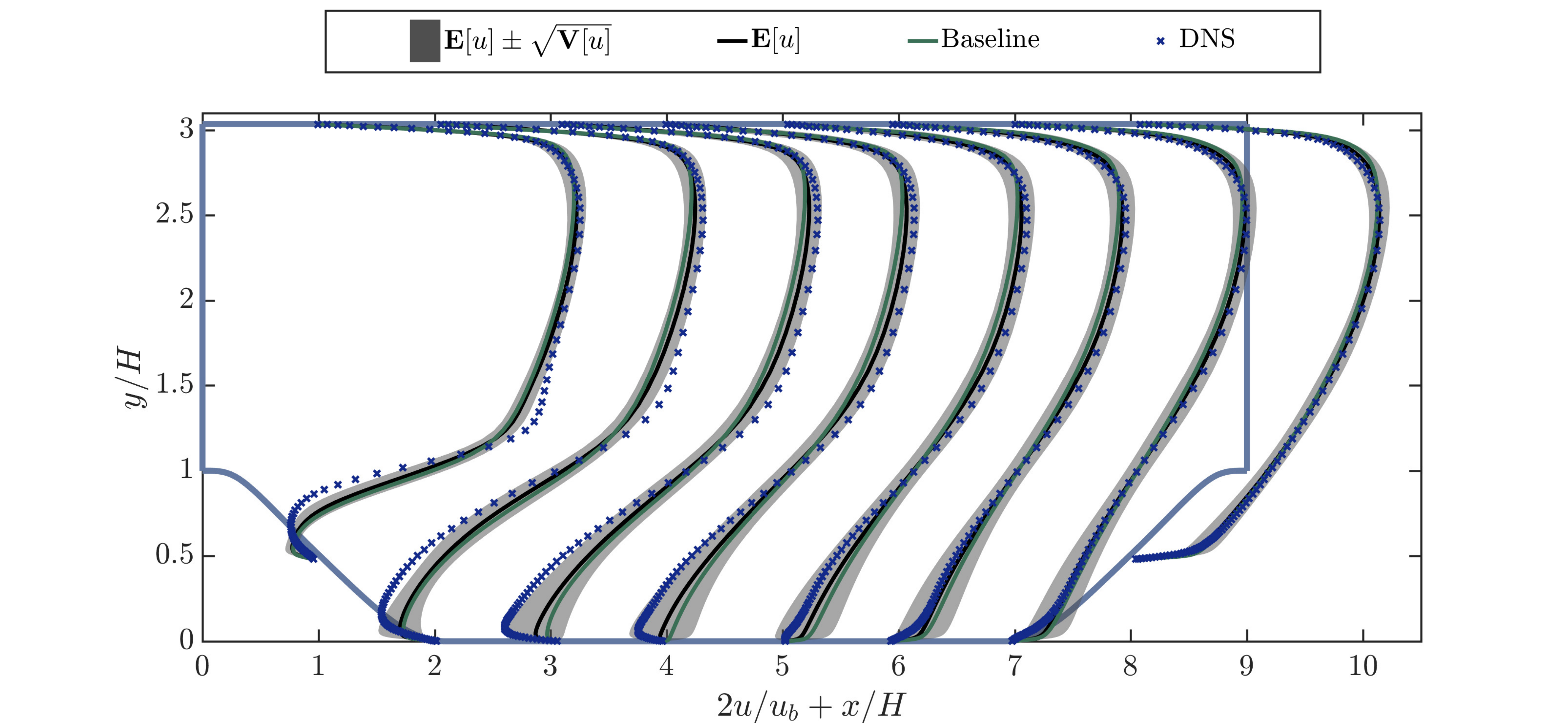} 
            \caption{Case 2}
        \end{subfigure}
\caption{Mean and variance of  the streamwise velocity computed using the 4-level estimator and comparison with DNS data at locations $x/H=1,2,3,...,8$. Velocities are scaled by a factor of two to facilitate visualization.}\label{mean_std_u_SM2}
\end{figure}
\begin{figure}[H]
        \begin{subfigure}[b]{0.98\textwidth}   
            \centering 
            {\includegraphics[clip, trim=2cm 0cm 2cm 0cm,width=0.9\textwidth]{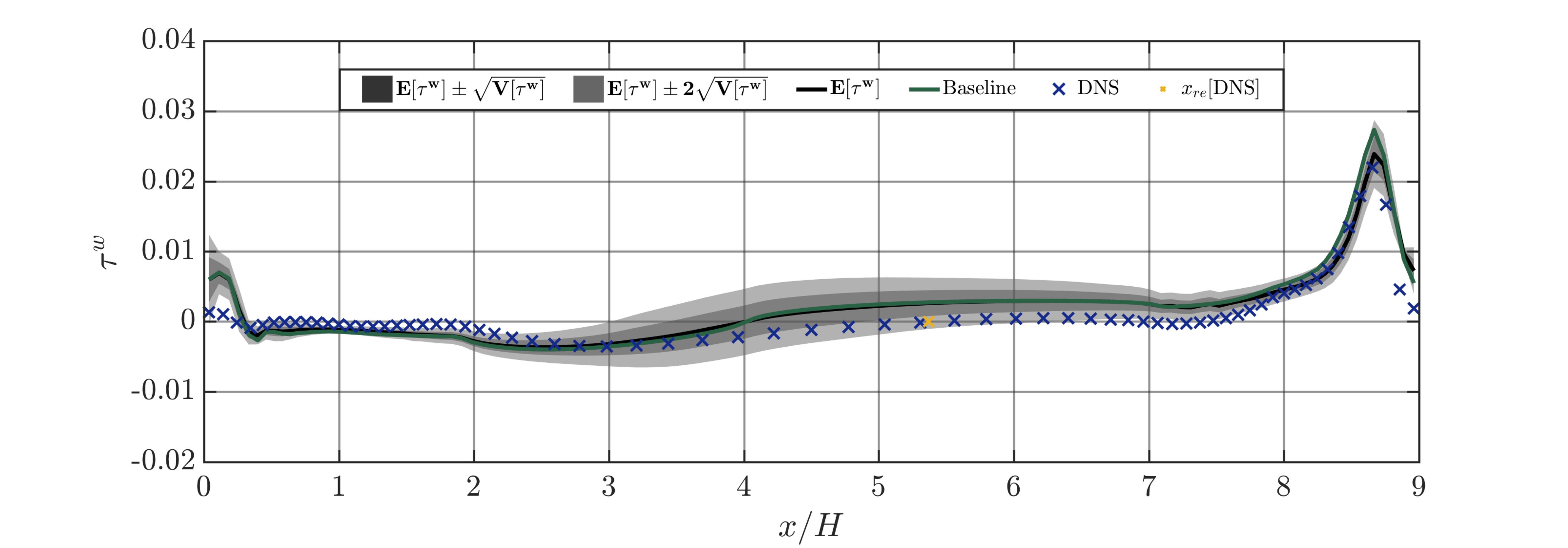}} 
            \caption{Case 1}
        \end{subfigure}
                
        \vspace{1cm}
        \begin{subfigure}[b]{0.98\textwidth}   
            \centering 
            {\includegraphics[clip, trim=2cm 0cm 2cm 0cm,width=0.9\textwidth]{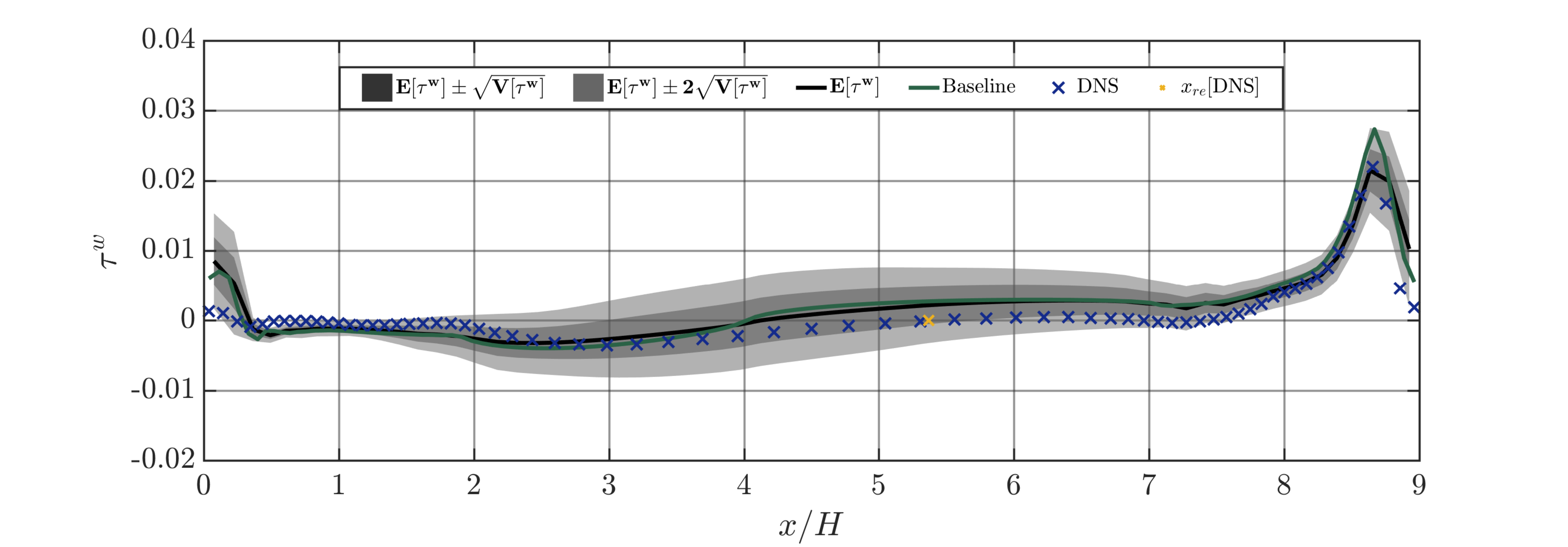}}
            \caption{Case 2}
        \end{subfigure}
\caption{Mean and variance of the wall shear stress $\tau^w$ computed using the 5-level estimator and comparison with DNS data.}\label{mean_std_wss_SM2}
\end{figure}

\section{Summary and conclusions}
In this paper, we undertook first steps towards solving high-dimensional stochastic formulations of RANS turbulence models using the multilevel Monte Carlo method.  We demonstrated the efficiency of the MLMC method using two stochastic models based on a perturbation of the baseline eddy viscosity field and the baseline Reynolds stress tensor field. The MLMC estimator proposed is only slightly more involved than the plain MC estimator but results in a large speedup. The savings afforded by employing coarser levels allowed us to incorporate finer meshes thereby enabling the computation of the mean and variance with higher accuracy. We demonstrated that for QoIs for which the level-dependent variance decays slower than the growth rate of computational cost with level (i.e. $\gamma>\beta$), an optimal MLMC estimator can be achieved. For two benchmarking problems, we utilized a nested and geometric hierarchy of grids. This is not a necessary criterium and a valid MLMC estimator can be constructed on any hierarchy that consists of levels with increasing cost and accuracy. A more sophisticated approach to obtain finer levels in the MLMC hierarchy can be based on adaptively refining the mesh in regions where a large numerical error is observed. We would like to point out that there is a negligible difference in terms of the computational cost between, the REV and RRST models, but the implementation of the latter model is more involved. Especially, obtaining a robust solver with respect to random Reynolds stress tensors is challenging. The continuation solver proposed in this paper is moderately successful but suffers from convergence issues on very fine grids as well as when the random tensors are sampled from high-variance parameter sets.

This article presented the MLMC method as an efficient uncertainty propagation tool without taking into account any available data. A natural extension would be the development of multilevel variants of the Markov Chain Monte Carlo (MCMC) method to obtain a data-informed prediction \cite{Dodwell}. For such algorithms, the random matrix approach can act as a better prior than the random eddy viscosity model as it circumvents the Boussinesq approximation. Currently, to the authors' knowledge, no data-based approach exists that takes into account the uncertainty in the full Reynolds tensor field. This idea will be actively explored in the upcoming works.
\section*{Acknowledements}
This research is funded by the Shell-NWO/FOM programme `Computational Sciences for Energy Research' (CSER) under the research grant 14CSER004. The authors are also grateful to Prof. C. W. Oosterlee for his insightful comments and suggestions.
\section*{Appendix}
\subsection*{A1:Projection of Reynolds Stresses on a Barycentric triangle}
Reynolds stresses can be divided into an isotropic part $\frac23k\delta_{ij}$ and an normalized anisotropic component given by
\begin{equation}
A_{ij} := \frac{R_{ij}}{2k} - \frac{\delta_{ij}}{3}, 
\end{equation}
\begin{equation}
\Rightarrow A_{ij}\in
\begin{cases}
[-1/3,2/3]\text{ for } i=j,\\
[-1/2,1/2]\text{ for } i\neq j,
\end{cases}
\end{equation}
forming the entries of a symmetric and deviatoric anisotropy tensor $\mathbf{A}$. Utilizing the eigenvalue decomposition, the anisotropy tensor $\mathbf{A}$, can be expressed as
\begin{equation}
\mathbf{A} = \mathbf{V}\boldsymbol{\Lambda}\mathbf{V}^T,
\end{equation}
where $\mathbf{V} = [v_1,v_2,v_3]$ with three mutually orthonormal eigenvectors $v_i$ and the corresponding eigenvalue matrix $\boldsymbol{\Lambda} = \text{diag}[\lambda_1,\lambda_2,\lambda_3]$ with $\lambda_1+\lambda_2+\lambda_3=0$ and ordering such that $\lambda_1>\lambda_2>\lambda_3$. 

In physical terms, quantities $k$,$\mathbf{V}$ and $\boldsymbol{\Lambda}$ represent the magnitude, shape and orientation of the Reynolds stress, respectively. The state of the turbulence anisotropy can be visualized using a barycentric triangle \cite{Banerjee}. This requires mapping the eigenvalues to the barycentric coordinates, $C_{1c},C_{2c},C_{3c}$, using linear relations:
\begin{equation}
C_{1c} = \lambda_1-\lambda_2,\qquad C_{2c} = 2(\lambda_2-\lambda_3),\qquad C_{3c} = 3\lambda_3+1,\quad\Rightarrow C_{1c}+C_{2c}+C_{3c}=1.
\end{equation}  
Reynolds stress anisotropy is said to attain a limiting state when one of these components equals 1. Therefore, $C_{1c}=1$ represents 1-component turbulence, $C_{2c}=1$ represents 2-component turbulence and $C_{3c}=1$ represents 3-component turbulence. One can express the anisotropy states in Cartesian coordinates using a barycentric triangle with the vertices $(x_{1c},y_{1c}),(x_{2c},y_{2c})$ and $(x_{3c},y_{3c})$, corresponding to the three limiting states. Now, any anisotropy tensor can be projected into barycentric triangle via the convex combination of the three limiting states: 
\begin{eqnarray}
x=x_{1c}C_{1c}+x_{2c}C_{2c}+x_{3c}C_{3c},\\
y=y_{1c}C_{1c}+y_{2c}C_{2c}+y_{3c}C_{3c}.
\end{eqnarray}
This transformation enables us to analyse the states of the Reynolds stresses generated using the random matrix approach. These perturbed Reynolds stresses should lie on, or within, this triangle to be physically realizable. The contours in Fig. \ref{bary_PDF} are generated by making bins of equal size inside the barycentric triangle and plotting the normalized frequency for each bin. 
\subsection*{A2: Spectral generator for Gaussian random fields and covariance upscaling}
As the random eddy viscosity field and the components of the random Reynolds stress tensor need to be sampled many times, a fast sampling algorithm is necessary to obtain an efficient (ML)MC estimator. There are a number of spectral generators available in the literature \cite{Ravalec2000,RF2,RF4} that exploit the efficiency of the FFT algorithm to achieve fast sampling of Gaussian random fields. We use the Fast Fourier Transform moving average (FFT-MA) technique from \cite{Ravalec2000}. 
Given a covariance matrix $\mathbf{C}_{\ell}$ computed on the mesh $\mathcal{D}_{\ell}$, a standard way to sample correlated Gaussian random vectors $\mathbf{z}_\ell(\omega)$ is via a Cholesky decomposition $\mathbf{C}_\ell  = \mathbf{L}_\ell\mathbf{L}_\ell^T$ and use $\mathbf{z}_\ell = \mathbf{L}\mathbf{y}_\ell$ where $\mathbf{y}_{\ell}$ is a vector of i.i.d. samples from the standard normal distribution. This procedure requires a large storage as well as an expensive matrix-vector product for each sample of $\mathbf{z}_\ell$. The FFT-MA method is based on a decomposition of the covariance function $C$ as a convolutional product of some function $S$ and its transpose $S'$ ($S'(x) = S(-x)$). We can express this decomposition as
\begin{equation}\label{conv2}
\mathbf{c}_\ell = \mathbf{s}_\ell* \mathbf{s}'_\ell,
\end{equation}
where $\mathbf{c}_\ell,\mathbf{s}_\ell$ are vectors obtained by evaluating $C$ and $S$, respectively at grid points of the mesh $\mathcal{D}_\ell$. A correlated random vector $\mathbf{z}_\ell$ can now be synthesized by using the convolution product
\begin{equation}\label{conv2}
\mathbf{z}_\ell= \mathbf{s}_\ell*\mathbf{y}_\ell.
\end{equation}
The key idea of the FFT-MA approach is to perform the above computations in the frequency domain. The first task is to extend the vector $\mathbf{c}_\ell$ to obtain a periodic signal, which is also real, positive and symmetric, see, for instance \cite{RF2}, for details. As a result $\mathbf{s}_\ell$ is also real, positive and symmetric and $\mathbf{s}_\ell = \mathbf{s}'_\ell$. As a convolution product is equivalent to component-wise product in the frequency domain, we  can use
\begin{equation}\label{product1}
\mathcal{F}(\mathbf{c}_\ell) = \mathcal{F}(\mathbf{s}_\ell)\cdot \mathcal{F}(\mathbf{s}_\ell) \implies \mathcal{F}(\mathbf{s}_\ell) = \sqrt{\mathcal{F}(\mathbf{c}_\ell)},
\end{equation}
where $\mathcal{F}$ denotes the discrete FFT and $\cdot$ denotes component-wise multiplication. Here, the component-wise square-root operation does not pose any problems as the power spectrum $\mathcal{F}(\mathbf{c}_\ell)$ is real, positive and symmetric. Next, we express the convolution product in \eqref{conv2} as a vector-vector product in frequency domain as
\begin{equation}\label{product2}
\mathcal{F}(\mathbf{z}_\ell) = \mathcal{F}(\mathbf{s}_\ell*\mathbf{y}_\ell)=\mathcal{F}(\mathbf{s}_\ell)\cdot\mathcal{F}(\mathbf{y}_\ell).
\end{equation}
Finally, the correlated random field is obtained by an inverse fast Fourier transform
\begin{equation}
\mathbf{z}_\ell = \mathcal{F}^{-1}(\mathcal{F}(\mathbf{s}_\ell)\cdot\mathcal{F}(\mathbf{y}_\ell)).
\end{equation} 
Note that due to the periodicity in the covariance vector $\mathbf{c}_\ell$, the resulting random field $\mathbf{z}_\ell$ is also periodic. Therefore, the part of the vector that does not correspond to the physical domain is discarded.

One of the advantages of the FFT-MA algorithm is that the entries of the vector $\mathbf{y}_\ell$ are associated with respective grid points, thus, coarser grid  realizations of the fine grid Gaussian random field $\mathbf{z}_\ell$ can be obtained by locally averaging of the fine grid normally distributed vector $\mathbf{y}_\ell$. As proposed in \cite{mishra2016multi}, an upscaled version $\mathbf{z}_{\ell-1}$ of the fine grid random field $\mathbf{z}_\ell$ can be derived by using multi-dimensional averaging of vector $\mathbf{y}^\ell$. For instance, in two dimensions for a cell-centred grid,
\begin{equation}\label{averaging}
\mathbf{y}^{i,j}_{\ell-1} = \frac{1}{2}(\mathbf{y}^{2i-1,2j-1}_{\ell}+\mathbf{y}^{2i-1,2j}_{\ell}+\mathbf{y}^{2i,2j-1}_{\ell}+\mathbf{y}^{2i,2j}_{\ell}),
\end{equation}
where $i,j$ is the cell index for the mesh $\mathcal{D}_{\ell-1}$. The scaling by a factor 2 is needed to obtain a standard normal distribution for the averaged quantity $\mathbf{y}_{i,j}^{\ell-1}$. The coarser random field  can now be simply assembled as
\begin{equation}\label{cov_upscale}
\mathbf{z}_{\ell-1} = \mathcal{F}^{-1}(\mathcal{F}(\mathbf{s}_{\ell-1})\cdot\mathcal{F}(\mathbf{y}_{\ell-1})).
\end{equation}
As the averaging in \eqref{averaging} smooths out high frequencies, the upscaled version $\mathbf{z}_{\ell-1}$ will also be slightly smoother compared to $\mathbf{z}_{\ell}$. 
%In Fig. \ref{upscaled}, a random realization of $R_{12}$ on $128\times192$ grid and the upscaled version on $64\times96$ grid is shown for comparison.
%\begin{figure}[H]
%\begin{subfigure}[b]{0.49\textwidth}
%\begin{center}
%\hbox{\hspace{1cm}\includegraphics[clip, trim=0cm 0cm 0cm 0cm,scale = 0.22]{R12_L4.pdf}}
%\caption{Sample on the fine grid $128\times192$}
%\end{center}
%\end{subfigure}
%\begin{subfigure}[b]{0.49\textwidth}
%\begin{center}
%\hbox{\hspace{1cm}\includegraphics[clip,  trim=0cm 0cm 0cm 0cm,scale= 0.22]{R12_L3.pdf}}
%\caption{Upscaled version on the coarse grid $64\times96$ }
%\end{center}
%\end{subfigure}
%\caption{Comparison of random realization of $R_{12}$ generated using parameters from Case 2.}\label{upscaled}
%\end{figure}
\section*{References}
%\bibliography{references}
\bibliography{\myreferences}
\end{document}